\documentclass[a4paper,11pt]{article}
\usepackage{jcappub}
\usepackage{amsmath}
\usepackage{subcaption}
\usepackage{mathtools}
\usepackage{bm}
\usepackage[normalem]{ulem}
\usepackage{multirow}
\usepackage{comment}
\usepackage{xfrac}
\usepackage{enumitem}
\usepackage[pscoord]{eso-pic}
\usepackage{url}
\usepackage{tikz}
\usetikzlibrary{shapes.geometric, arrows, positioning}
\usepackage{ragged2e}
\usepackage{blindtext}
\usepackage{circuitikz}
\tikzstyle{process} = [rectangle, rounded corners, minimum width=3cm, minimum height=1.5cm, text centered, draw=black, fill=blue!20]
\tikzstyle{arrow} = [thick,->,>=stealth]
\tikzstyle{title} = [rectangle, text centered, minimum height=1cm]

\newcommand{\placetextbox}[3]{
	\setbox0=\hbox{#3}
	\AddToShipoutPictureFG*{
		\put(\LenToUnit{#1\paperwidth},\LenToUnit{#2\paperheight}){\vtop{{\null}\makebox[0pt][c]{#3}}}
	}
}

\newcommand{\Neff}{N_\text{eff}}

\newcommand{\eqsp}{\;\,}

\title{Photo- and Hadrodisintegration constraints on massive relics decaying into neutrinos\\ \vspace{-.64cm}}

\author[a]{S.~Bianco,}
\author[b]{P.~F.~Depta,}
\author[a,c]{J.~Frerick,}
\author[d]{T.~Hambye,}
\author[d]{M.~Hufnagel,}
\author[a]{and K.~Schmidt-Hoberg}

\affiliation[a]{Deutsches Elektronen-Synchrotron DESY, Notkestr.~85, 22607 Hamburg, Germany}
\affiliation[b]{Max-Planck-Institut für Kernphysik, Saupfercheckweg 1, 69117 Heidelberg, Germany}
\affiliation[c]{Dipartimento di Fisica, Sapienza Università di Roma, Piazzale Aldo Moro 5, 00185, Roma, Italy}
\affiliation[d]{Service de Physique Th\'{e}orique, Universit\'{e} Libre de Bruxelles, C.P. 225, B-1050 Brussels, Belgium}

\emailAdd{sara.bianco@desy.de}
\emailAdd{jonas.frerick@uniroma1.it}
\emailAdd{thomas.hambye@ulb.be}
\emailAdd{marco.hufnagel@ulb.be}
\emailAdd{kai.schmidt-hoberg@desy.de}

\abstract{We perform a detailed study of the cosmological constraints on the decay of a relic particle $\phi$ into neutrinos, $\phi \rightarrow \nu \bar{\nu}$, in particular those arising from the observed light-element abundances in the early Universe. Several processes are relevant, including final-state radiation associated with the decay, as well as subsequent interactions of the injected neutrinos with the thermal background neutrinos or between themselves. All processes generically contribute to the production of electromagnetic and often also hadronic material and may therefore induce late-time photodisintegration and hadrodisintegration reactions, i.e.~the destruction of light elements that have previously been formed during BBN. Here, we examine this scenario with a Monte-Carlo inspired probabilistic approach rather than Boltzmann techniques, taking into account all of these different reactions as well as their interplay. We find the resulting constraints to be very significant, covering a broad range of previously unexplored masses and lifetimes of the relic source particle.}

\begin{document}

\vspace*{0.1 cm}
\placetextbox{0.85}{0.97}{\small DESY-25-056, ULB-TH/25-03}

\maketitle
\flushbottom

\section{Introduction}
\label{sec:intro}

Many beyond the Standard Model (SM) scenarios suggest the existence of new unstable particles that were present during the early Universe. This is, for example, the case in dark-sector (DS) models, where in addition to dark matter (DM), other particles are present, often mediating interactions between the DM and the SM or within the DS itself. Specific examples include DM scenarios where communication with the SM proceeds via a portal interaction~\cite{Batell:2009di}, or where the exchange of a light mediator induces sizeable DM self-interactions~\cite{Feng:2009hw,Buckley:2009in,Loeb:2010gj,Tulin:2013teo,Bringmann:2016din, Hambye:2019tjt}.

These additional particles will in general significantly impact the cosmological evolution and the associated observables. In particular, if they are unstable and decay with a lifetime shorter than the age of the Universe, they will generally affect predictions for Big Bang Nucleosynthesis (BBN) as well as for the Cosmic Microwave Background (CMB), resulting in constraints on their abundance, mass and lifetime.
The strongest constraints typically arise if the decay proceeds into electromagnetic (EM) material (such as photons or charged leptons) and/or hadronic material (such as quarks or gluons), see e.g.~\cite{Slatyer:2012yq,Poulin:2016anj}. Thus, to avoid too strong constraints in various scenarios (see e.g.~\cite{Bringmann:2016din,Hambye:2019tjt}), one could assume that the decay instead proceeds predominantly into neutrinos.
However, this does not imply that the constraints will become completely irrelevant, because a fraction of the neutrino energy will unavoidably still be converted into electromagnetic and/or hadronic material.
In fact, it is known that heavy new particles decaying into neutrinos, inevitably produce additional on- or off-shell EW gauge bosons on top of the primary neutrinos. The effects of the resulting shower have been studied before, in particular for the EM case (see e.g.~\cite{Kanzaki:2007pd,Slatyer:2016qyl,Stocker:2018avm,Liu:2019bbm,Cirelli2011PPPC,Hambye:2021moy}).
Other works that study the direct or indirect injection of neutrinos into the primordial plasma, include~\cite{Acharya:2020gfh,Ovchynnikov:2024xyd,Akita:2024nam,Akita:2024ork}.

In this work, we perform a detailed study of the decay of a relic particle $\phi$ into neutrinos, $\phi \rightarrow \nu \bar{\nu}$, carefully evaluating the limits arising from the observed light-element abundances, while also including effects that have previously been neglected in the literature. We concentrate on lifetimes larger than about $10^4\,\mathrm{s}$, focusing in particular on the late-time disintegration of the light elements previously synthesised during BBN. In addition to the effect of the electroweak (EW) shower from final-state radiation (FSR) mentioned above, we also properly take into account the effect of subsequent interactions of the injected neutrinos, specifically for masses below or around the EW scale. Of particular relevance are interactions with a cosmic background neutrino, which we refer to as \emph{thermal scattering} (see \cite{Kawasaki:1994bs,Kanzaki:2007pd,Acharya:2020gfh} for early work), as well as interactions among the injected neutrinos themselves, which we refer to as \emph{non-thermal scattering}. 
While thermal scatterings may readily induce electromagnetic material, non-thermal scatterings typically have a larger energy available and therefore also result in hadronic injections. All processes may therefore induce late-time photodisintegration reactions, while FSR and non-thermal scatterings additionally result in hadrodisintegration as well. 
Here,  we carefully examine this scenario, using the most recent nuclear abundances and rates, and taking into account all the different reactions as well as their interplay and their impact on the cosmological parameters. In particular, we track the evolving phase-space distribution of the injected neutrinos with a Monte-Carlo inspired probabilistic approach and also take into account that for masses $m_\phi \lesssim m_{W,Z}$, the primary decay and the emission process in FSR no longer factorise. We point out that non-thermal scatterings, while being computationally rather challenging and therefore largely ignored in the literature, can be very relevant and dominate the resulting limits in significant regions of parameter space. 
We also stress that hadrodisintegration reactions induced by FSR often improve the resulting limits by many orders of magnitude compared to those from photodisintegration largely considered in the literature. Overall we find that, even for the case where neutrinos are injected into the primordial plasma --- which are after all electromagnetically and colour neutral particles ---  the resulting constraints can be very stringent, covering a broad range in mass and lifetime of the relic source particle.

This work is structured as follows:
In section~\ref{sec:basics}, we begin with an introductory discussion of the various processes that are relevant for neutrino injections, and of the corresponding physical effects. 
We then proceed in section~\ref{sec:details} with a more detailed description of the thermal and non-thermal scattering processes, both contributing to the neutrino cascade. The aim of that section is to gain an intuitive understanding of the processes and scales at play during the injection of neutrinos into the primordial plasma. In section~\ref{sec:photodis}, we briefly discuss some BBN basics, including photo- and hadrodisintegration.
The more technical aspects of our work are covered in section~\ref{sec:MC}. There, we describe our numerical framework and explain in detail how we perform the actual calculations. We interface our code with the extended version \texttt{v2.0.0-dev} (currently on a different branch on \texttt{GitHub}) of the publicly available code \texttt{ACROPOLIS}~\cite{Depta:2020mhj}, which not only enables the calculation of the evolution of the light-element abundances resulting from photodisintegration reactions but also takes into account the effects from hadrodisintegration (so far only for specific scenarios and certain assumptions). We dedicate section~\ref{sec:results} to the discussion of the resulting limits, showing several parameter planes and the effect of a mixed branching ratio into $e^+e^-$ before concluding in section~\ref{sec:conc}. For the interested reader, we provide several appendices with more details.
Throughout this work, we use natural units $\hbar=c=k_\mathrm{B}=1$.

\section{Basics of neutrino injections}\label{sec:basics}

In this work, we perform a detailed study of late-time neutrino injections into the primordial plasma and evaluate the corresponding limits that arise from changes to the light-element abundances produced during BBN. Specifically, we consider a relic particle $\phi$ with mass $m_\phi$ that decays into neutrinos with a lifetime $\tau_\phi \gtrsim 10^4\,\mathrm{s}$, meaning that -- at the time of the decay -- the light-element abundances have already reached their asymptotic values as predicted by standard, i.e.~\textit{thermal}, BBN. 
However, even for such lifetimes, the latter values may still deviate from those in the SM, since the expansion history of the Universe is altered due to the presence of the relic $\phi$. Additionally, once thermal BBN has concluded at $t < 10^4\,\mathrm{s} \lesssim \tau_\phi$, the subsequent decay of $\phi$ and the following neutrino cascade then induce additional photo- and hadrodisintegration reactions, which lead to a second phase of \textit{non-thermal} nucleosynthesis, which may further alter the previously fixed abundances.
We postpone a detailed discussion of decays with shorter lifetimes, i.e.~$\tau_\phi \lesssim 10^4\,$s, to future work.\footnote{The case of short lifetimes for the decay $\phi\to \nu\bar{\nu}$ has been studied with some simplifying assumptions in \cite{Chang:2024mvg}. In particular, interconversions of protons into neutrons are considered, which alter the abundances of the light nuclei already during thermal nucleosynthesis. However, for the relic masses considered in our work, such interactions would be dominated by deep inelastic scattering between nucleons and neutrinos \cite{Formaggio:2012cpf}, adding significant complications to the process. We are not aware of such a study.}
Such early injections will generically already change the reactions \emph{during thermal} BBN, which adds an additional complication to the calculation.

In this section, we present the basic processes that may lead to a change of the light-element abundances, e.g.~via the injection of electromagnetic and/or hadronic material, as well as their implications.

\subsection{Relevant processes}\label{sec:processes}

\begin{figure}[t]
    \centering
    \includegraphics[width=0.7\textwidth]{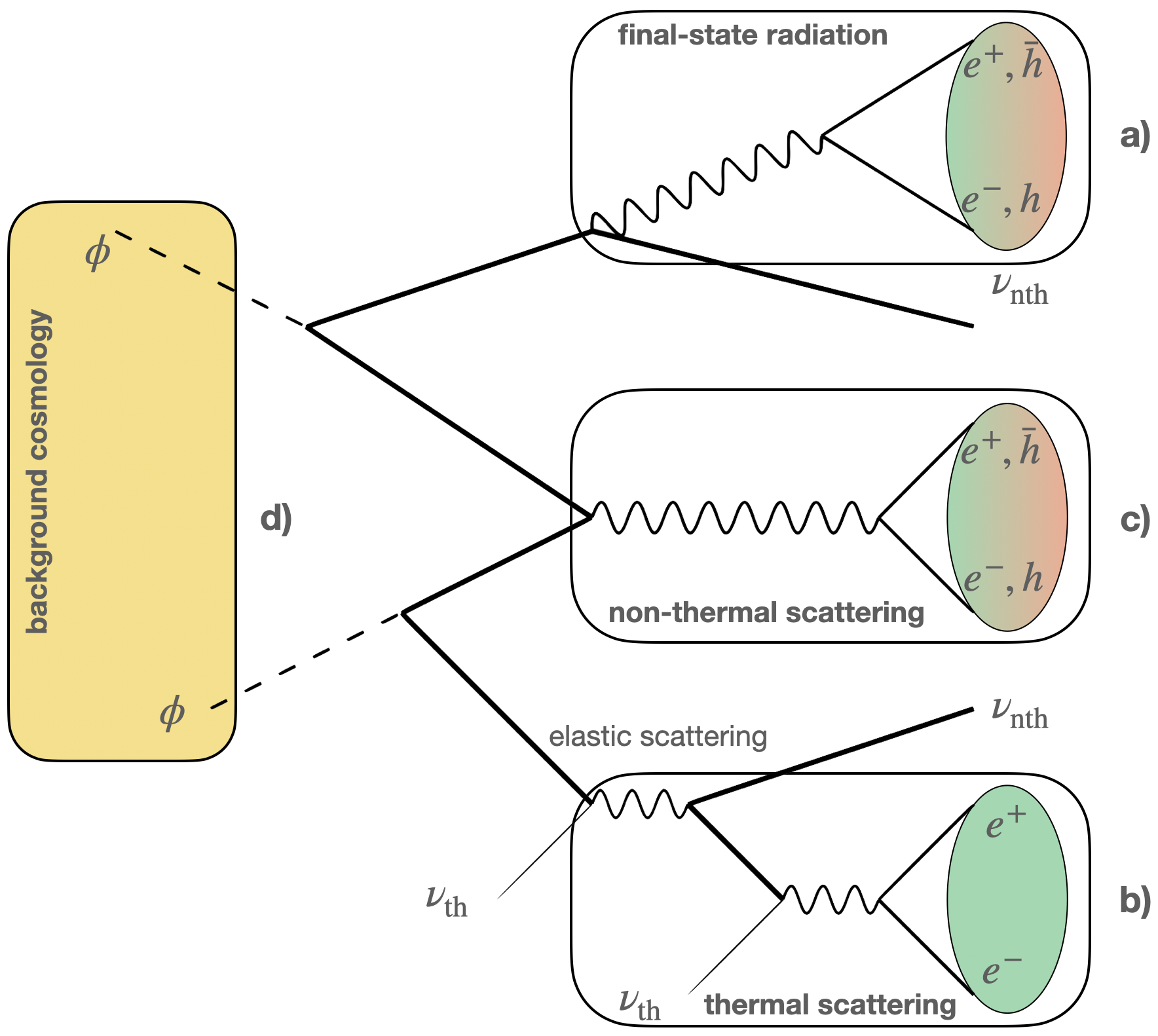}
    \caption{Graphical representation of a possible neutrino cascade, where a fraction of the energy ends up in electromagnetic and hadronic material. We briefly describe the physics of the various sub-processes in the text.}
    \label{fig:nu_shower}
\end{figure}
The processes that are relevant for our analysis are summarised in figure~\ref{fig:nu_shower}, where we show a possible realisation of a neutrino cascade with different sub-processes. We will discuss them one-by-one in the following:

\textbf{a) Final-state radiation} \quad As depicted in sub-process a) of figure~\ref{fig:nu_shower}, in order to produce electromagnetic and/or hadronic material, a neutrino is not necessarily required to interact with the medium. Instead, the decay of the source particle can produce a virtual neutrino emitting a $W$- or $Z$-boson, which then leads to an EW shower that may produce photons and electromagnetically charged particles, as well as hadrons such as pions, protons, and neutrons. These radiative corrections, which are most relevant for larger $\phi$ masses, are already well known and have previously been studied for electromagnetic final states. Notably, the spectrum of final-state particles can be determined using \texttt{PYTHIA8.3}~\cite{Sjostrand:2014zea,Bierlich:2022pfr} as presented in \cite{Hambye:2021moy}. Here, we follow the latter work but also extract the hadronic component of the shower using the same approach. Given the range of possible final states, this process can lead to both photo- and hadrodisintegration. For smaller masses, FSR becomes suppressed and we find that thermal and/or non-thermal scattering processes (see below) are more relevant. Note that in \cite{Kanzaki:2007pd}, the authors also consider both components of the FSR. However, they treat the hadronic branching ratio as a free parameter instead of using a robust SM prediction.

\textbf{b) Thermal scattering} \quad As depicted in sub-process b) of figure~\ref{fig:nu_shower}, each injected neutrino can also interact with a neutrino of the cosmic neutrino background (C$\nu$B) to produce a pair of light, charged SM particles, in particular $e^+ e^-$ pairs. For the lifetimes and thus temperatures considered in this work, the production of heavier particles such as muons or pions is kinematically suppressed for relic masses below the TeV-scale, meaning that we do not consider them here.
In addition, each injected neutrino may scatter elastically, once or several times, with any C$\nu$B neutrino.
As we argue below, these scatterings do not typically affect the distribution of final-state electrons and positrons.

For most of the relevant parameter space, we find that the thermal interaction rate of the injected neutrinos is well below the Hubble rate, in which case a fairly simple analytical description of these reactions can be achieved (cf.~section~\ref{sec:eeonly}). In particular, one can show that the injected neutrinos typically scatter at most once during their propagation through the thermal background, which constitutes an important simplification for our numerical treatment in section~\ref{sec:MC}. Earlier works have already considered this effect in various ways~\cite{Kawasaki:1994bs,Kanzaki:2007pd,Acharya:2020gfh}.

\textbf{c) Non-thermal scattering} \quad  As depicted in sub-process c) of figure~\ref{fig:nu_shower}, two injected neutrinos can also undergo interactions among each other, thereby producing a pair of SM particles. 
In this case, both initial neutrinos have an energy much larger than the background temperature, unlike in the case of thermal scattering. This implies that for injected neutrinos with an energy of the order of the EW scale, all possible final states are usually available.
After their production, the produced SM particles may either decay or hadronise, and the resulting shower can again be simulated using \texttt{PYTHIA8.3}. In this work, we assume that all unstable particles except for neutrons decay immediately, which we find to be a good approximation. After the shower, we therefore end up with \textit{(i)} electromagnetic material, which induces subsequent photodisintegration reactions, \textit{(ii)} secondary neutrinos, which are at much smaller energies than the initial ones and can therefore be neglected, and \textit{(iii)} protons and neutrons, which induce subsequent hadrodisintegration reactions.

Let us stress that, when aiming for a conservative limit, it is generally important to take into account both hadronic and EM final states. This is because
hadrodisintegration tends to overproduce deuterium via the destruction of helium-4, while photodisintegration often leads to a direct destruction of deuterium. Consequently, both effects go in different directions and therefore may cancel each other. Prior works~\cite{Frieman:1989vt,Gratsias:1991,Scherrer:1993} also discussed the effect of non-thermal neutrino scatterings. However, they make use of a significantly simplified setup while relying on long-outdated observations.

\textbf{d) Background cosmology} \quad
As depicted in sub-process d) of figure~\ref{fig:nu_shower}, and in addition to the previously discussed effects leading to photo- and hadrodisintegration, the extra energy density associated with the long-lived relic $\phi$ and the injected neutrinos also leads to a change of the cosmological evolution, which in turn affects BBN and the resulting light-element abundances. On a similar note, this also leads to a change in the effective number of relativistic degrees of freedom, generally parameterised by the effective number of neutrinos $\Neff$, which in turn can be constrained by CMB measurements.

\subsection{A first look at the resulting limits}\label{sec:first_look}

\tikzstyle{process} = [rectangle, rounded corners, minimum width=3cm, minimum height=1.5cm, text centered, draw=black, fill=blue!20]
\tikzstyle{arrow} = [thick,->,>=stealth]
\tikzstyle{title} = [rectangle, text centered, minimum height=1cm]

\begin{figure}[!t]
\centering
\resizebox{1\textwidth}{!}{%
\begin{circuitikz}
\tikzstyle{every node}=[font=\normalsize]
\draw [ color={rgb,255:red,23; green,20; blue,33} , line width=0.2pt , dashed] (8.75,25.5) rectangle  node {\LARGE $\phi \rightarrow \nu \bar{\nu}$}  (13,24.5);
\draw [->, >=Stealth] (13,25) -- (14,25);
\draw [ fill={rgb,255:red,248; green,223; blue,129} , rounded corners = 14.4] (14,25.5) rectangle  node {\normalsize Change in background cosmology} (20.25,24.5);

\draw (10.75,24.5) to[short] (10.75,23.75);
\draw (7.25,23.75) to[short] (14.5,23.75);
\draw [->, >=Stealth] (7.25,23.75) -- (7.25,22.75);
\draw [->, >=Stealth] (14.5,23.75) -- (14.5,22.75);
\draw  (5.5,22.75) rectangle  node {\normalsize Scattering} (9.25,22);
\draw  (12.5,22.75) rectangle  node {\normalsize Final-state radiation} (16.5,22);
\draw [short] (7.25,22) -- (7.25,21.25);
\draw [short] (4.25,21.25) -- (10.25,21.25);
\draw [->, >=Stealth] (4.25,21.25) -- (4.25,20.25);
\draw [->, >=Stealth] (10.25,21.25) -- (10.25,20.25);
\draw  (2.25,20.25) rectangle  node {\normalsize Thermal} (6,19.5);
\draw  (8.25,20.25) rectangle  node {\normalsize Non-thermal} (12,19.5);
\draw [short] (4.25,19.5) -- (4.25,17.5);
\draw [->, >=Stealth] (4.25,17.5) -- (7.5,17.5);
\draw  (7.5,18) rectangle  node {\normalsize EM shower} (11,17);
\draw  (13.75,18) rectangle  node {\normalsize Hadronic shower} (17.25,17);
\draw [short] (10.25,19.5) -- (10.25,19);
\draw [short] (14.5,22) -- (14.5,19);
\draw [short] (10.25,19) -- (14.5,19);
\draw [short] (12.25,19) -- (12.25,18.75);
\draw [short] (9.25,18.75) -- (15.5,18.75);
\draw [->, >=Stealth] (9.25,18.75) -- (9.25,18);
\draw [->, >=Stealth] (15.5,18.75) -- (15.5,18);
\draw [->, >=Stealth] (9.25,17) -- (9.25,16.25);
\draw [->, >=Stealth] (15.5,17) -- (15.5,16.25);
\draw [ fill={rgb,255:red,150; green,217; blue,175} , rounded corners = 14.4] (7,16.25) rectangle  node {\normalsize Photodisintegration} (11.5,15.25);
\draw [ fill={rgb,255:red,246; green,170; blue,144} , rounded corners = 14.4] (13.25,16.25) rectangle  node {\normalsize Hadrodisintegration} (17.75,15.25);
\end{circuitikz}
}%
\caption{Flowchart of the processes considered in this work.}

\label{fig:flowchart}
\end{figure}
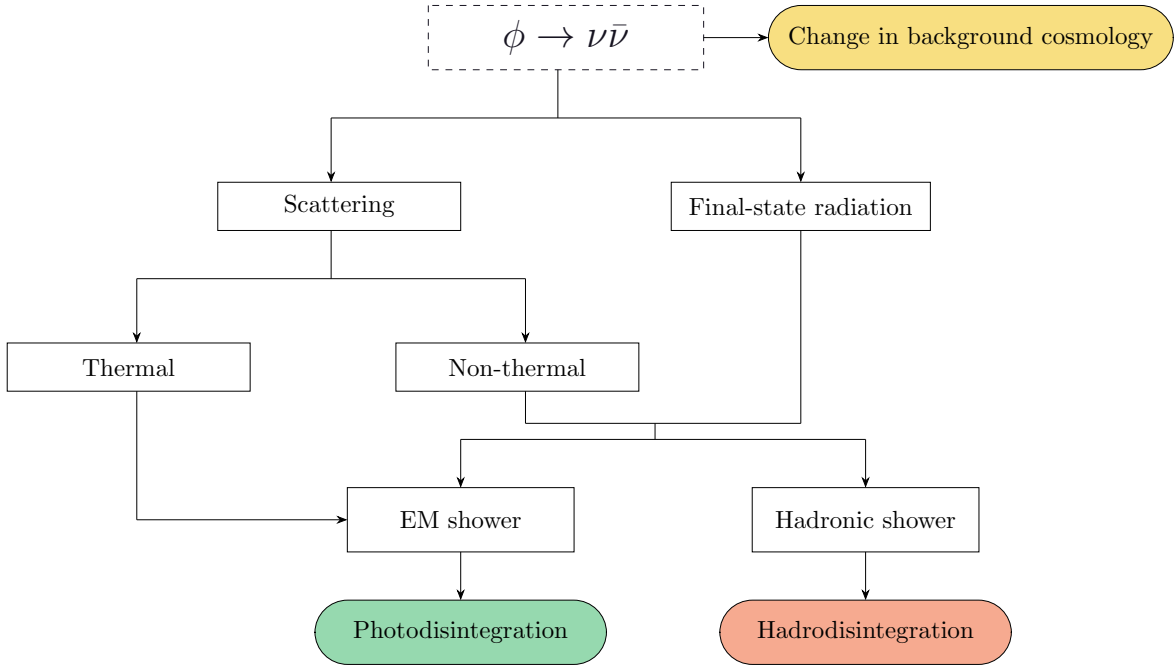

\begin{figure}[!t]
    \centering
    \includegraphics[width=1.
\linewidth]{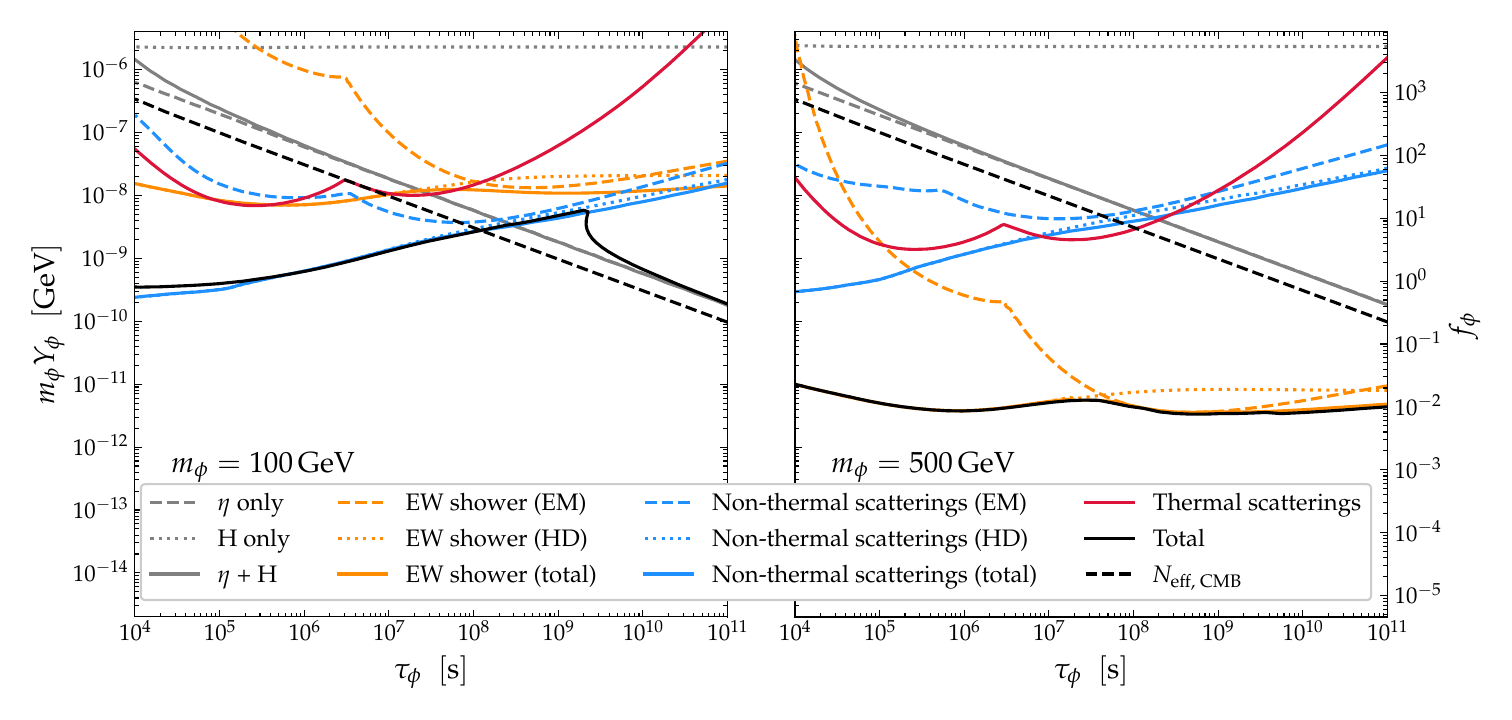}
\caption{\textbf{Left:} BBN limits for $m_\phi = 100\,\mathrm{GeV}$ in the $\tau_\phi-m_\phi Y_\phi$ parameter plane. Besides the overall limit (solid black), we also show the individual contributions that arise from the effects of the EW shower, i.e.~FSR, (orange), thermal scattering (red), non-thermal scattering (blue), and the modified background cosmology (grey). For comparison, we also indicate the CMB $\Neff$ limit (dashed black). \textbf{Right:} Same, but for $m_\phi = 500\,\mathrm{GeV}$.}
\label{fig:flags_early}
\end{figure}

In figure~\ref{fig:flowchart}, we provide a condensed visualisation of the procedure that we use to calculate the limits on neutrino injections. As part of this flowchart, we emphasise the individual branches that we discussed in the previous section, but also the fact that there is an interference between the different effects as indicated by the convergence of the tree before the boxes labelled "EM shower" and "Hadronic shower".
Before entering into the details of the analysis, it is useful to qualitatively discuss the physical effects of the various processes. Here, we do so by showing in advance the results obtained in this work for two fixed masses, i.e.~$m_\phi=100\,$GeV and $m_\phi=500\,$GeV. By fixing the mass to one of these values and scanning over the abundance and the lifetime of $\phi$, we obtain the limits presented in figure~\ref{fig:flags_early}. These figures show the total limits that we obtain by fully including all effects, as well as the partial limits that we obtain by retaining only individual effects of the various processes above.\footnote{This separation is somewhat artificial, since a true separation is basically impossible due to the rich interplay between the different effects.} In the process, we also separate the effects of EM and hadronic injections. Note that this section heavily relies on various concepts that are only explained briefly, but will be discussed in more detail later. However, whenever this is the case, we provide forward or literature references to aid the reader.

We start by discussing the results for  $m_\phi=100\,$GeV.
For this choice of mass, one of the least important contributions comes from the EM component of the EW shower, originating from \textit{final-state radiation} during the decay process (cf.~e.g.~\cite{Hambye:2021moy}) (dashed orange). This is not surprising, as this contribution is suppressed for values of $m_\phi/2$ below the EW scale, i.e.~when the emitted gauge boson is forced to be off-shell (cf.~figure~\ref{fig:EW_shower}). These constraints are particularly weak for short lifetimes, since any EM injections that happen too early will additionally be shut off as all photons with energy $E > m_e^2/(22T)$ get rapidly depleted due to $e^+e^-$ pair production on the background photons (cf.~e.g.~\cite{Kawasaki:1994sc,Hufnagel_2018} and eq.~\eqref{eq:universal}). This becomes particularly evident from the sharp decrease in sensitivity towards $\tau_\phi \sim 10^4\,$s. 
Overall, the resulting (photodisintegration) constraints are mainly driven by an underproduction of deuterium ($\tau_\phi \lesssim 3\times 10^6\,\mathrm{s}$) or an overproduction of helium-3 ($\tau_\phi \gtrsim 3\times 10^6\,\mathrm{s}$). In addition to this, the hadrons produced in the shower may induce sufficient hadrodisintegration reactions (dotted orange), which in-turn significantly effect the light-element abundances. For small lifetimes, $\tau_\phi < 10^7\,$s, the corresponding limits turn out to be orders of magnitude more stringent than the ones originating purely from EM material.
This is due to efficient hadrodisintegration induced by energetic neutrons. In fact, neutrons are completely stopped by their interactions with the background photons only for injections at $t\lesssim 10^2\,$s \cite{Kawasaki:2005}, i.e.\ much earlier than any lifetime considered in this work. Consequently, at the temperatures we are interested in, neutrons experience only negligible energy loss compared to other particles (cf.~e.g.~\cite{Kawasaki:2005}). This allows them to disintegrate helium-4 at earlier times leading to deuterium overproduction (cf.~section~\ref{sec:hadro} and appendix~\ref{app:hadro}), when other particles would dissipate their energy too quickly and thus impose only very weak constraints.
For large lifetimes, $\tau_\phi > 10^7\,$s, the hadrodisintegration effects turn out to be comparable to the EM ones. In both cases, the limits stem mainly from an overproduction of deuterium, and there is no cancellation between both effects. In conclusion, the total FSR limit (solid orange) is stronger than the individual ones. Note that for such large lifetimes, the effect of the hadronic part is plateauing around $\tau_\phi \sim10^9$~s.

When it comes to \textit{thermal scattering}, hadronic injections are irrelevant, as the corresponding final states are 
kinematically impossible due to the small energy of thermal neutrinos (cf.~section~\ref{sec:thresholds}). However, the same is not true for EM injection (solid red). For large lifetimes, the corresponding constraint rapidly gets much weaker than the one from FSR (solid orange).
This behaviour is expected, as scattering processes depend quadratically on the number density of the initial-state particles and hence are suppressed quadratically due to cosmic expansion, whereas FSR is only suppressed linearly.
For $\tau_\phi\lesssim 10^7\,$s, these injections lead to similar constraints than the ones from hadrons injected via FSR (dotted orange). Like in the case of FSR, the overall limits mainly stem from either an underproduction of deuterium ($\tau_\phi \lesssim 3\times 10^6\,\mathrm{s}$) or an overproduction of helium-3 ($\tau_\phi \gtrsim 3\times 10^6\,\mathrm{s}$).

Regarding \textit{non-thermal scattering}, the corresponding bounds from photodisintegration (dashed blue) are very similar to the ones from thermal scattering for $\tau_\phi < 10^7\,$s. For longer lifetimes, however, they do not decouple as fast as the thermal ones and thus become much stronger. While interactions between two non-thermal neutrinos also suffer from the aforementioned dilution suppression, the non-relativistic relic which sources the neutrinos, dilutes slower than the background neutrinos, giving a relative boost to the impact of the non-thermal scattering. The same is true for the hadronic component (dotted blue) when considering large lifetimes. Similar to FSR, the contributions from EM and hadronic injections add up in this regime, thus leading to slightly stronger bounds (solid blue). For lifetimes $\tau_\phi < 10^8\,$s, the hadrodisintegration limits from non-thermal scattering are significantly stronger than any other limit. As a result, they dominate the overall limit over a large range of parameter space and therefore are essentially identical to the combined limit (solid black). For short lifetimes, $\tau_\phi \lesssim 10^5\,\mathrm{s}$, we further observe that the individual limit due to non-thermal scattering is even stronger than the total limit. However, this effect is rather artificial and stems from the (non-perfect) separation of the different effects: when thermal scatterings are turned off, less neutrinos scatter thermally into $e^+e^-$ or $\nu\nu$, which increases the pool of potential particles that can undergo non-thermal scattering. Since the latter mechanism is dominant in this regime, this leads to somewhat stronger limits.

Finally, the limits are also affected by the modified \textit{background cosmology}, which can manifest in the form of two different effects. On the one hand (dashed grey), a change in $\Neff$ leads to a different best-fit value for the baryon-to-photon ratio $\eta$ (cf.~\eqref{eq:def_eta}), which has a direct influence on the deuterium abundance, which decreases (increases) for larger (smaller) values of $\eta$ (cf.~e.g.~section~\ref{sec:abundances}). On the other hand (dotted grey), the additional energy density of $\phi$ and the non-thermal neutrinos leads to a change of the Hubble rate $H$, which has a direct effect on the BBN products (cf.~e.g.~\cite{Hufnagel:2017dgo}). For the lifetimes considered in this work, i.e.~those that lie sufficiently past the end of thermal BBN, the resulting bound can effectively be interpreted as an $\Neff$ bound from BBN, since it probes the additional energy density (as encoded in $H$) that is present during this time. In general, this effect is comparatively weak, and its inclusion changes the overall limit that originates from both cosmological effects (solid grey) only below $\tau_\phi \sim 10^6\,$s. Notably, the total limit becomes weaker in this region, due to a cancellation between an $\eta$-driven deuterium underproduction and an $H$-driven deuterium overproduction.
Overall, the limits arising from the modified background cosmology are the dominant BBN limits for very large lifetimes, $\tau_\phi > 2\times 10^9\,\mathrm{s}$. Thus, the full BBN constraint (solid black), which takes into account all effects simultaneously, goes from a regime where it is dominated by the non-thermal hadrodisintegration effect at small lifetimes (see above) to a regime where it is dominated by the cosmological BBN limit at large lifetimes.
Interestingly, in between these two regimes, the full BBN limit follows the one from non-thermal scattering for almost one order of magnitude in lifetime at around $\tau_\phi \sim 10^9\,\mathrm{s}$. This is a prime example of a cancellation between different contributions which would have been invisible when only showing each contribution individually. In this region, the aforementioned underproduction of deuterium due to the change in $\eta$ is (over-)compensated for by the overproduction of deuterium induced by non-thermal scattering reactions.

For comparison, we also show the CMB limit that arises due to the presence of an additional radiation component during recombination, parameterised in terms of the effective number of neutrinos $\Neff$ (dashed black). Numerically, we find that the resulting constraint is well parameterised by $f_\phi (\tau_\phi / 10^4\,\mathrm{s}) \simeq 680$ (also cf.~appendix~\ref{app:neff}). Deviations from this naive formula only arise for large masses, when the non-thermal neutrinos lose a significant amount of their energy due to the injection of EM/hadronic material. Overall, we find that the CMB limits are stronger than the one from BBN only for lifetime above $10^8\,$s. By also considering BBN constraints, it is therefore possible to exclude much larger parts of parameter space.

So far we have discussed the resulting constraints for $m_\phi = 100\,\mathrm{GeV}$. For smaller masses, the behaviour is qualitatively similar. In particular, the total BBN constraint displays the same general behaviour, dominated by the same two regimes just discussed above with a similar cancellation feature in between (cf.~figure~\ref{fig:mphi_all}). However, for larger masses, important differences arise, as illustrated in right panel of figure~\ref{fig:flags_early} for $m_\phi = 500\,\mathrm{GeV}$.
Most importantly, both constraints arising from EM and hadronic FSR injections become much more stringent. While the qualitative behaviour remains similar to the one for $m_\phi=100\,\mathrm{GeV}$, the strength of the limit increases by as much as three orders of magnitude. As a result, the limits from FSR injections come to dominate all other constraints. In particular, the hadronic contribution (dotted orange) dominates for lifetime below $\sim 10^7$~s, while becoming comparable to the EM one (dashed orange) above. Notably, taking into account the effects from hadrodisintegration improves the constraints by several orders of magnitude for small lifetimes. Overall, the combined constraint (solid black) depends only mildly on the lifetime. Finally, most other limits do not change qualitatively. While the bounds arising from the background cosmology (solid grey) remain very similar, the ones from thermal scattering (solid red) increase significantly but, as before, are never dominant.

\section{Details of neutrino injections}\label{sec:details}

In this section, we provide a more rigorous description of the simplified picture given in section~\ref{sec:processes}. This includes, among other aspects, discussions of neutrino oscillations, kinematic thresholds, and details on the various scattering processes. We want to provide, as far as possible, some analytical intuition for the processes and some of the interesting effects we found, before we introduce our numerical framework in section~\ref{sec:MC}. Note that in this section, we do not give an explicit discussion of FSR, details of which can be found in \cite{Hambye:2021moy}.

\subsection{Neutrino oscillations\label{sec:osc}}

Once produced, and potentially before undergoing further interactions, the injected neutrinos may oscillate.
Depending on the temperature at the time of injection, the timescale for a neutrino to oscillate can be much shorter than the one for scattering. Indeed, for the range of lifetimes considered in this work, the oscillation time (see e.g.~\cite{Escudero:2018mvt})
\begin{equation}\label{eq:osc_scale}
t_\text{osc}\sim \frac{T}{\Delta m^2}\sim 10^{-5}\, \text{s}\left(\frac{T}{\text{MeV}}\right)
\end{equation}
is much shorter than both the Hubble time and the interaction time (also cf.~eqs.~\eqref{tH} and \eqref{tth} below). The only timescale smaller than this, is the one for FSR.
For the present discussion, the important conclusion is that neutrino oscillations are basically instantaneous, so that all flavours will be present basically immediately after injection / FSR, even if the initial decay produces only one flavour. 
We therefore utilise the following prescription: if a neutrino of a given flavor $\alpha\in\{e,\mu,\tau\}$ is injected into the plasma, it is instead treated as a superposition of all three flavor eigenstates $\beta\in\{e,\mu,\tau\}$ weighted by the oscillation probabilities encoded in the PMNS matrix (a more detailed description can be found in section~\ref{sec:MC})
\begin{align}\label{eq:osc}
    (\mathbf{M}_{\text{osc}})_{\alpha\beta}=\sum_{i=1}^3 |U_{\alpha i}|^2|U_{\beta i}|^2\approx\begin{pmatrix} 0.55&0.17&0.28\\0.17&0.45&0.37\\0.28&0.37&0.35\end{pmatrix}\eqsp.
\end{align}
This matrix can be applied to the spectrum of decay products for arbitrary branching ratios, but also to the spectrum resulting from elastic scattering reactions to determine the distribution of outgoing neutrinos in each case. In this work, we largely concentrate on injections of the form $\phi \to \bar{\nu}_e \nu_e$ for concreteness (cf.~figure~\ref{fig:flags_early} above), noting that other choices will hardly affect the final results (cf.~section~\ref{sec:results}).

\subsection{Thermal scattering of the form $\nu \nu_\mathrm{th} \rightarrow e^+ e^-$}
\label{sec:eeonly}

\subsubsection{Thresholds and timescales}
\label{sec:thresholds}
We begin the discussion of thermal interactions with an investigation of the kinematic thresholds and the interaction rates, before providing a more rigorous analytical treatment.

The threshold energy of injected neutrinos $E_{ee}$, required to produce an electron-positron pair while scattering off a background neutrino at temperature $T$, is approximately given by (here, we use that $T_\nu \sim \mathcal{O}(1) T$)
\begin{align}
    E_{ee}\sim \frac{m_e^2}{T}\gtrsim 26\,\text{MeV}\left(\frac{10\,\text{keV}}{T}\right)\eqsp,\label{eq:ee_threshold}
\end{align}
with $m_e$ being the electron mass. Consequently, since we focus on neutrino injections at $T\lesssim 10$~keV, the injected neutrinos must have a kinetic energy well above the one of the background neutrinos in order for this production to be feasible. For smaller energies, non-thermal neutrinos can no longer annihilate into electron-positron pairs, meaning that they no longer need to be tracked for the neutrino cascade. However, they do still contribute to $\Neff$. For the production of heavier particles, e.g.~$\mu^+ \mu^-$ or $\pi^+ \pi^-$, the respective threshold energies $E_{\mu\mu}$ and $E_{\pi\pi}$ must instead be significantly larger
\begin{align}\label{eq:mu_pi}
    E_{\mu\mu/\pi\pi}\sim \frac{m_{\mu/\pi}^2}{T}\gtrsim \left(1.1/1.8\right)\,\text{TeV}\left(\frac{10\,\text{keV}}{T}\right)\eqsp,
\end{align}
with similar values (up to factors of $\mathcal{O}(1)$) being obtained also for the mixed production of, e.g.~$e^+ \mu^-$.
For such high energies, the overall limit is dominated by FSR (cf.~sub-process (a) in figure~\ref{fig:nu_shower}), meaning that these processes are not relevant to determine the overall limit. We will therefore not consider them further in the following.

For the purpose of this work, any neutrino that undergoes scattering with the neutrino background can therefore either produce an electron-positron pair or a pair of (potentially different) non-thermal neutrinos (cf.~section \ref{subsec:el_scattering}). In general, we find that -- for $T \lesssim 10\,\mathrm{keV}$ -- the corresponding centre-of-mass energy $s$ is always small compared to the EW scale as described by the $Z$-boson mass $m_Z$, i.e.~$s\ll m_Z^2$. This implies that the various processes can be described well within the effective Fermi theory without the need to invoke the full EW theory. 
Within this approximation, the cross-section for any of the available scattering reactions can be parameterised as
\begin{align}
    \sigma(s) = \Sigma(s) \frac{G_F^2s}{6\pi}\eqsp.
    \label{eq:def_sigma}
\end{align}
Here, $G_F$ is the Fermi constant and $\Sigma(s)$ is a numerical coefficient, which differs by process (cf.~table~\ref{tab:fermirates}). For $s\gg m_e^2$, which is true for most parts of parameter space, $\Sigma(s) \simeq \text{const.}$ holds, meaning that all cross-sections are approximately linear in $s$. 

\begin{table}
    \centering
    \begin{tabular}{c|c}
   \textbf{Process} & \textbf{Cross-section coefficient} $\Sigma(s)$\\
   \hline\hline
    $\nu_i\bar{\nu}_i\rightarrow \nu_j\bar{\nu}_j$   & $1$ \\
    $\nu_i\nu_i\rightarrow \nu_i\nu_i$  & $6$\\
    $\nu_i\nu_j\rightarrow \nu_i\nu_j$   & $3$ \\
    $\nu_i\bar{\nu}_j\rightarrow \nu_i\bar{\nu}_j$ & $1$\\
    $\nu_j\bar{\nu}_j\rightarrow \nu_j\bar{\nu}_j$   & $4$\\ \hline
     $\nu_e\bar{\nu}_e\rightarrow e\bar{e}$  & $\left[\left(1+2\frac{m_e^2}{s}\right)\left(1+4s_w^2+8s_w^4\right)-3\frac{m_e^2}{s}\right]\sqrt{1-\frac{4m_e^2}{s}}$\\
    $\nu_\mu\bar{\nu}_\mu\rightarrow e\bar{e}$  & $\left[\left(1+2\frac{m_e^2}{s}\right)\left(1-4s_w^2+8s_w^4\right)-3\frac{m_e^2}{s}\right]\sqrt{1-\frac{4m_e^2}{s}}$\\
    $\nu_\tau\bar{\nu}_\tau\rightarrow e\bar{e}$  & $\left[\left(1+2\frac{m_e^2}{s}\right)\left(1-4s_w^2+8s_w^4\right)-3\frac{m_e^2}{s}\right]\sqrt{1-\frac{4m_e^2}{s}}$
    \end{tabular}
    \caption{Cross-section coefficients $\Sigma(s)$ (cf.~eq.~\eqref{eq:def_sigma}) for all available processes in the small energy regime (Fermi theory). These expressions have been adapted and generalised from table~1 in \cite{Chang:2024mvg}. Here, $i, j \in \{e, \mu, \tau\}$ with $i\neq j$, and we use the shorthand notation $\sin\theta_w\equiv s_w$ for the sine of the Weinberg angle.}
    \label{tab:fermirates}
\end{table}

Depending on the energy of the injected neutrino, the interaction time $t_\text{th}$ corresponding to thermal scattering can be either faster or slower than the Hubble time $t_H$. In general, these times can be approximated as
\begin{align}
    t_H&\sim \frac{1}{H(T)} \sim \frac{m_P}{T^2}\sim 1\, s \left(\frac{T}{\text{MeV}}\right)^{-2}\label{tH} \eqsp, \\
    t_\text{th}&\sim \frac{1}{G_F^2 T^4 E}\sim 10^{-3}\,\text{s}\left(\frac{T}{\text{MeV}}\right)^{-4}\left(\frac{E}{10\,\text{GeV}}\right)^{-1}\eqsp,\label{tth}
\end{align}
with the Planck mass $m_P$. For the last expression, we used that roughly $t_\text{th} \sim 1/(\sigma_\text{th} \bar{n}_\nu)$ with the number density $\bar{n}_\nu \sim T^3$ of thermal neutrinos and the thermal scattering cross-section $\sigma_\text{th} \propto G_F^2 s \propto G_F^2 E T_\nu$ with $T_\nu \propto T$ within effective Fermi theory (also cf.~eqs.~\eqref{eq:rate_def_gen}-\eqref{eq:rate_me0} below for a more detailed calculation). Notably, the interaction time $t_\text{th}$ depends on the energy $E$ of the non-thermal neutrino, meaning that for certain choices of $E$, the interaction time can be much shorter than the Hubble time. Specifically, $t_\text{th}<t_H$ is true for
\begin{equation}
E \gtrsim 100\,\mathrm{GeV}\left( \frac{10\,\mathrm{keV}}{T} \right)^2\eqsp. \label{eq:hubblevsscat}
\end{equation}
Consequently, a significant amount of scatterings per Hubble time only occurs for heavy $\phi$ with lifetimes $\tau_\phi \sim 10^4\,\mathrm{s}$ corresponding to a neutrino injection at $T \sim 10\,\mathrm{keV}$. In this case, the injected neutrinos basically scatter instantaneously, i.e.~as soon as they are produced. However, this does not imply that the neutrinos efficiently thermalise, since each scattering reaction reduces the energy of the final-state neutrino, making thermal scattering progressively less efficient and potentially spoiling eq.~\eqref{eq:hubblevsscat} before thermal equilibrium can be established.

For most of the relic masses and lifetimes of interest, the thermal interaction rate is below the Hubble rate and interactions will generally not take place instantaneously (if at all), which is why we concentrate on this regime in the analytical discussion below.

\subsubsection{A simplified analytical approach}\label{sec:simplified}

In this subsection, we expand on the previous discussion by providing a simplified analytical approach for handling EM injection due to reactions of the form $\nu \nu_\mathrm{th} \to e^+ e^-$, with a non-thermal (i.e.~injected) neutrino $\nu$ and a thermal  neutrino $\nu_\mathrm{th}$. In the process, we provide a simplified analytical treatment to gain some intuition for this process and its implications. Afterwards, we discuss the shortcomings of this treatment, thus motivating the more sophisticated treatment outlined in section~\ref{sec:MC}, which is used for the final analysis.

Let us start by considering a single neutrino injected at time $t_\mathrm{inj}$ corresponding to the temperature $T_\mathrm{inj}$.\footnote{For simplicity, we neglect neutrino oscillations in the following (simplified) analytical discussion.}
Assuming a reaction $X+Y\to Z$ among massless initial-state particles, the interaction rate $\Gamma_Z$ describing the number of reactions happening per time, is given by (this expression is obtained by simplifying the corresponding collision operator in the Boltzmann equation)
\begin{equation}
    \Gamma_{Z}(T,E) = \frac{g_Y}{16\pi^2E^2}\int_{0}^\infty \text{d}\epsilon\; f_{Y}(\epsilon)\int_{0}^{4E\epsilon}\text{d}s \; s \cdot \sigma_{Z}(s)\eqsp . \label{eq:rate_def_gen}\\
\end{equation}
Here, $\sigma_Z(s)$ is the cross-section of the reaction, $E$ is the energy of $X$ and $\epsilon$ is the energy of $Y$, with the latter particles featuring a phase-space distribution $f_Y$ and $g_Y$ degrees of freedom.
Consequently, the corresponding rate $\Gamma_{ee}(T,E)$ for the process $\nu \bar{\nu}_\text{th} \rightarrow e^+ e^-$ with $X = \nu$, $Y = \bar{\nu}_\text{th}$ and $Z = e^+ e^-$ is given by
\begin{align}
    \Gamma_{ee}(T,E)& = \frac{g_\nu}{16\pi^2E^2}\int_{0}^\infty \text{d}\epsilon\; f_{\nu,\text{th}}(\epsilon)\int_{0}^{4E\epsilon}\text{d}s\; s \cdot \sigma_{ee}(s)\eqsp, \label{eq:rate_def}
\end{align}
where $\sigma_{ee}(s)$ is the cross-section defined in eq.~\eqref{eq:def_sigma} with the correct expression for $\Sigma(s)$ from table~\ref{tab:fermirates}. Assuming for simplicity that $m_e = 0$, in which case $\Sigma(s) = \text{const.} \equiv \Sigma_\infty$, we have $\sigma_{ee}(s) \equiv \Sigma_\infty G_F^2 s /(6\pi)$ and consequently
\begin{align}
    \Gamma_{ee}(T, E) \overset{m_e = 0}{\simeq} \frac{g_\nu}{16\pi^2E^2}\int_{0}^\infty \text{d}\epsilon\; f_{\nu,\text{th}}(\epsilon)\int_{0}^{4E\epsilon}\text{d}s\;
    \frac{\Sigma_\infty G_F^2s^2}{6\pi}
    = \frac{7\pi}{90} \Sigma_\infty G_F^2 E T_\nu^4\eqsp. \label{eq:rate_me0}
\end{align}
Note that in this expression, $\Sigma_\infty$ assumes different values depending on the flavours of the initial-state neutrinos participating in the reaction (cf.~table \ref{tab:fermirates}).

As already stated above -- in order to  gain a simplified analytical understanding --, we consider the case where the Hubble rate defines the shortest timescale of the problem (when ignoring FSR and neutrino oscillations).
Under these assumptions, the energy of each injected neutrino is modified mainly by redshift, meaning that at a given redshift $z$, the remaining energy of the neutrino is given by $E(z)=E_\mathrm{inj}(z/z_\mathrm{inj})$, where $E_\mathrm{inj}$ and $z_\mathrm{inj}$ refer to the neutrino energy and the redshift at the time of injection. By enforcing the condition $E(z) T \lesssim m_e^2$, we therefore find that a neutrino can no longer produce electron-positron pairs at temperatures $T \lesssim (m_e^2/E_\mathrm{inj})(z_\mathrm{inj}/z)$. This is also the point at which $\sqrt{s} \sim 2 \,m_e$, meaning that our assumption in eq.~\eqref{eq:rate_me0}, i.e.~$\Sigma(s)\sim {\rm const.}$, breaks down. However, we find that -- for sufficiently heavy relics -- this problem is of little relevance, since the corresponding temperatures are small enough to no longer allow for efficient energy injection anyway, e.g.~because of dilution effects or because the relic has already fully decayed.
Nevertheless, it is worth noting that the analytical treatment of this subsection technically only applies to sufficiently heavy relics.

To calculate the total fraction of energy $\zeta_\text{em}(t_\text{inj})$ that is transferred into the plasma by a neutrino injected at time $t_\text{inj}$, let us consider an infinitesimal time step $\text{d}t$. In this case, the differential energy fraction $\text{d}\zeta_\text{em}$ injected during the interval $\text{d} t$ is given by
\begin{align}
    \text{d}\zeta_\text{em}&\simeq \frac{E(t)}{E_\text{inj}}\Gamma_{ee}(t)\text{d} t\eqsp.
\end{align}
Here, $\Gamma_{ee}(t) \text{d} t$ is the probability for a non-thermal neutrino with energy $E(t)$ to scatter into $e^+ e^-$ during the interval $t \rightarrow t + \text{d} t$, in which case it transfers a fraction $E(t)/E_\text{inj}$ of its original energy $E_\text{inj}$ into EM material. Using the time-temperature relation $\text{d} T / \text{d} t = -H(T) T$, we then find
\begin{align}
    \frac{\text{d}\zeta_\text{em}}{\text{d}T}&=\frac{\text{d}\zeta_\text{em}}{\text{d}t}\times\frac{\text{d}t}{\text{d}T}=-\frac{E(T)}{E_\text{inj}}\Gamma_{ee}(T) \times \frac{1}{H(T)T}= -\left(\frac{\Gamma_{ee}}{H}\right)_\text{inj}\frac{T^3}{T_\text{inj}^4}\eqsp,\label{eq:diff_inj}
\end{align}
In the last step, we have used $\Gamma_{ee}(T) \propto E(T) T^4$ according to eq.~\eqref{eq:rate_me0}, $H(T) \propto T^2$ and $E(T) = E_\text{inj}(z/z_\text{inj}) = E_\text{inj} (T/T_\text{inj})$.
By integrating this expression from $T = T_\text{inj}$ to $T \rightarrow 0$, we therefore find
\begin{align}
    \zeta_\mathrm{em}(T_\text{inj}) = \int_{T_\mathrm{inj}}^0 \text{d}T\,\frac{\text{d}\zeta_\mathrm{em}}{\text{d}T}\simeq\int_0^{T_\mathrm{inj}} \text{d}T\;\left(\frac{\Gamma_{ee}}{H}\right)_\mathrm{inj}\frac{T^3}{T_\mathrm{inj}^4}= \frac{1}{4}\left(\frac{\Gamma_{ee}}{H}\right)_\mathrm{inj}\eqsp.\label{eq:em_tot_inj}
\end{align}
Notably, this result demonstrates that the total fraction of the injected energy in Eq.~(\ref{eq:em_tot_inj}) is smaller than the naive estimate $(\Gamma_{ee}/H)_\mathrm{inj}$ by a factor 4, courtesy to the energy loss due to redshifting before scattering. Additionally, given that the scattering reactions of the neutrinos usually happen on timescales larger than the Hubble time (see above), the explicit polynomial dependence on $T$ in eq.~\eqref{eq:diff_inj} -- contrary to an expression $\propto \delta(T - T_\text{inj})$ --, shows that the energy transfer into EM material is \emph{not instantaneous}, but rather happens over a prolonged period of time; a key feature that is unique to neutrinos. In contrast, particles that are either electromagnetically or colour charged, would dump their energy instantaneously, thus leading to an immediate change of the light-element abundances. Non-thermal neutrinos, on the other hand, can still be around until well-after the lifetime of the relic to induce disintegration processes.

\begin{figure}
    \centering
    \includegraphics[width=0.7\linewidth]{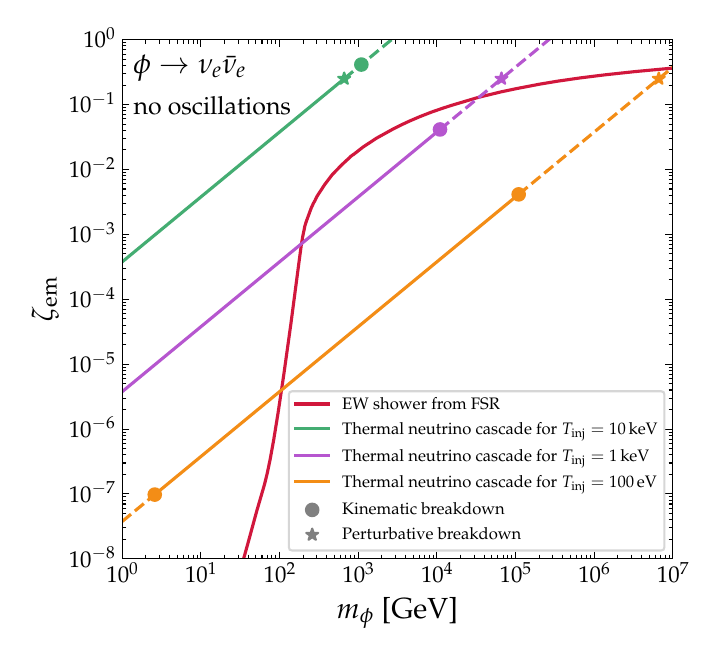}
    \caption{Comparison of the fraction $\zeta_\text{em}$ of energy that is injected in the form of EM material as a function of the relic mass $m_\phi$ within \textit{(i)} the context of a relic $\phi$ that decays into two neutrinos, which subsequently cause an EW shower due to FSR (red, cf.~\cite{Hambye:2021moy}), and \textit{(ii)} our simplified analytical description of the neutrino cascade after injection with $T_\text{inj} = 10\,\mathrm{keV}, 1\,\mathrm{keV}, 100\,\mathrm{eV}$ (green, purple, orange). We further indicate the values of $m_\phi = 2 E_\text{inj}$ for which electron-positron production becomes kinematically inaccessible (lower circle), muons become kinematically available (upper circle), or the annihilation rate approaches the Hubble rate (star). In all of these cases, our simplified analytical description breaks down.}
    \label{fig:EW_shower}
\end{figure}

In figure~\ref{fig:EW_shower}, we visualise these results by showing the energy fraction  $\zeta_\text{em}$ that was derived within this framework using the electron-neutrino annihilation cross-section of table~\ref{tab:fermirates} in the $m_e = 0$ limit.
We find that $\zeta_\text{em}$ scales linearly with the mass of the relic $\phi$ as expected from eq.~\eqref{eq:em_tot_inj}. For comparison, we also indicate the values of $\zeta_\text{em}$ that are obtained when considering only the EW shower from FSR as calculated in~\cite{Hambye:2021moy}. Notably, for small masses or early injections, thermal scattering dominates and therefore is expected to lead to a strengthening of the resulting constraints. Additionally, we also indicate the kinematic threshold (lower circle) below which electron-positron production becomes inaccessible according to eq.~\eqref{eq:ee_threshold}. While we did not take into account the kinematic suppression just above the threshold, we find that this would have an effect only for low-energy, late-time injections as shown in the figure for $T_\text{inj}=100\,$eV. However, there also exists a second kinematic breakdown (upper circle), which is driven by the production of particles heavier than electrons, specifically muons. As outlined in eq.~\eqref{eq:mu_pi}, this effect only plays a role for very heavy particles or very early injections. Therefore, this effective cut-off never becomes dominant in our work.

In addition to the kinematic breakdowns, there also exists a perturbative breakdown for our description (star), which we estimate by solving $\Gamma_{ee}(T_\text{inj})/H(T_\text{inj})=1$ for $T_\text{inj}$. Beyond this threshold, eq.~(\ref{eq:em_tot_inj}) is no longer valid, as it does not account for multi-scatterings. However, we find that for masses beyond $1\,$TeV, the limits are always dominated by the EW shower from FSR, meaning that neither the perturbative nor the upper kinematic threshold play any significant role.

Note that -- besides the shortcomings already discussed above and the fact that we only consider a single neutrino species -- for the purpose of describing the non-thermal interactions (see below), we need information on the full energy distribution of the high-energy neutrinos, which is impossible to obtain within this simplified approach. Thus, we will ultimately need to abandon this analytical description for a more rigorous numerical treatment. A thorough discussion of this approach can be found in section~\ref{sec:MC}.

\subsection{Thermal scattering of the form $\nu \nu_\mathrm{th} \rightarrow \nu\nu$}
\label{subsec:el_scattering}

So far, we  have assumed that -- before scattering into $e^+ e^-$, the non-thermal neutrinos only lose energy due to redshift. However, they may also scatter with a C$\nu$B to produce a pair of (potentially different) secondary neutrinos instead. In this case, both neutrinos redistribute their initial energy between each other, meaning that the thermal background neutrino becomes non-thermal. At first sight, it seems that such reactions could have a large effect on the subsequent production of $e^+ e^-$ pairs, since table~\ref{tab:fermirates} shows that scattering into neutrinos is about 5 to 10 times faster than scattering into $e^+e^-$. However, a closer look reveals that this is not the case. In fact, the annihilation rate only depends linearly on the energy of the individual particles, which implies that -- at a given temperature -- the annihilation rate of both final-state, non-thermal neutrinos is the same as the annihilation rate that the single initial-state neutrino would have had, if it did not scatter before, i.e.~(cf.~eq.~\eqref{eq:rate_me0} with $\Sigma = \text{const.} \equiv \Sigma_\infty$ from table~\ref{tab:fermirates})
\begin{equation}
      \Gamma_{ee}(T,E_1)+\Gamma_{ee}(T,E_2)=\frac{7 \pi}{90} \Sigma_\infty G_F^2 T_\nu^4\left(E_1+E_2\right)=\Gamma_{ee}(T,E_1+E_2)\eqsp.
\label{eq:redist}
\end{equation}
Consequently, for thermal scattering, the total energy that is injected in the form of $e^+e^-$ pairs does not depend on how that energy is distributed among the different neutrinos.\footnote{As discussed in the previous section, this result is only valid well above the $e^+e^-$ production threshold.} Since we assume that interactions of the form $\nu\nu_\text{th} \rightarrow \nu\nu$ do not significantly alter the thermal neutrino abundance, it is reasonable to disregard these interactions. In other words, it is justified to consider the energy injection induced by a single neutrino, as discussed in section~\ref{sec:eeonly}. However, the introduction of non-thermal collisions will void this argument due to the breakdown of Fermi theory, as discussed in the next section.

\subsection{Non-thermal scattering between two high-energy neutrinos}
\label{sec:non-thermal}

Naively, one could expect that scattering reactions between two injected, high-energy neutrinos are negligible due to their comparatively small number density  (this has been assumed e.g.~in \cite{Kawasaki:1994bs,Kanzaki:2007pd,Acharya:2020gfh}).
However, as already emphasised in section~\ref{sec:basics}, in the context of this work, we instead find that such interactions are highly relevant over large parts of parameter space due to their larger centre-of-mass energy, thus leading to scattering that occurs on resonance with much higher cross-section. In this work, we therefore utilise a rather complete treatment of non-thermal neutrino scattering to calculate the resulting constraints. 

\begin{figure}
    \centering
    \includegraphics[width=0.7\linewidth]{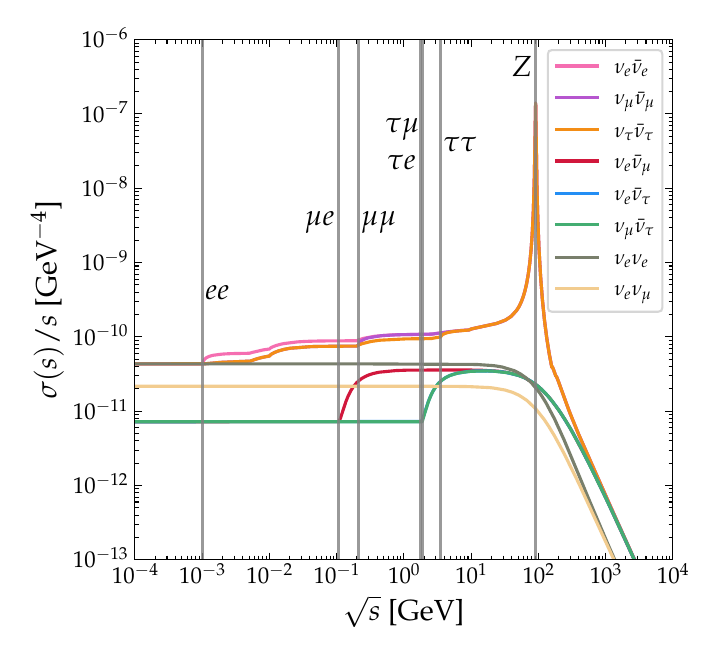}
    \caption{The total cross-section for different initial-states and all possible two-body final states. Note that only interactions between neutrinos and anti-neutrinos can go on resonance, while the remaining processes consistently drop below the naive Fermi scaling at high energies. All data has been extracted by running \texttt{MadGraph5}~\cite{Alwall:2011uj}. Note that the contributions for $\nu_e \bar\nu_\tau$ (blue) and $\nu_\mu \bar\nu_\tau$ (green), while strictly speaking slightly different, are virtually indistinguishable in the plot.}
    \label{fig:xsec}
\end{figure}

While the small number density of high-energy neutrinos suppresses interactions among them, this suppression can be compensated for by a significant increase in the scattering cross-section, since the larger centre-of-mass energy (compared to thermal scattering) makes the $Z$-resonance in the s-channel diagrams of the annihilation processes kinematically available for large parts of parameter space. 
The larger centre-of-mass energy also implies the existence of additional final-states (as opposed to only $\nu\nu$ and $e^+e^-$),
leading, in particular, to hadronic particles, which have a strong influence on the resulting limits. To the best of our knowledge, the inevitable injection of high-energy hadrons from non-thermal scatterings has not yet been considered in the literature, but still provides the leading constraints for relatively small lifetimes and masses below $\sim 100\,$GeV (see above). In figure~\ref{fig:xsec}, we plot the total cross-sections (including all possible two-body final states) as a function of the centre-of-mass energy $\sqrt{s}$ for various initial-state neutrino flavours. As expected, the resulting curves approach the sum of all cross-sections in table~\ref{tab:fermirates} asymptotically for small $\sqrt{s}$.\footnote{Note that the given cross-sections have been calculated at leading order, which is why processes with charge conjugated initial states, like e.g.~$\nu_e\nu_e$ and $\bar\nu_e \bar\nu_e$, feature the same cross-section. Also, again at leading order, the cross-sections for $\nu_e \nu_\mu$, $\nu_e \nu_\tau$ and $\nu_\mu \nu_\tau$ are identical. We therefore do not show all combinations.} In the figure, we further indicate the mass thresholds for the charged lepton, as well as the $Z$-boson mass, thus highlighting the sharp transitions at threshold, e.g.~at $\sqrt{s} = 2m_e$ and $\sqrt{s} = m_Z$. In fact, for each lepton $l_i$, there exist two thresholds, i.e.~the standard annihilation threshold from $\nu_i\bar{\nu}_i\rightarrow l_i\bar{l}_i$ at $\sqrt{s} = 2m_{l_i}$ and the mixed threshold(s) from $\nu_i\bar{\nu}_j\rightarrow l_i\bar{l}_j$ at $\sqrt{s} = (m_{l_i}+m_{l_j})$.

\begin{figure}
    \centering
    \includegraphics[width=0.7\linewidth]{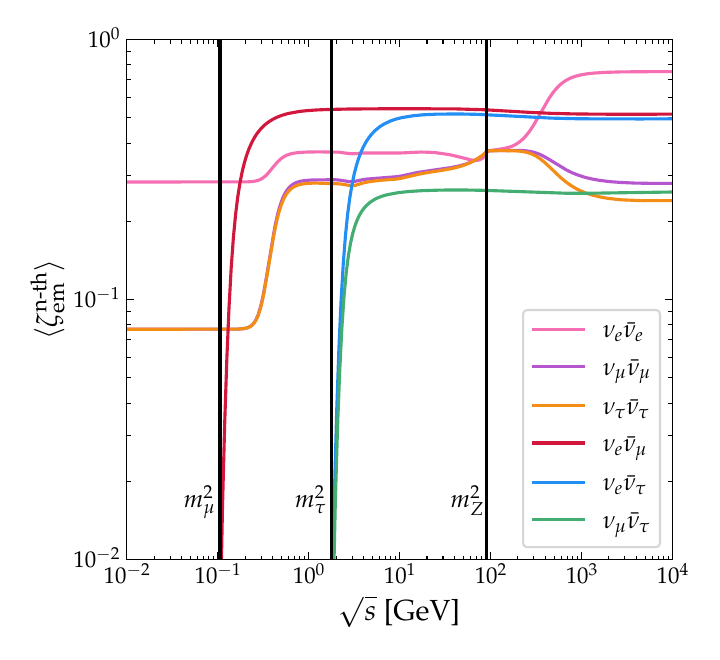}

    \caption{The average fraction of energy injected in the form of EM material from collisions of two high-energy neutrinos, as calculated via \texttt{PYTHIA8.3}. We only show neutrino-antineutrino combinations, since -- at leading order -- only those combinations feature non-neutrino final states due to lepton-number conservation.
}
    \label{fig:avg_zeta}
\end{figure}

To quantify the effect of the non-thermal interactions, in figure~\ref{fig:avg_zeta}, we additionally show the average fraction $\langle \zeta_\text{em}^\text{n-th}\rangle$ of EM energy that is injected into the plasma due to one reaction happening at a centre-of-mass energy $\sqrt{s}$. These results have been obtained by utilising \texttt{PYTHIA8.3} to simulate the particles shower induced by the final-state particles (see section~\ref{sec:MC} below for more information).\footnote{Note that this fraction is the same in the cosmic rest frame and the centre-of-mass frame due to the isotropic distribution of the injected particles.} We find that in the low-energy regime, $\langle \zeta_\text{em}^\text{n-th}\rangle$ reaches a plateau in agreement with our discussion on thermal scattering (cf.~table~\ref{tab:fermirates}).\footnote{We have interpolated between the \texttt{PYTHIA8.3} results and those from Fermi theory for $\sqrt{s} < 1\,$GeV. Moreover, in the plot, we only consider combinations that can have non-neutrino final states at leading order.} Overall, we find that once kinematically allowed, the amount of EM material injected by neutrino collisions of different flavour (red, blue, green), does not vary much with energy and ultimately converges to a constant value at very high energies. This result is expected, since such combinations of neutrinos only undergo $t$-channel scattering and hence do not feel the $Z$-resonance. For interactions with identical flavours in the initial state, there instead exist small variations with energy. Interestingly, all corresponding lines (orange, purple, pink) converge around the resonance, at which point the final states are dominated by $Z$-boson decay, irrespective of the initial-state configuration. For higher energies, the lines diverge again, since the large phase-space favours non-resonant interactions over the kinematically suppressed resonance.

For centre-of-mass energies around the $Z$-pole, the resonance provides the dominant contribution for non-thermal scattering.  In particular, for a collision of two high-energy neutrinos, the resonance effectively selects the scattering angle in such a way to put the interaction on resonance (as the probabilities for other configurations are much smaller). However, in this work, we still take into account the full energy distribution (which we assume to be isotropic) and the exact cross-sections, without relying on a narrow-width approximation. In particular, this means that scattering is also allowed, albeit suppressed, in the energy range below the resonance (cf.~section \ref{sec:MC} for more details). 

Regarding the injection of hadronic material, the distribution of final states obtained by \texttt{PYTHIA8.3} reveals that non-thermal interactions on average inject less than one nucleon per scattering, with an average kinetic energy of a few percent of the sum of the initial neutrino energies. At first glance, this seems to be a negligible contribution. However, in section~\ref{sec:hadro} we will emphasise that hadronic injections should still be handled with care. In fact, we find that this contribution yields the dominant constraints for significant parts of parameter space. We provide more details on this contribution in section~\ref{sec:MC}.

At this point, it is worth noting that the inclusion of non-thermal scattering reactions significantly complicates the calculation. Many of the arguments that were used for the simplified approach in section \ref{sec:simplified} rely on the linearity of the underlying processes, among other assumptions. However, many of these simplifications are no longer justified for non-thermal scattering, since \textit{(i)} non-thermal, high-energy collisions inject both EM and hadronic material, \textit{(ii)}
the corresponding interaction rate is inherently quadratic in the amount of non-thermal neutrinos, and \textit{(iii)} Fermi theory breaks down for high-energy collisions at and beyond the resonance. As a result, the amount of injected EM (and hadronic) material is no longer independent of whether the neutrinos underwent $\nu\nu_\text{th} \rightarrow \nu\nu$ scatterings before, as no equivalent of eq.~(\ref{eq:redist}) exists for non-thermal scattering. Therefore, it is crucial to take into account the temporal evolution of the energy distribution of non-thermal neutrinos. However, for the parameter space considered in this work, we find that $H\gg\Gamma_{\nu\nu_\text{th}\rightarrow \nu\nu}$, meaning that two injected neutrinos usually only scatter once and therefore are not much affected by previous (non-existent) scattering reactions. However, scatterings of the form $\nu\nu_\text{th}\rightarrow \nu\nu$ will still deplete the number of high-energy neutrinos. In practice, we remove each neutrino from the spectrum that scatters in this way, as a second scattering of the final-state particles is unlikely. This approximation, which considerably simplifies the simulation, is justified as long as the aforementioned relation between interaction rate and Hubble rate is true.


\subsection{Interactions with other particles}

Collisions between the injected neutrinos and the relic $\phi$ (such as $\nu\phi \rightarrow \nu \phi$) are highly suppressed due to the small coupling corresponding to lifetimes $\tau_\phi > 10^4\,\mathrm{s}$ and can therefore be neglected in our calculation. Additionally, neutrinos do not interact directly with the background photons (apart from loop-corrections), and their interaction with the background electrons/positrons, like e.g.~$\nu e^\pm \rightarrow \nu e^\pm$, is suppressed by the small baryon-to-photon ratio. The same argument applies to direct interactions between neutrinos and any light nuclei $N$, i.e.~neutrinodisintegration of the form $\nu N \rightarrow \dots$ (cf.~appendix~\ref{app:deplete} for a numerical demonstration). However, the neutrino-induced interconversion between neutrons and protons during and shortly after thermal BBN, e.g.~$\nu p \rightarrow n e^+$, can play an important role \cite{Chang:2024mvg}. Nevertheless, we find that for the relatively long-lived relics considered in this work, the limits presented in \cite{Chang:2024mvg} are subdominant, meaning that the inclusion of this effect would not change our results.


\section{A brief overview of thermal and non-thermal BBN}\label{sec:photodis}
In this section, we provide a concise review of the standard approach for deriving bounds on BSM models using BBN results. We begin by introducing the observational light-element abundances and by briefly discussing the correlation between $\Neff$ and $\eta$. Next, we summarise the main features of photodisintegration and present the basic framework in which \texttt{ACROPOLIS} operates. Finally, we describe the hadrodisintegration formalism, which we have incorporated into \texttt{v2.0.0-dev} of \texttt{ACROPOLIS} for the purpose of this study.

\subsection{The light-element abundances}\label{sec:abundances}
In this work, we use the latest recommended values for the observed light-element abundances for  $\mathcal{Y}_p$, D$/^1$H as reported in \cite{ParticleDataGroup:2024cfk} and for $^3$He/D (only as an upper limit) as reported in \cite{Geiss2003}:
\begin{align}
    \mathcal{Y}_p &= (2.45 \pm 0.03) \times 10^{-1}\eqsp,\\
    \text{D}/^1\text{H} &=(2.547 \pm 0.029)\times 10^{-5}\eqsp,\label{eq:DH_obs} \\
    ^3\text{He}/\text{D} &= (8.3 \pm 1.5) \times 10^{-1}\eqsp.
\end{align}
Following the usual convention for the light-elements abundance, we use $N/N' \equiv Y_N/Y_{N'}$ with $N,\,N' \in \{ ^1\text{H},\,\text{D},\,^3\text{He},\,^7\text{Li} \}$, $Y_N \equiv n_N/n_b$, and $n_b$ being the baryon number density. For $^4\text{He}$, we instead use the common parameterisation $\mathcal{Y}_p \equiv 4Y_{^4\text{He}}$. In addition to the elements mentioned above, the primordial lithium abundance is inferred to be \cite{ParticleDataGroup:2024cfk}
\begin{align}
    ^7\text{Li}/^1\text{H} = (1.6 \pm 0.3) \times 10^{-10}\eqsp. 
\end{align}
However, this measurement is known to be in contradiction with the value predicted by standard BBN, a discrepancy commonly referred to as the (cosmological) lithium problem.\footnote{Currently, an astrophysical solution to this discrepancy appears most likely \cite{Korn:2024gel, Gao:2020}.} In this work, we do not try to resolve this tension 
and therefore (conservatively) take into account only the limits from the other abundances.

To calculate the evolution of the different abundances theoretically during thermal BBN (potentially with a modified background cosmology), we utilise a modified version of \texttt{AlterBBN} \cite{Arbey:2011nf,Arbey:2018zfh}, denoted as \texttt{AlterAlterBBN} in the following,\footnote{For the corresponding source code, see \url{https://github.com/hep-mh/alteralterbbn}.} which can handle arbitrary background cosmologies and further incorporates the most recent set of nuclear reaction rates \cite{Gariazzo:2021iiu,Pisanti:2020efz,Tisma:2019acf,2014ApJ...785...96T,Mossa:2020gjc}. As part of the calculation, we utilise the value of the baryon-to-photon ratio $\eta$ obtained by the latest Planck measurements \cite{Planck:2018vyg}. Notably, it is well known that $\eta$ has a strong correlation with $\Neff$, which we incorporate by following the prescription presented in \cite{Depta:2020wmr}. More precisely, the latest Planck data, i.e.~the 95\% confidence region ellipse (Planck TT, TE, EE+lowE+lensing+BAO) in the $\Omega_bh^2 - \Neff$ plane in figure 26 of \cite{Planck:2018vyg}, implies that for a given value of $\Neff$, the best-fit value of $\eta$ is given by
\begin{align}
    \eta_{\Neff} = \overline{\eta} + r\sigma_\eta \frac{\Neff - \overline{N}_\text{eff}}{\sigma_{\Neff}}\eqsp,\label{eq:def_eta}
\end{align}
with
\begin{align}
    \overline{\eta}=6.128\times10^{-10}\eqsp, \quad \sigma_\eta = 4.9 \times 10^{-12}\eqsp, \quad \overline{N}_\text{eff} = 2.991\eqsp, \quad \sigma_{\Neff}=0.169\eqsp, \quad r=0.677\eqsp.
\end{align}
Since D/$^1$H exhibits a rather strong dependence on $\eta$, the above uncertainty further introduces an additional error on the deuterium abundance, and consequently the total experimental uncertainty $\sigma^\text{exp}_{\text{D}/^1\text{H}}$ becomes
\begin{align}
\sigma_{\mathrm{D}/^1\mathrm{H}}^{\eta} &= \left| \frac{\mathrm{d}(\mathrm{D}/^1\mathrm{H})}{\mathrm{d}\eta} \sigma_\eta \sqrt{1 - r^2} \right|_{\eta = \eta_{N_{\text{eff}}}} \approx 0.024 \times 10^{-5}\eqsp, \\
\Rightarrow \quad \sigma_{\mathrm{D}/^1\mathrm{H}}^{\mathrm{exp}} &= \sqrt{\left( \sigma_{\mathrm{D}/^1\mathrm{H}}^{\mathrm{obs}} \right)^2 + \left( \sigma_{\mathrm{D}/^1\mathrm{H}}^{\eta} \right)^2} \approx 0.035 \times 10^{-5}\eqsp.
\end{align}
with $\sigma_{\mathrm{D}/^1\mathrm{H}}^{\mathrm{obs}} = 0.029 \times 10^{-5}$ from eq.~\eqref{eq:DH_obs} above.
For the other elements,  the uncertainty induced by $\eta$ is much smaller and thus negligibly compared to the one from observations. In this work, we use these measurements to calculate constraints at 95\% C.L., identical to the procedure used in \cite{Depta:2020zbh} (also cf.~this paper for more information).

\subsection{A quick look at photodisintegration}

The particles originating from neutrino collisions, e.g.~among others the $e^+ e^-$ pairs discussed in section~\ref{sec:eeonly}, feature an energy $E_0$ that is much larger than the one of thermal species, i.e.~$E_0 \gg T$. If the injected particles are electromagnetic, i.e.~electrons, positrons and photons, they can efficiently scatter on the background photons, thus inducing an EM cascade, which ultimately produces a spectrum of non-thermal photons. If the energy of the particles initiating this cascade is above the electron pair-creation threshold, i.e.~$ E_0 > E_{e e}^\text{th} \simeq m_e^2/(22T)$, any photon produced during this cascade with an energy above this threshold is rapidly depleted. In this case, it is well known that the cascade produces a universal spectrum of non-thermal photons, which is well approximated by \cite{Kawasaki:1994sc}\footnote{We made sure that the universal spectrum is always a good approximation for the parameters considered in this work. Therefore, we do not have to explicitly solve the equations governing the EM cascade to obtain the spectrum (cf.~e.g.~\cite{Poulin:2015woa,Poulin:2015opa})}
\begin{align}\label{eq:universal}
    \text{f}_{\gamma,\text{univ}}(T, E) \simeq \frac{S_\text{em}(T)}{\Gamma_\gamma(T, E)} \times \begin{cases}
        K_0 (E/E_X)^{-3/2}\quad &,\quad E<E_X\eqsp,\\
        K_0 (E/E_X)^{-2}\quad &,\quad E_X<E< E^\text{th}_{ee}\eqsp,\\
        0\quad &,\quad E>E^\text{th}_{ee}\eqsp.
    \end{cases}
\end{align}
Here, $K_0=E_0 E_X^{-2}\left[2+\ln(E^\text{th}_{ee}/E_X)\right]^{-1}$, $E_X=m_e^2/(80T)$, $\Gamma_\gamma$ is the total scattering rate of non-thermal photons with the background, and $S_\text{em}$ is the source-term describing the total amount of injected EM material.

Note that in the context of photodisintegration, the distribution function $\text{f}_\gamma$ is usually defined to be differential in energy instead of momentum $f_\gamma$, with both of them being related via
\begin{align}
    \text{f}_\gamma(T,E)=g_\gamma\frac{Ep}{2\pi^2}f_\gamma(T,p)\eqsp.
\end{align}
Here, $g_\gamma = 2$ is the number of photon degrees of freedom.

To evaluate eq.~\eqref{eq:universal} for our scenario, we have to compute the source term $S_\text{em}$ originating from the neutrino cascade. If all particles are injected with the same energy $E_0$,
\begin{align}
    S_{\text{em}}(T)=\sum_{x = e^\pm, \gamma} \frac{\text{d} n_x^\text{inj}}{\text{d} t} = \sum_{x = e^\pm, \gamma} \frac{1}{E_0}\frac{\text{d} \rho_x^\text{inj}}{\text{d} t}\label{eq:def_Sem}\eqsp,
\end{align}
where the different $\text{d} n_x^\text{inj}/\text{d} t$ ($\text{d} \rho_x^\text{inj}/\text{d} t$) denote the number (energy) densities of injected particles per time interval originating from the neutrino cascade. Notably, the first expression for $S_\text{em}$ in eq.~\eqref{eq:def_Sem} is indeed only true if all particles are injected with the same energy $E_0$; however, the second expression remains valid even if particles of different energies $E_0' < E_0$ contribute to the same universal spectrum with maximal injection energy $E_0$, which is the case for our scenario with $E_0 = m_\phi/2$. A more detailed calculation of this quantity is presented in section~\ref{sec:MC} below. At this point, let us note that $S_\text{em}$ receives independent contributions from all relevant processes, meaning that in our case
\begin{align}
    S_\text{em}(T) = S_\text{em}^\text{dcy}(T) + S_\text{em}^\text{fsr}(T) + S_\text{em}^\text{th}(T) + S_\text{em}^\text{n-th}(T)\eqsp.\label{eq:def_S_any}
\end{align}
Here, $S_\text{em}^\text{fsr}(T)$, $S_\text{em}^\text{th}(T)$, and $S_\text{em}^\text{n-th}(T)$ are the source terms originating solely from FSR, thermal, and non-thermal scattering, respectively (a detailed calculation of the different source terms is provided in section~\ref{sec:MC} below). Finally, $S_\text{em}^\text{dcy}(T)$ encodes potential direct contributions from the decay of the relic, e.g.~if the latter also decays into electron-positron pairs with a certain branching ratio $\text{BR}_{ee}$. We explore this possibility in more detail below.

After the EM cascade, the non-thermal photons originating from the EM cascade, i.e.~the ones encoded in  $\text{f}_{\gamma, \text{univ}}$, will initiate photodisintegration reactions, e.g.~$\gamma \text{D} \rightarrow n p$ among others. The Boltzmann equations governing these reactions can be written as~\cite{Depta:2020mhj} (dropping the $t/T$ dependence for convenience)
\begin{align}
\left[\frac{\text{d} n_X}{\text{d} t}\right]_\text{photo} = \sum_{j} n_{j} N_{j\gamma\rightarrow X} \int_{0}^{\infty}\text{d} E\;\text{f}_\gamma(E)\sigma_{j\gamma \rightarrow X}(E) - n_X \sum_{j'} \int_{0}^{\infty} \text{d} E\; \text{f}_\gamma(E)\sigma_{X\gamma \rightarrow j'}(E)\eqsp.
\label{eq:y_pdi}
\end{align}
Here, $\text{f}_\gamma = \text{f}_{\gamma, \text{univ}}$, $X$ is any of the light nuclei under consideration, $N_{j\gamma\rightarrow X}$ is the number of $X$ nuclei that is produced in the reaction $j\gamma\rightarrow X$, and $\sigma_{j\gamma\rightarrow X}$ ($\sigma_{X\gamma\rightarrow j'}$) are the cross-sections for the different reactions creating (destroying) $X$, which can be found in~\cite{Depta:2020mhj}. Given an expression for $S_\text{em}(T)$, we could in principle use the most recent stable version \texttt{v1.3.2} or the updated version \texttt{v2.0.0-dev} of \texttt{ACROPOLIS} to calculate $\text{f}_{\gamma, \text{univ}}$ and to solve eq.~\eqref{eq:y_pdi} in order to obtain the final light-element abundances after photodisintegration. However, the neutrino cascade also injects hadronic material and the hadrodisintegration reactions usually happen at similar timescales as the photodisintegration reactions. Consequently, both effects are generally intertwined and eq.~\eqref{eq:y_pdi} must be extended by additional reactions in order to handle both effects simultaneously. 

\subsection{A quick look at hadrodisintegration}\label{sec:hadro}

Overall, the hadrons produced in the neutrino cascade lead to significant effects, even though their production is generically subject to a stronger kinematic suppression. 

As already mentioned above, non-thermal scattering reactions (but also final-state radiation) usually produce less than one nucleon per interaction, with kinetic energies at the GeV-scale, i.e.~large enough to destroy $^4$He in a nucleon-nucleus interaction. This usually leads to a \textit{production} of deuterium. However, since photodisintegration mainly causes a \textit{destruction} of deuterium, there might be cancellations between both effects, which is why it is imperative to consider both of them simultaneously.

In fact, despite the smaller number of injected hadrons (compared to EM material), hadrodisintegration is still important due to the following two reasons:
\begin{enumerate}
    \item The cross-section for the destruction of $^4$He via a proton or neutron ($p/n + {}^4\text{He} \rightarrow \text{D} + \dots$) is about two orders of magnitude bigger than the one for the destruction of D via a photon ($\gamma + \text{D} \rightarrow \dots$)
    \item $^4$He is more than three orders of magnitude more abundant than D, which further boosts the relative importance of the hadrodisintegration rate.
\end{enumerate}
Therefore, even if the ratio between injected hadrons and photons is $\mathcal{O}(10^{-5})$, hadrodisintegration effects can still be as significant as photodisintegration effects.
In addition, for small lifetimes, $\tau_\phi \lesssim 10^4\,\mathrm{s}$, high-energy photons are efficiently depleted before undergoing any disintegration reactions, which is not true for neutrons.

The effect of injecting hadrons during or after thermal BBN has already been extensively discussed in \cite{Kawasaki:2005, Jedamzik:2006}, and more recently in \cite{Kawasaki:2018,Angel:2025dkw}. However, since there does not exist any public code, in this work, we perform an independent calculation based on these publications. Further details on our implementation can be found in appendix~\ref{app:hadro}, where we present a concise overview of the most important calculational steps. In this appendix, we also point out slight discrepancies between our calculation and the one in the literature, as well as how this changes the final disintegration processes.

While the details of our calculation are delegated to the appendix, let us point out how eq.~\eqref{eq:y_pdi} is modified in the presence of hadrodisintegration reactions. To this end, we introduce two parameters $\xi_{X}^{n/p}[n_j](T, K)$ for each element $X \in \{p, n, \text{D}, \text{T}, {}^3\text{He}, {}^4\text{He}\}$ (cf.~appendix \ref{app:hadro} for details on how to calculate these parameters). This quantity encodes the amount of $X$ particles that are produced (or destroyed if negative) due to the injection of a single neutron/proton with kinetic energy $K$ at a given temperature $T$. Moreover, it depends explicitly on the abundance of the present light elements, as indicated by the argument $n_j$. Given this quantity, the Boltzmann equation governing hadrodisintegration can be written as (again dropping the $t/T$ dependence for convenience)
\begin{align}
    \left[ \frac{\text{d} n_X}{\text{d} t} \right]_\text{hadro} = \sum_{y = n, p}\int_0^\infty \text{d}K\; \xi_{X}^y[n_j](K) \frac{\text{d}^2 n_y^\text{inj}}{\text{d} t \text{d} K}(K)\eqsp.
    \label{eq:y_hdi}
\end{align}
Here, $\text{d}^2 n_{n/p}^\text{inj}/(\text{d} t \text{d} K)$ is the number density of neutrons/protons that are injected per time and kinetic-energy interval. Combining this equation with eq.~\eqref{eq:y_pdi}, the total Boltzmann equation governing non-thermal BBN (NBBN) is consequently given by
\begin{align}
    \left[ \frac{\text{d} n_X}{\text{d} t} \right]_\text{NBBN} = \left[ \frac{\text{d} n_X}{\text{d} t} \right]_\text{photo} + \left[ \frac{\text{d} n_X}{\text{d} t} \right]_\text{hadro}\eqsp.
    \label{eq:y_nth}
\end{align}
At this point, let us note that in order to solve this equation, we further make the following two simplifications:
\begin{enumerate}
    \item Both processes that inject hadrons, i.e.~non-thermal scattering and FSR, inject an equal amount of protons and neutrons. We checked that this statement holds true within our setup to a reasonable accuracy.
    \item Both processes inject hadrons of a single, \emph{average} kinetic energy, which however still depends on the initial energy and the time of injection (cf.\ figure~\ref{fig:aux_out} below for details). This is in line with other simplifications we make in section~\ref{sec:MC}.
\end{enumerate}
Under these assumptions, eq.~\eqref{eq:y_hdi} simplifies to
\begin{align}
    \left[ \frac{\text{d} n_X}{\text{d} t} \right]_\text{hadro} \simeq \sum_{y = n, p} \left( \xi_{X}^{y}[n_j](K_\text{hd}^\text{fsr}) \frac{\text{d} n_\text{hd}^\text{fsr}}{\text{d} t} + \xi_{X}^{y}[n_j](K_\text{hd}^\text{n-th}) \frac{\text{d} n_\text{hd}^\text{n-th}}{\text{d} t} \right) \eqsp.
    \label{eq:y_hdi_app}
\end{align}
Here, $\text{d} n_\text{hd}^\text{n-th}/\text{d} t$ ($\text{d} n_\text{hd}^\text{fsr}/\text{d} t$) is the number density of injected hadrons per time interval that are produced with an average kinetic energy $K_\text{hd}^\text{n-th}$ ($K_\text{hd}^\text{fsr}$) due to non-thermal scattering (FSR). A detailed calculation of these quantities is presented in section~\ref{sec:MC} below. Note again, that the right-hand side of this equation explicitly depends on the light-element abundances $n_j$ via the parameters $\xi_X^y$. Finally, let us note that similar to eq.~\eqref{eq:def_S_any}, we can also define a hadronic source term
\begin{align}
    S_\text{hd}(T) = S_\text{hd}^\text{fsr}(T) + S_\text{hd}^\text{n-th}(T)\eqsp,\label{eq:def_Shd_any}
\end{align}
which receives contributions from FSR and non-thermal scattering reactions. While not directly entering the corresponding Boltzmann equation, this quantity is still relevant for the handling of the neutrino cascade below (cf.~eq.~\eqref{eq:rhonu_upd}).

For the parameters considered in this work, eq.~\eqref{eq:y_nth} for NBBN decouples from the Boltzmann equation for thermal, i.e.~standard, BBN (SBBN). In practice, we therefore \textit{(i)} track the neutrino cascade to calculate the modified Hubble rates as well as the quantities $S_\text{em}$, $K_\text{hd}^\text{fsr}$, etc. (see section~\ref{sec:MC} below for more information), \textit{(ii)} use the modified background cosmology as an input to solve the equations for SBBN with the help of \texttt{AlterAlterBBN}, and \textit{(iii)} use the resulting abundances as initial conditions for our updated version of \texttt{ACROPOLIS}, which then solves eq.~\eqref{eq:y_nth} with the expressions from eqs.~\eqref{eq:y_pdi} and \eqref{eq:y_hdi}, the latter of which has been simplified specifically for our scenario. 

At this point, let us also briefly comment on how we handle the theoretical errors entering eq.~\eqref{eq:y_nth}. Specifically, we consider \textit{(i)} the errors on the nuclear reaction rates used in \texttt{AlterAlterBBN}, and \textit{(ii)} the errors on our calculation of $\xi_X^y$, which amount to $\sim 20\% \equiv \epsilon_\xi$ (cf.~appendix~\ref{app:hadro}). Regarding \textit{(i)}, we run \texttt{AlterAlterBBN} three times with the high, mean and low values of the nuclear reaction rates, thus obtaining three sets of final SBBN abundances $Y_X^{\Gamma\;\text{high},0}$, $Y_X^{\Gamma\;\text{mean},0}$, and $Y_X^{\Gamma\;\text{low},0}$, respectively. We then solve eq.~\eqref{eq:y_nth} once for each set of these initial abundances,
leading to the final abundances $Y_X^{\Gamma\;\text{high}}$, $Y_X^{\Gamma\;\text{mean}}$, and $Y_X^{\Gamma\;\text{low}}$. Regarding \textit{(ii)}, we solve eq.~\eqref{eq:y_nth} with the initial condition $Y_X^{\Gamma\;\text{mean},0}$ and $\xi_X^y \{ (1+\epsilon_\xi), 1, (1-\epsilon_\xi) \}$ to obtain three sets of final abundances  $Y_X^{\xi\;\text{high}}$, $Y_X^{\xi\;\text{mean}}$, and $Y_X^{\xi\;\text{low}}$. We then calculate the theoretical error on the different abundances $Y_X$ as the minimal difference between all six values presented above. Finally, we combine the theoretical error calculated in this way with the observational error on the abundances presented in section~\ref{sec:abundances} to obtain the overall exclusion limits.

\vspace{0.2cm}\emph{A note on anti-nucleons.}
In our analysis, we do not include the effect of anti-baryons, which are subleading. Hadrodisintegration reactions caused by anti-nucleons have been studied in 
\cite{Kawasaki:2018} (henceforth \hyperlink{cite.Kawasaki:2018}{KKMT18}). Using figure~6 of \hyperlink{cite.Kawasaki:2018}{KKMT18}, we find that the effect of anti-neutrons (denoted by ``$\xi_{H_i,\overline{n}}$'') becomes bigger than the effect of protons (denoted by ``$\xi_{H_i,p}$'') only at very high energies, which are not relevant for the parameters considered in this work. Since all of our interactions conserve baryon number, high-energy anti-baryons and baryons are always produced in equal numbers. In this work, we explicitly take into account the fact that the non-thermal collisions respect baryon-number conservation. In practice, we subtract the injected baryons from the final abundance of hydrogen to make sure that it does not bias the baryon-to-photon ratio.

\vspace{0.2cm}\emph{A note on proton-neutron conversions.}
Protons and neutrons are not the only hadrons injected via the cascade. As shown in \cite{Kawasaki:2005} (henceforth \hyperlink{cite.Kawasaki:2005}{KKM05}), injected pions also contribute to a change in the light-element abundances. However, this happens more indirectly via the interconversion of protons and neutrons during SBBN, i.e.\ for $T\gtrsim 0.1\,$MeV or $t\lesssim 100\,$s, which is below the lifetimes we consider in this work.

\subsection{Interlude: The CMB $\Neff$ constraint}
In the literature, constraints from BBN are often compared to the one originating from $\Neff$ at the time of recombination. In the context of this work, the contribution of the additional non-thermal neutrinos to the effective number of neutrinos at the time $t_\text{rec}$ of recombination can be written as 
\begin{align}
    \Neff=\frac{\rho_{\nu}^\text{th}(t_\text{rec})+\rho_{\nu}^\text{n-th}(t_\text{rec})}{2\frac{7}{8}\frac{\pi^2}{30}\left(\frac{4}{11}\right)^{4/3}T(t_\text{rec})^4}\equiv \left[3+\Delta \Neff (t_\text{rec})\right]\left(\frac{11}{4}\right)^{4/3}\left(\frac{T_\nu(t_\text{rec})}{T(t_\text{rec})}\right)^4\eqsp.
    \label{eq:neff}
\end{align}
In this expression, which defines $\Delta \Neff$, we have split the energy density of the neutrinos $\rho_\nu$ into a thermal and a non-thermal part, with the former one being sourced by standard cosmology while the latter one arises solely from the decay of the dark-sector particles.\footnote{Naively, additional relativistic neutrinos will always increase $\Neff$, provided that interactions with the thermal bath (or even with other non-thermal neutrinos) do not substantially deplete their energy density. However, this is not necessarily the case, as shown in \cite{Boyarsky:2021yoh}, where interactions of neutrinos with the thermal neutrino background were considered. These interactions result in the reheating of the neutrino bath, which, in turn, partially rethermalises with the electromagnetic (EM) sector, leading to a decrease in $\Neff$. However, in our work this effect does not play a role.}
Like the one for BBN from $H$ and $\eta$, the limits considered in this section also derive from a change in the background cosmology, with the Planck mission \cite{Planck:2018vyg} constraining $\Delta \Neff$ to $\Delta \Neff(t_\text{rec})<0.29$. In appendix~\ref{app:neff}, we provide a simplified parameterisation of the resulting constraint in the context of this work.

\section{Overview of the numerical treatment}\label{sec:MC}

\subsection{Introductory remarks}
This section is dedicated to a thorough discussion of the computational approach that is used to track the neutrino cascade, as well as the resulting injections of EM and hadronic material. As already mentioned above, since we only consider lifetimes sufficiently beyond the end of thermal BBN, we can effectively ``factorise'' the calculation into three different steps:
\begin{itemize}
    \item \textbf{Neutrino cascade:}\\ In a first step, we track the evolution and decay of the relic $\phi$, as well as the secondary production of EM and hadronic material due to the injected neutrinos. Here, we use the properties of the relic, e.g.~its mass $m_\phi$ and its lifetime $\tau_\phi$, as input, while returning as an output \textit{(i)} the parameters describing the modified background cosmology, e.g.~$H$, $\text{d}T/\text{d}t$ etc., and \textit{(ii)} the parameters quantifying the EM and hadronic injections, e.g.~$S_\text{em}$, $K_\text{hd}^\text{fsr}$, etc. A summary of the different input and output parameters can be found in table~\ref{tab:input_output}.
    \item \textbf{Thermal nucleosynthesis (SBBN):}\\ In a second step, the modified cosmological history, specifically $H$ and $\text{d}T/\text{d}t$, is used as an input for \texttt{AlterAlterBBN}, which solves the Boltzmann equations for SBBN and returns the resulting light-element abundances.
    \item \textbf{Non-thermal nucleosynthesis (NBBN):}\\ In a final step, the modified background cosmology, the parameters describing the EM and hadronic injections, as well as the abundances obtained from \texttt{AlterAlterBBN}, are used as an input for our modified version of \texttt{ACROPOLIS}, which calculates the late-time modification of the previously produced light elements due to photo- and hadrodisintegration reactions. Note that, in order to use \texttt{ACROPOLIS} (cf.~\cite{Depta:2020mhj}), it is required to implement a model encoding the desired setup. As to not unnecessarily clutter the codebase of \texttt{ACROPOLIS}, we implement this model as part of an independent code called \texttt{Xena},\footnote{For the corresponding source code, see \url{https://github.com/hep-mh/xena}.} which can be used in conjunction with \texttt{v.2.0.0-dev} of \texttt{ACROPOLIS}.
\end{itemize}

In this section, we mainly focus on the first step of the calculation, i.e.~the neutrino cascade, since details on the other steps can be found in the literature. Specifically, \textit{(i)} \cite{Arbey:2011nf} discusses the equations governing SBBN, as well as details on their solution within \texttt{AlterBBN}, which is used as the basis for \texttt{AlterAlterBBN}, \textit{(ii)} \cite{Depta:2020mhj} discusses the inner workings of \texttt{ACROPOLIS}, including details on how to handle the process of photodisintegration, and \textit{(iii}) appendix~\ref{app:hadro} -- which is largely based on \cite{Kawasaki:2005,Kawasaki:2018} -- thoroughly discusses the hadronic cascade leading to hadrodisintegration as implemented in the modified version \texttt{v.2.0.0-dev} of \texttt{ACROPOLIS}, further including several references to the literature.

To determine the modified background cosmology as well as the parameters describing the EM/hadronic injections, we need to track the temporal evolution of both the relic particles $\phi$ and the injected neutrinos. Usually, this is done by solving the appropriate set of coupled Boltzmann equations. In simple scenarios or under simplifying assumptions, this set of equations can be solved semi-analytically. However, in more general cases, this task may be difficult to achieve even numerically. This is also true for the scenario at hand, which features coupled integro-differential equations. We therefore refrain from using the usual Boltzmann approach. Instead, we employ a \emph{probability-based} ansatz for tracking the evolution of the particles participating in the neutrino cascade. Regarding the handling of thermal scattering processes, we find that a regular Monte-Carlo approach works well, since the target particles feature a predefined thermal distribution (also cf.~\cite{Ovchynnikov:2024xyd,Ovchynnikov:2024rfu} for a Monte Carlo approach focusing on smaller lifetimes and smaller masses, i.e.\ $\tau_\phi\lesssim 1\,$s and $m_\phi\lesssim 2\,$GeV). However, when also considering non-thermal scattering reactions, which are not considered in \cite{Ovchynnikov:2024xyd,Ovchynnikov:2024rfu}, we find that a full Monte-Carlo approach is computationally unfeasible. Therefore, we instead opt for the procedure described in this section, which is still inspired by a Monte-Carlo solution, but instead utilises averages over the distributions of certain quantities in order to reduce the computationally expensive task of drawing each quantity from a full distribution.

In a nutshell, we therefore proceed as follows: We discretise the time evolution over small individual time steps $\text{d}t$, which are calculated based on the most relevant time-scales of the problem, i.e.~the Hubble scale and the time-scales for (non-)thermal scattering. In every step, we calculate the probability for the relic to decay as well as the probabilities for (non-)thermal scattering to occur. Based on these probabilities, we then handle the EW shower from FSR as well as the various scattering reactions in order to calculate the amount of injected EM/hadronic material using averaged quantities, while redshifting the abundance and energy of all neutrinos that have not scattered within the given time step.\footnote{As a conservative assumption, we neglect rescattered particles. See also the discussion in section~\ref{sec:non-thermal}.}
This way, we also track the temporal evolution of the relic particles and the non-thermal neutrinos, thus quantifying their influence on the background cosmology.
\begin{table}[t]
    \centering
    \renewcommand{\arraystretch}{1.2}
    \begin{tabular}{|c|c|c|}
        \hline
        \textbf{Variable} & \textbf{Description} & \textbf{Unit} \\
        \hline
        \multicolumn{3}{|c|}{\textbf{Input parameters}} \\
        \hline
        $Y_\phi$ & Abundance of $\phi$  & - \\
        $m_\phi$ & Mass of $\phi$ & MeV \\
        $\tau_\phi$ & Lifetime of $\phi$ & s \\
        BR$_{ee}$ & Branching ratio into $e^+e^-$ & -\\
        \hline
        \multicolumn{3}{|c|}{\textbf{One-dimensional output parameters}} \\
        \hline
        $f_\phi$ & Fractional abundance of $\phi$, $\frac{\Omega_\phi}{\Omega_{\text{DM}}}$ & - \\
        $N_{\text{eff}}$ & Effective number of neutrinos & - \\
        $\eta$ & Baryon-to-photon ratio & - \\
        \hline
        \multicolumn{3}{|c|}{\textbf{Two-dimensional output parameters}} \\
        \hline
        $t$ & Time & s \\
        $T$ & Temperature & MeV \\
        $\frac{\text{d}T}{\text{d}t}$ & Time-temperature relation & MeV$^2$ \\
        $T_\nu$ & Neutrino temperature & MeV \\
        $H$ & Hubble rate & MeV \\
        $n_\gamma$ & Photon number density & MeV$^3$ \\
        $S_\text{em}$ & Electromagnetic source term, eq.~\eqref{eq:def_Sem} & MeV$^4$ \\
        $S_\text{hd}$ & Hadronic source term  & MeV$^4$  \\
        $\frac{\text{d}n_\text{hd}^\text{n-th}}{\text{d}t}$ & Injected hadrons per time from non-th. scattering, eq.~\eqref{eq:y_hdi} & MeV$^4$ \\
        $K_\text{hd}^\text{n-th}$ & Avg. kinetic energy of hadrons from non-th. scattering, eq.~\eqref{eq:y_hdi} & MeV \\
        $\frac{\text{d}n_\text{hd}^\text{fsr}}{\text{d}t}$ & Injected hadrons per time interval from FSR, eq.~\eqref{eq:y_hdi} & MeV$^4$ \\
        $K_\text{hd}^\text{fsr}$ & Avg. kinetic energy of hadrons from FSR, eq.~\eqref{eq:y_hdi} & MeV \\
        \hline
    \end{tabular}
    \caption{Description of the input/output parameters used for our numerical treatment. Some of the output quantities are also shown for a specific benchmark point in figure~\ref{fig:aux_out}.}
    \label{tab:input_output}
\end{table}

As shown in table~\ref{tab:input_output}, the input for the simulation comprises the mass $m_\phi$, the lifetime $\tau_\phi$, and the abundance $Y_\phi = n_\phi/s$ of the relic, with the latter quantity featuring the total entropy density $s$, which is approximately equal to the one of the SM $s_\text{sm}$.
Additionally, we may also parameterise the abundance of $\phi$ via the parameter $f_\phi \equiv \Omega_\phi^0/\Omega_\mathrm{DM}^0$. Here, $\Omega_\phi^0$ is the fraction of energy relative to the critical energy density that $\phi$ would have today if it would not decay, and $\Omega_\mathrm{DM}^0=0.26$. In the case of negligible entropy production, there exists a linear, one-to-one relation between $m_\phi Y_\phi$ and $f_\phi$. This is the case for all relevant parts of parameter space, which is why in the following, we often use $m_\phi Y_\phi$ and $f_\phi$ interchangeably. Using the above quantities, the number of $\phi$ particles per unit volume is given by 
\begin{equation}
   n_\phi \simeq Y_\phi s_\text{sm}=\tilde{n}_\phi e^{-t/\tau_\phi}  = \frac{\Omega_{\phi}^0\rho_\text{crit}^0}{(m_\phi R^3)} e^{-t/\tau_\phi}\eqsp.
   \label{nphit}
\end{equation}
Here, $\tilde{n}_\phi$ is the `would-be' number density of $\phi$ if it was not decaying, $R$ is the scale factor (we fix the scale factor to be equal to $1$ today), and $\rho_\text{crit}^0$ is the critical energy density. Additionally, we also consider the case of a non-vanishing branching ratio of $\phi$ directly into electron-positron pairs, which introduces an additional input parameter $\text{BR}_{ee}$. We discuss the impact of this parameter in more detail in section~\ref{sec:br}. Finally, let us note that for the evaluation of other cosmological quantities, we assume no additional dark-sector contributions beyond $\phi$. Regarding the different input parameters, in the following, we present a concise summary of the parameter space covered in this work:
\begin{itemize}
    \item $\bm{m_\phi}$: We restrict our calculation to $m_\phi \in [1\,\text{GeV},10\,\text{TeV}]$. The lower boundary is motivated by the fact that the limits weaken significantly with mass and for $m_\phi < 1\,\mathrm{GeV}$ are only determined by the modified cosmology. Specifically, the kinematic threshold for the thermal production of electrons ($ET\gtrsim m_e^2$) is usually not met for sub-GeV relics.
    \item $\bm{\tau_\phi}$: We focus on lifetimes $\tau_\phi \in [10^4,10^{11}]\,$s. Most notably, the lower limit is motivated by the breakdown of our simplifying assumptions, i.e.~neutrinos may also scatter on background electrons or into muons, SBBN and NBBN no longer factorise, etc. Instead, the upper limit is motivated by the fact that for larger lifetimes, other measurements provide more powerful bounds, e.g.~those from CMB anisotropies or spectral distortions \cite{Hambye:2021moy,Slatyer:2012yq,Poulin:2016anj}.
    \item $\bm{f_\phi\,|\,Y_\phi}$: 
    The upper limit of values to consider for $f_\phi$ depends on $\tau_\phi$ as it is determined by the $\Neff$ constraints, which gives the dominant limit for sufficiently large values of $f_\phi$, thus making other constraints irrelevant. In this context, we have confirmed that our numerical results, i.e.~specifically the resulting light-element abundances, are calculated consistently well above the $\Neff$ limit.
    \item $\bm{\textbf{BR}_{ee}}$: While we consider the full range $\text{BR}_{ee} \in [0, 1]$, our main focus remains on pure neutrino injections, i.e.~$\text{BR}_{ee} = 0$.
\end{itemize}

\subsection{Algorithmic implementation of the neutrino cascade}
\tikzstyle{process} = [rectangle, rounded corners, minimum width=3cm, minimum height=1.5cm, text centered, draw=black, fill=blue!20]
\tikzstyle{arrow} = [thick,->,>=stealth]
\tikzstyle{title} = [rectangle, text centered, minimum height=1cm]

\begin{figure}[!ht]
\centering
\resizebox{\textwidth}{!}{%
\begin{circuitikz}
\tikzstyle{every node}=[font=\large]
\node [font=\normalsize, color={rgb,255:red,45; green,148; blue,200}] at (11.5,34) {Input quantities};
\draw  (8.25,16.25) -- (15.25,16.25) -- (14.5,15) -- (7.5,15) -- cycle;
\draw [ fill={rgb,255:red,247; green,247; blue,255} , line width=0.6pt , rounded corners = 12.0] (2.25,30.75) rectangle (20.25,16.5);
\draw [ fill={rgb,255:red,255; green,238; blue,238} , rounded corners = 9.0, dashed] (16.25,24.25) rectangle  (19.75,19.5);
\draw  (10,34.5) -- (13.75,34.5) -- (13,33.25) -- (9.25,33.25) -- cycle;
\draw  (7.5,32.75) rectangle (15,31);
\draw [fill={rgb,255:red,238; green,240; blue,255}] (6.75,30) rectangle (15.75,28.25);
\draw [fill={rgb,255:red,238; green,240; blue,255}] (6.5,19) rectangle (16,17.25);
\draw [fill={rgb,255:red,238; green,240; blue,255}] (6.5,21.25) rectangle (16,20);
\draw [fill={rgb,255:red,238; green,240; blue,255}] (6.5,24) rectangle (16,22.25);
\draw [fill={rgb,255:red,238; green,240; blue,255}] (6.5,27.25) rectangle (16,25);
\node [font=\normalsize] at (11.5,33.5) {$m_\phi, \tau_\phi, Y_\phi, \text{BR}_{ee}$};
\node [font=\normalsize, color={rgb,255:red,45; green,148; blue,200}] at (11.25,32.25) {Initial conditions};
\node [font=\normalsize] at (11.25,31.75) {Start computing at $t_0 = \tau_\phi/\Lambda$, with $\Lambda \gg 1$};
\node [font=\normalsize] at (11.5,31.25) {and evaluate background cosmology.};
\node [font=\normalsize, color={rgb,255:red,45; green,148; blue,200}] at (11.25,29.5) {Decay of the relic};
\node [font=\normalsize] at (11.25,29) {Handle $\phi$ decay into $\nu\nu$ and $e^+e^-$, cf.~eqs.~\eqref{eq:med_decay}-\eqref{eq:Dnu_wrong}.};
\node [font=\normalsize] at (11.25,25.5) {};
\node [font=\normalsize, color={rgb,255:red,45; green,148; blue,200}] at (11.25,26.75) {FSR and neutrino oscillations};
\node [font=\normalsize] at (11.25,26.25) {Compute EM and HD injections, cf.~eqs.~\eqref{eq:en_shower}-\eqref{eq:dnhd_dt}.};
\node [font=\normalsize] at (11.25,25.75) { Apply neutrino oscillations, cf.~eq.~\eqref{eq:nu_osc}.};
\node [font=\normalsize] at (11,25.25) {Compute the $\nu$ spectra, cf.~eq.~\eqref{eq:nu_spectrum}.};
\node [font=\normalsize] at (11.25,28.5) {Define time step $\Delta t$, cf.~eq.~\eqref{eq:t_step}.};
\node [font=\normalsize, color={rgb,255:red,45; green,148; blue,200}] at (11.25,23.5) {Thermal scattering};
\node [font=\normalsize] at (11.25,23) {Compute EM injections, cf.~eqs.~\eqref{eq:gamma_thermal}-\eqref{eq:Sem_th}.};
\node [font=\normalsize, color={rgb,255:red,45; green,148; blue,200}] at (11.25,20.75) {Non-thermal scattering};
\node [font=\normalsize] at (11.25,20.25) {Compute EM and HD injections, cf.~eqs.~\eqref{eq:nn_XX_nth_simp}-\eqref{eq:nth_avgkin}.};
\node [font=\normalsize, color={rgb,255:red,45; green,148; blue,200}] at (11.25,18.5) {Redshift};
\node [font=\normalsize] at (11.25,18) {Adjust the $\nu$ spectra, cf.~eq.~\eqref{eq:Dnu_rd}.};
\node [font=\normalsize] at (16.5,12.25) {};
\node [font=\normalsize] at (13.5,14) {};
\node [font=\normalsize, color={rgb,255:red,45; green,148; blue,200}] at (11.25,15.75) {Output quantities};
\node [font=\normalsize] at (11.25,15.25) {$H$, $S_\text{em}$, etc. (cf.~table~\ref{tab:input_output})};
\node [font=\normalsize] at (11.25,13.5) { and \texttt{ACROPOLIS} \cite{Depta:2020mhj} / \texttt{Xena}};
\node [font=\normalsize] at (11.25,17.5) {Update background quantities for next iteration.};
\node [font=\LARGE] at (11.75,24) {};
\node [font=\normalsize] at (11.25,14) {Input for \texttt{AlterAlterBBN}};
\draw  (7.5,14.5) rectangle (15,13.25);
\draw [ rounded corners = 9.0, dashed] (16.75,24) rectangle  node {\normalsize EM injections}  (19.25,22.25);
\draw [ rounded corners = 9.0, dashed] (16.75,21.5) rectangle  node {\normalsize HD injections}  (19.25,19.75);
\draw [fill={rgb,255:red,238; green,240; blue,255}] (17,29) -- (18,28) -- (19,29) -- (18,30) -- cycle;
\node [font=\normalsize] at (18,29) {BR$_{ee} \neq 0$};
\draw [dashed] (15.75,29) -- (17,29);
\draw [dashed] (18,19.5) -- (18,15.75);
\draw [->, >=Stealth, dashed] (18,15.75) -- (15,15.75);
\node [font=\normalsize] at (11.25,22.5) {Remove secondary $\nu$'s, cf.~paragraph below eq.~\eqref{eq:Sem_th}};
\draw [ color={rgb,255:red,30; green,98; blue,133}, line width=1.5pt, short] (11.25,19.75) -- (6.25,19.75);
\draw [ color={rgb,255:red,30; green,98; blue,133}, line width=1.5pt, short] (11.25,19.5) -- (5.25,19.5);
\draw [ color={rgb,255:red,30; green,98; blue,133}, line width=1.5pt, short] (5.25,19.5) -- (5.25,24.5);
\draw [ color={rgb,255:red,30; green,98; blue,133}, line width=1.5pt, short] (6.25,19.75) -- (6.25,21.75);
\draw [ color={rgb,255:red,30; green,98; blue,133}, line width=1.5pt, ->, >=Stealth] (5.25,24.5) -- (11.25,24.5);
\draw [ color={rgb,255:red,30; green,98; blue,133}, line width=1.5pt, ->, >=Stealth] (6.25,21.75) -- (11.25,21.75);
\draw [ color={rgb,255:red,30; green,98; blue,133}, line width=1.5pt, short] (11.25,16.75) -- (4.25,16.75);
\draw [ color={rgb,255:red,30; green,98; blue,133}, line width=1.5pt, short] (4.25,16.75) -- (4.25,30.5);
\draw [ color={rgb,255:red,30; green,98; blue,133}, line width=1.5pt, ->, >=Stealth] (4.25,30.5) -- (11.25,30.5);
\node [font=\small, color={rgb,255:red,30; green,98; blue,133}, rotate around={90:(0,0)}] at (6,20.75) {Loop over $E_j$};
\node [font=\small, color={rgb,255:red,30; green,98; blue,133}, rotate around={90:(0,0)}] at (5,22) {Loop over $E_i$};
\node [font=\small, color={rgb,255:red,30; green,98; blue,133}, rotate around={90:(0,0)}] at (4,23.5) {Loop over $t$};
\node [font=\normalsize] at (4.25,31) {\textbf{Neutrino cascade}};
\draw [->, >=Stealth] (11.25,20) -- (11.25,19);
\draw [->, >=Stealth] (11.25,15) -- (11.25,14.5);
\draw [->, >=Stealth] (11.25,33.25) -- (11.25,32.75);
\draw [->, >=Stealth] (11.25,31) -- (11.25,30);
\draw [->, >=Stealth] (11.25,28.25) -- (11.25,27.25);
\draw [->, >=Stealth] (11.25,25) -- (11.25,24);
\draw [->, >=Stealth] (11.25,22.25) -- (11.25,21.25);
\draw [ color={rgb,255:red,45; green,148; blue,200}, , line width=1.5pt](11.25,17.25) to[short] (11.25,16.75);
\draw [ color={rgb,255:red,13; green,113; blue,18}, line width=1.5pt, ->, >=Stealth, dashed] (18,28) -- (18,24);
\draw [ color={rgb,255:red,13; green,113; blue,18}, line width=1.5pt, ->, >=Stealth, dashed] (16,23) -- (16.75,23);
\draw [ color={rgb,255:red,13; green,113; blue,18}, line width=1.5pt, ->, >=Stealth, dashed] (16,25) -- (16.75,24);
\draw [ color={rgb,255:red,13; green,113; blue,18}, line width=1.5pt, ->, >=Stealth, dashed] (16,21.25) -- (16.75,22.25);
\draw [ color={rgb,255:red,165; green,29; blue,45}, line width=1.5pt, ->, >=Stealth, dashed] (16,20.5) -- (16.75,20.5);
\draw [ color={rgb,255:red,165; green,29; blue,45}, line width=1.5pt, dashed] (16,26.25) -- (20,26.25);
\draw [ color={rgb,255:red,165; green,29; blue,45}, line width=1.5pt, dashed] (20,26.25) -- (20,20.5);
\draw [ color={rgb,255:red,165; green,29; blue,45}, line width=1.5pt, ->, >=Stealth, dashed] (20,20.5) -- (19.25,20.5);
\node [font=\small] at (16.5,29.25) {if};
\end{circuitikz}
}%

\caption{Flowchart of the code developed in the context of this study to determine the injection of EM and hadronic material from the (neutrino) cascade induced by the decay of a relic particle.}
\label{fig:flowchart_code}
\end{figure}
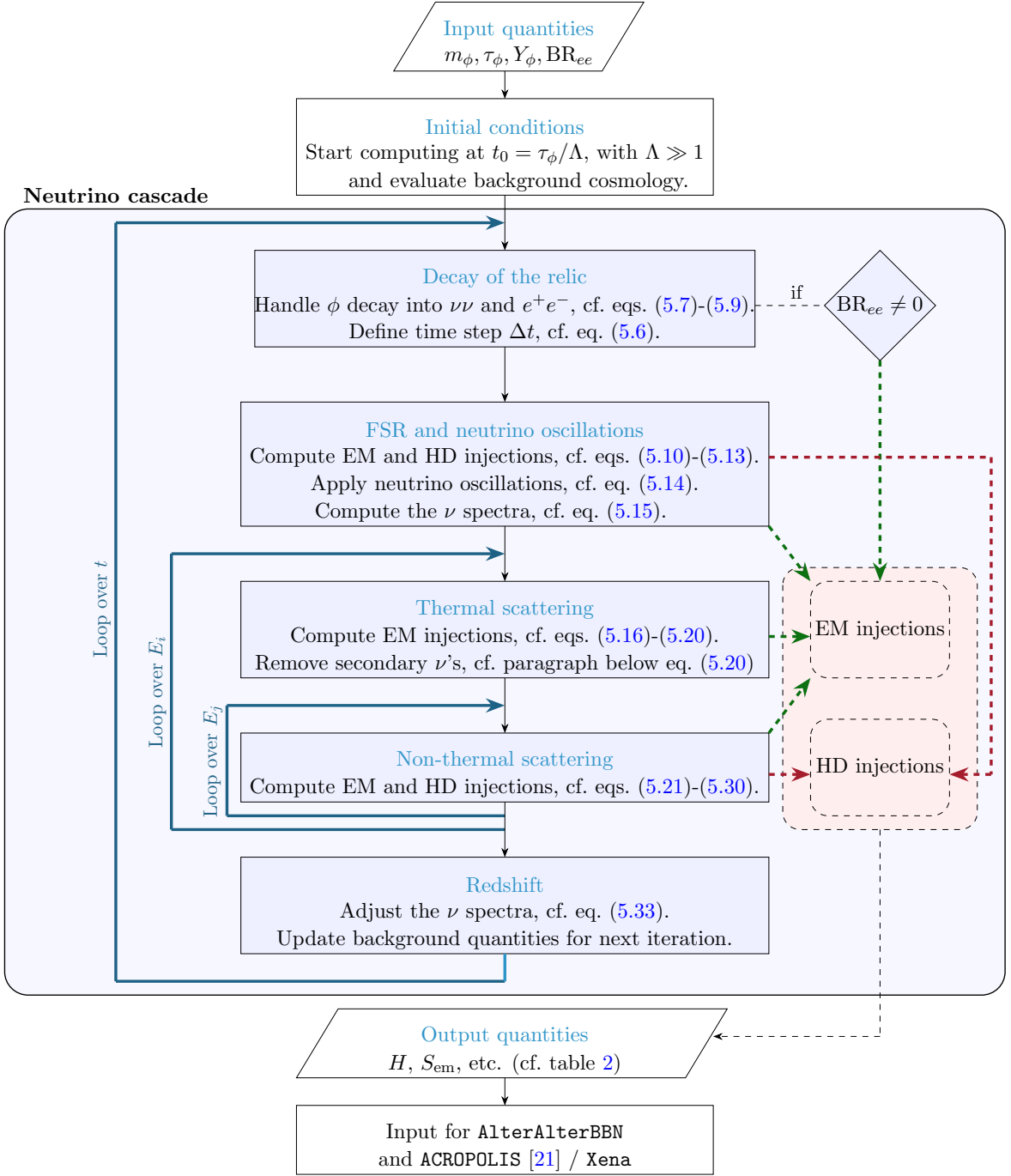
\emph{To aid the reader, in figure~\ref{fig:flowchart_code} we provide a visual representation of the algorithmic implementation discussed on this section. It is instructive to consult this figure while reading through this section.}\\

\noindent For the purpose of this section, we describe the abundance of any particle $x$ at a given time $t$ via the parameter $a_x(t)\equiv N_x(t)/N_\phi^\text{tot}$, where $N_x(t)$ is the number of $x$ particles at time $t$ inside a comoving volume element, and $N_\phi^\text{tot}$ is the total number of $\phi$ particles prior to any decay processes, again within a comoving volume element. This definition is convenient, since $a_x(t)$ is not affected by Hubble expansion. By choosing some reference time $t_0 \ll \tau_\phi$ --- see the initial conditions below --- with the scale factor $R_0 \equiv R(t_0)$ and $n_{\phi, 0} \equiv n_\phi(t_0)$, the number density $n_x(t)$ of $x$ at any other time $t \neq t_0$ can then be calculated from $a_x(t)$ via the relation
\begin{align}
    n_x(t) = a_x(t) n_{\phi, 0} [R_0 / R(t)]^3\eqsp.\label{eq:def_nx}
\end{align}
For the neutrinos, we therefore end up with six different quantities, which we conveniently write in vector notation (here and in the following, a bar under a quantity denotes a vector in neutrino space)
\begin{align}
    \underline{a}_\nu(t) \equiv (a_{\nu_e}, a_{\nu_\mu}, a_{\nu_\tau},a_{\bar\nu_e}, a_{\bar\nu_\mu}, a_{\bar\nu_\tau})^T(t)\eqsp.\label{eq:def_vec_a}
\end{align}
Using eq.~\eqref{eq:def_nx}, we therefore find $\underline{n}_\nu(t) = \underline{a}_\nu(t) n_{\phi,0} [R_0/R(t)]^3$. At this point, let us note that we ultimately want to quantify the effects of the different neutrino interactions. Regarding the non-thermal scattering processes specifically (cf.~eq.~\eqref{eq:rate_def_gen} with $f_Y = f_\nu$),\footnote{Some interactions such as FSR and neutrino oscillations are independent of the spectrum.} this requires the knowledge of the neutrino spectra $\underline{f}_\nu(t, E)$ at any given time $t$. 
As part of the code, we encode said spectrum in terms of a list $D_\nu(t)$, which --- for a given time $t$ --- stores discrete pairs of the form $[E: \underline{a}_\nu(t, E)]$. Consequently, $D_\nu(t)$ can usually be represented as\footnote{For performance reasons, we assume that two energies $E$ and $E'$ are equal if $|E'-E|/E \ll 1$. In this case, we only add one element with energy $E\sim E'$ and an abundance vector that is the sum of the two individual vectors.}
\begin{align}
    D_\nu(t) = \bigg\{ \big[E_1: \underline{a}_{\nu}(t, E_1)\big], \big[E_2: \underline{a}_{\nu}(t, E_2)\big], \cdots\bigg\}\eqsp.
\end{align}
Such a list encodes a spectrum of particles with discrete energies $E_i$ and associated abundances $\underline{a}_\nu(t, E_i)$, i.e.
\begin{align}
\underline{f}_\nu(t, E) = 2\pi^2\sum_i \underbrace{\underline{a}_{\nu}(t, E_i) n_{\phi, 0} [R_0/R(t)]^3}_{\equiv \underline{n}_{\nu}(t, E_i)}\frac{\delta(E - E_i)}{E_i^2}\eqsp.\label{eq:def_spec}
\end{align}
Here, $\underline{n}_{\nu}(t, E_i)$ are the individual number densities of neutrinos with fixed energy $\sim E_i$ at time $t$, which, when summed over $E_i$, yield the total number densities of neutrinos $\underline{n}_{\nu}(t) = \int \underline{f}_\nu(t, E) \text{d}^3p /(2\pi)^3 = \sum_{E_i} \underline{n}_\nu(t, E_i)$. In order to reduce visual clutter, we drop the time dependence of most quantities in the following. 

\begin{figure}
    \centering
    \includegraphics[width=0.49\linewidth]{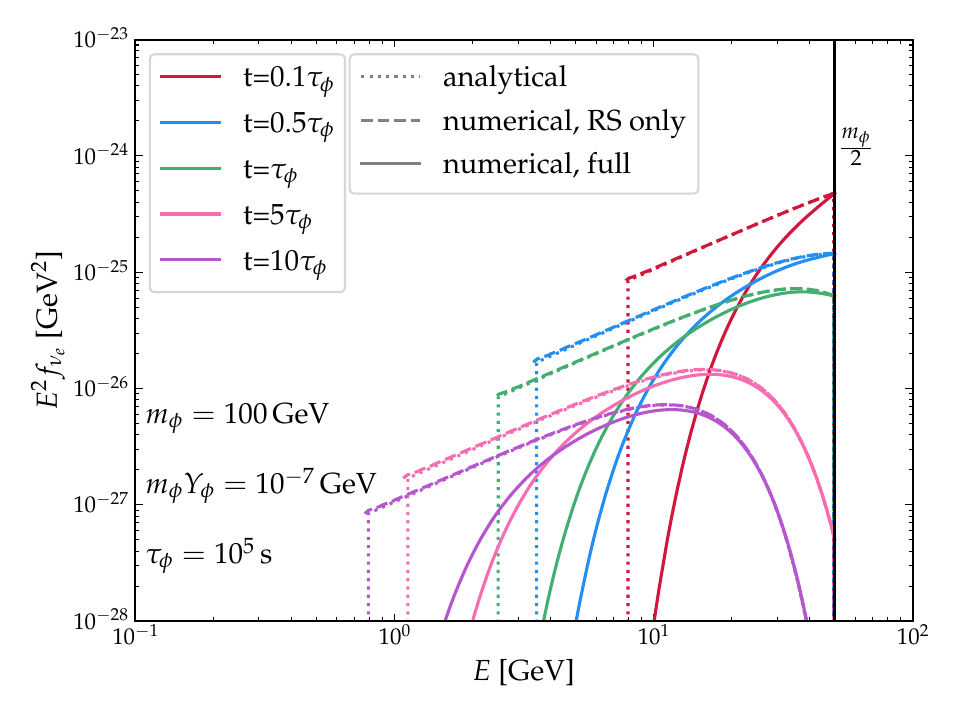}
    \includegraphics[width=0.49\linewidth]{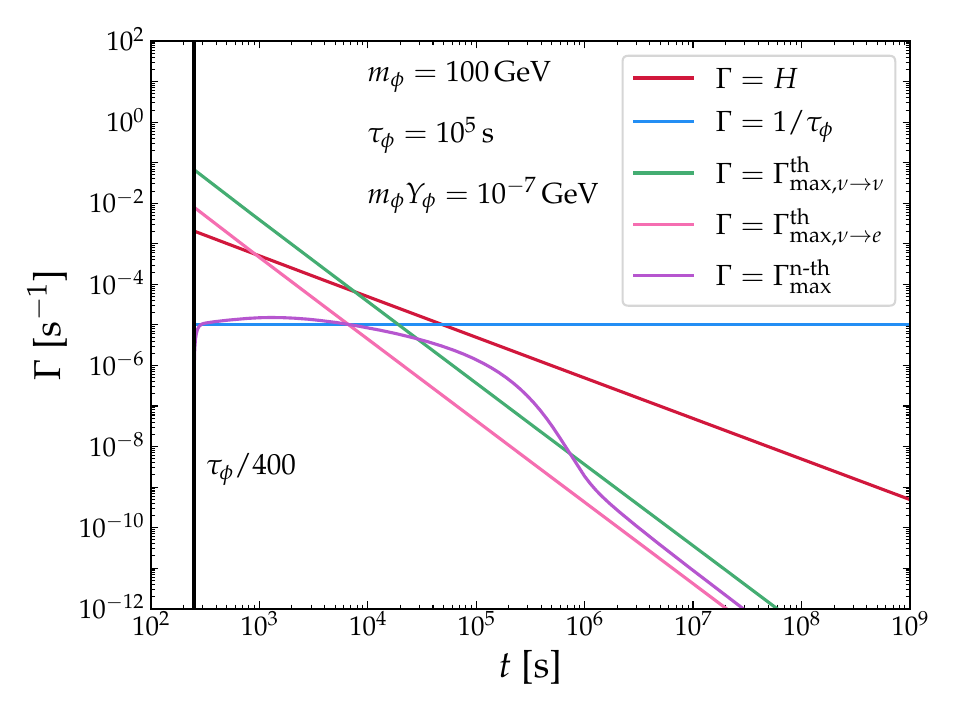}
    \caption{\textbf{Left:} Phase-space distribution function of the injected neutrinos from $\phi \rightarrow \nu_e \bar{\nu}_e$ for different ratios $t/\tau_\phi \in \{0.1, 0.5, 1, 5, 10\}$, taking into account all effects (solid), or only the redshift (dashed). For comparison, the latter case is also compared to the corresponding analytical approximation from appendix~\ref{app:ana} (dotted). Here, we conservatively only show those particles that have never undergone any scattering reactions up to the given time. \textbf{Right:} Time evolution of a selection of rates. Note that all scattering rates are taken for $\nu_e$ and we use their maximum values as introduced in eq.~\eqref{eq:t_step}.}
    \label{fig:spectra}
\end{figure}

For visualisation, in the left panel of figure~\ref{fig:spectra} (left) we show the time evolution of the neutrino spectrum for a benchmark point close to the exclusion limit, i.e.~$m_\phi=100\,$GeV, $\tau_\phi=10^5\,$s and $m_\phi Y_\phi=10^{-7}\,$GeV. Here, we focus on the temporal evolution close to the lifetime of $\phi$, i.e.~at times $t/\tau_\phi \in \{0.1, 0.5, 1, 5, 10\}$, in order to capture the most interesting effects. Specifically, we compare the full numerical evolution (``full'', solid line) with the spectrum that is obtained from only $\phi$ decay and redshifting (``RS only'', dashed line). Moreover, we compare the latter result with a corresponding analytical calculation, which has been obtained under the assumption of a vanishing scattering rate (dotted line, cf.~appendix~\ref{app:ana}). Overall, we find that the analytical calculation captures the full evolution very well, as long as neutrino interactions can be neglected. However, we also find that the neutrino spectrum is depleted for those particles that have been injected at an early time (left part of the curve). Note that in our approximation, we remove all neutrinos from the spectrum that have previously scattered. However, in reality these neutrinos would not vanish completely, but their energy would be moved to lower energy bins (cf.~section~\ref{subsec:el_scattering}). Numerically, we find that for $t=\tau_\phi/10$, roughly 40\% of all injected neutrinos did scatter before. Considering later times, the depletion is less efficient, e.g.~for $t=10\tau_\phi$ almost 90\% of all injected neutrinos did not yet scatter and hence are still present in the spectrum. Note that the cutoff in the spectrum towards small energies comes from the fact that we only track the spectrum starting from some initial time $t_0$, which is discussed in the next section. However, due to the suppression of the spectrum for such energies, this approach is fully justified.

\paragraph{Initial conditions}
Ultimately, we want to start our calculation at some time $t_0 = \tau_\phi/\Lambda$ with $\Lambda \gg 1$, i.e.~before the heavy relic starts to decay in significant amounts. Given a value of $\Lambda$, we therefore have to first determine the corresponding temperatures $T_0 \equiv T(t_0)$ and $T_{\nu, 0} \equiv T_\nu(t_0)$. To this end, let us note that $\rho_\phi \propto R^{-3}$ and $\rho_\text{sm} \propto R^{-4}$, meaning that for large temperatures, say $T \gg T_\text{eq}$ with $T_\text{eq}$ being the temperature at which $\rho_\text{sm} = \rho_\phi$, the energy density of the relic becomes negligible compared to the one of the SM. Consequently, above some temperature $T_i > \max\{m_t, T_\text{eq}\}$ with the mass $m_t$ of the top quark, the Hubble rate obeys $H(T) \propto \sqrt{\rho_\text{sm}(T)} \propto T^2$. For the parameters considered in this work, we find that $T_i = 10^4\,\mathrm{GeV}$ is a valid choice. At this temperature, $T = T_\nu$, $H\propto T^2$ and $\text{d}T/\text{d}t = -H T$, which implies $t_i = t(T_i) = [2H(T_i)]^{-1}$ as well as $T_{\nu,i} = T_\nu(T_i) = T_i$. Finally, to go from $t_i$ to $t_0$, we implement the following procedure, starting from the temperature $T=T_i$:
\begin{enumerate}[label=\textit{(\roman*)}]
    \item Calculate the energy density of the relic.
    \item Calculate the Hubble rate $H(T) = \sqrt{\rho_\phi(T) + \rho_\text{sm}(T)}/(\sqrt{3}M_\text{pl})$.
    \item Define a small time step $\Delta t = \tilde{\varepsilon}/H$ with $\tilde\varepsilon \ll 1$ (we find that $\tilde{\varepsilon} = 10^{-4}$ gives a good compromise between speed and accuracy).
    \item Calculate the respective time-temperature relations $\text{d} T/\text{d} t$ and $\text{d} T_\nu/\text{d} t$ with the previously determined Hubble rate and use them to update the time $t \rightarrow t + \Delta t$, the temperature $T \rightarrow T +  (\text{d}T/\text{d} t) \Delta t$, as well as the neutrino temperature $T_\nu \rightarrow T_\nu + (\text{d}T_\nu/\text{d} t) \Delta t$.
    \item Repeat until $t \geq t_0$.
\end{enumerate}
By following this procedure, we obtain a consistent set of initial conditions $(t_0, T_0, T_{\nu,0})$ with $t_0 \simeq \tau_\phi / \Lambda$. In this work, we specifically set $\Lambda = 400$, noting that larger values do not significantly change our results but are computationally much more expensive. The value of $t_0$ is then used as the reference value for $n_{\phi, 0}$ and $R_0$ from the previous paragraph, which implies $a_\phi(t_0) \simeq 1$ and $D_\nu(t_0) = \varnothing$. Setting $a_\phi(t_0) \simeq 1$ again insinuates that we neglect all potential decays until this time, which we find to be a good approximation, since the fraction of particles that decay before, e.g.~$|1-e^{1/\Lambda}| \sim 0.3\%$ for $\Lambda = 400$, is sufficiently small.

\begin{figure}[!t]
 \centering
    \begin{subfigure}{0.49\textwidth}
      \centering
      \includegraphics[width=\textwidth]{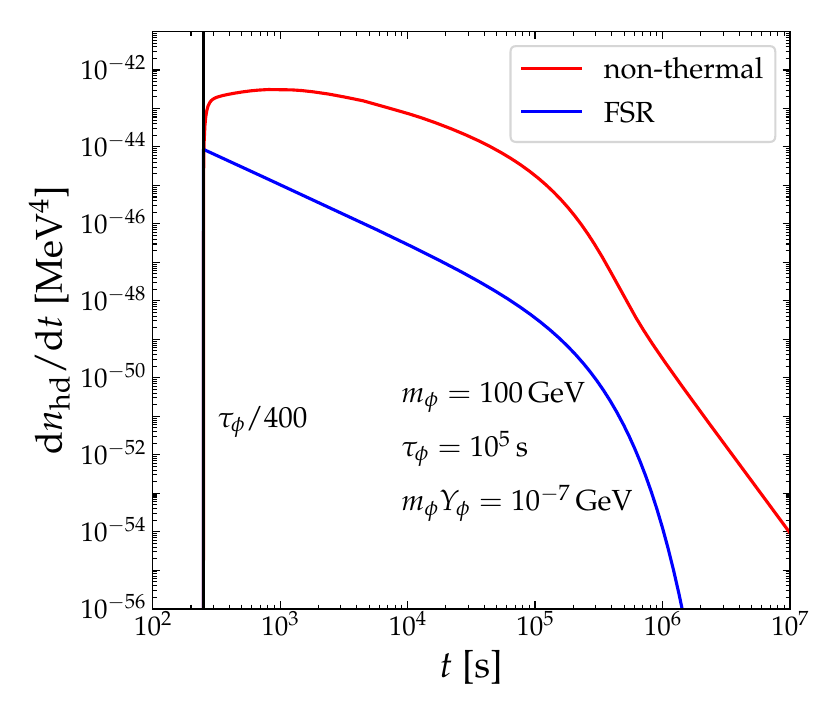}
    \end{subfigure}
    \begin{subfigure}{0.49\textwidth}
      \centering
      \includegraphics[width=\textwidth]{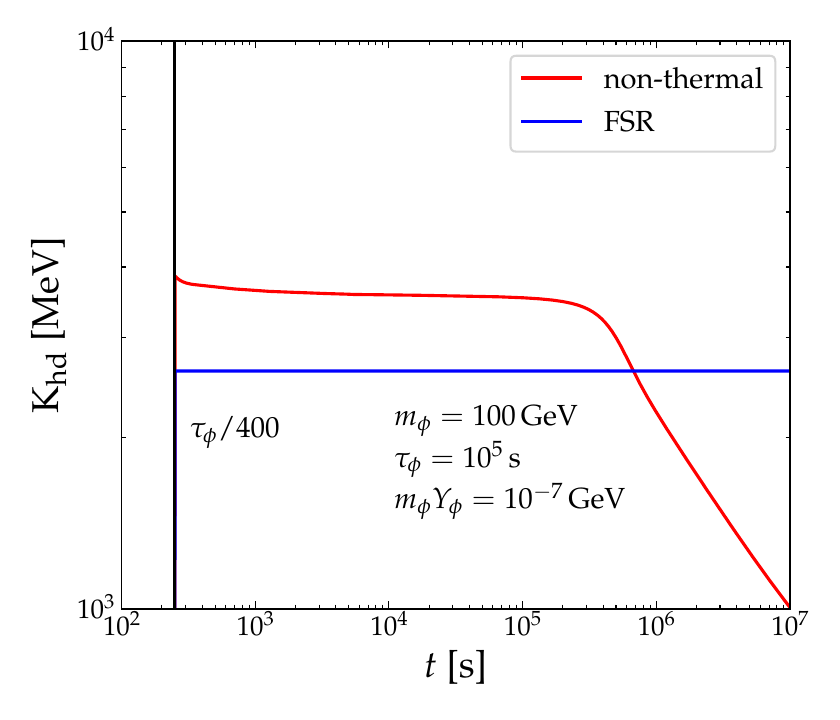}
    \end{subfigure}
    \begin{subfigure}{0.49\textwidth}
      \centering
      \includegraphics[width=\textwidth]{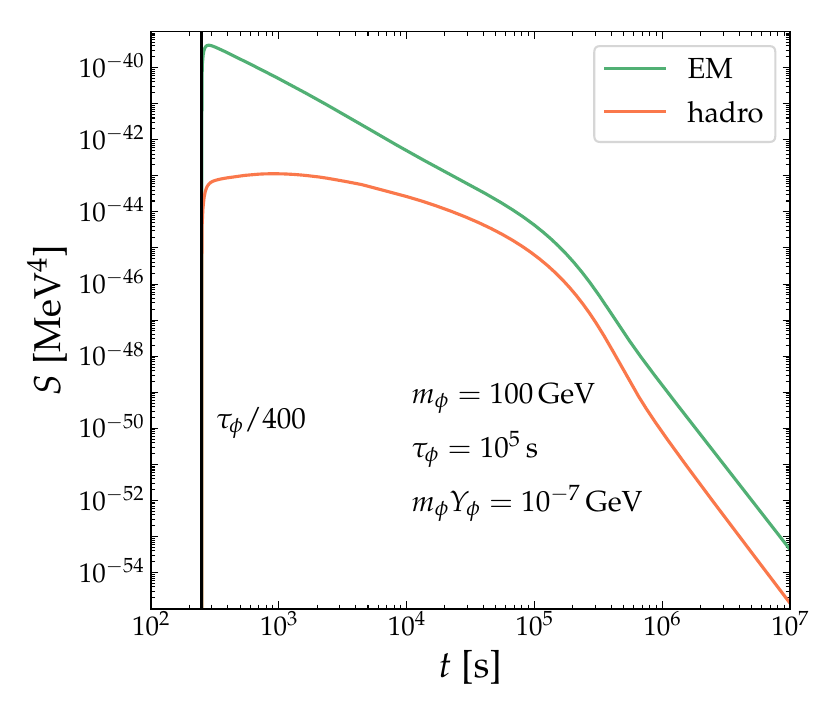}
    \end{subfigure}
    \begin{subfigure}{0.49\textwidth}
      \centering
      \includegraphics[width=\textwidth]{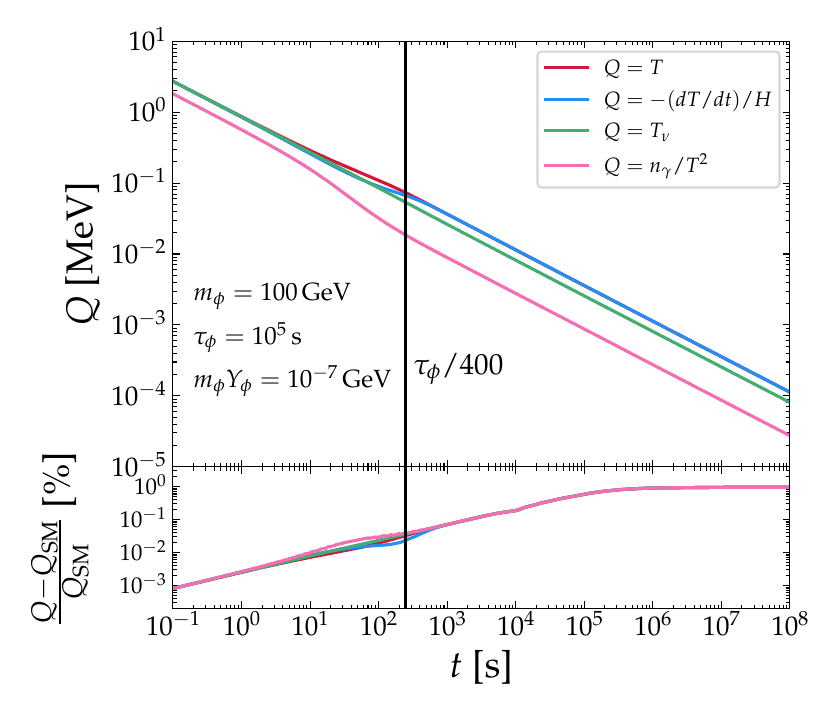}
    \end{subfigure}

\caption{Time dependence of the main quantities resulting from the neutrino cascade. This includes \textit{(i)} the injection rate of hadrons $\frac{\text{d}n_\text{hd}}{\text{d}t}$ (upper left) and the average kinetic energy of these particles $K_\text{hd}$ (upper right) from non-thermal scattering (red) and FSR (blue), and \textit{(ii)} the EM (green) and hadronic (orange) source terms $S_{\text{em/hd}}$ (lower left). Additionally, we show several quantities $Q$ describing the modified background cosmology (lower right), specifically by comparing them to their values $Q_\text{SM}$ in the SM.}
\label{fig:aux_out}
\end{figure}

\paragraph{Interlude} As part of the following steps, we calculate the relevant parameters describing the EM and hadronic injections, i.e.~$S_\text{em}$ in eq.~\eqref{eq:universal}, $K_\text{hd}^\text{fsr}$ in eq.~\eqref{eq:y_hdi}, etc. To provide the reader with a graphical representation of these parameters, in figure~\ref{fig:aux_out}, we show the resulting quantities (top, bottom left) for a single benchmark point with $m_\phi = 100\,\mathrm{GeV}$, $\tau_\phi = 10^5\,\mathrm{s}$ and $m_\phi Y_\phi = 10^{-7}\,\mathrm{GeV}$. Note that the initial increase of the different quantities close to the initial condition at $t = t_0 = \tau_\phi/400$ (black line) is not physical, but instead stems from the fact that we do not calculate these quantities for earlier times, i.e.~$S_\text{em}(t < t_0) \equiv 0$, etc. Therefore, each of them first needs to initialise to their correct physical value for $t > t_0$. However, we checked that this artefact has no influence on our results, since the light-element abundances during NBBN only start to significantly change at $t \sim \tau_\phi$ when $1/(2t) \simeq H \simeq \Gamma_\phi = 1/\tau_\phi$, at which point all quantities have already assumed their actual values. In fact, for $t < \tau_\phi$, the terms on the right-hand sides of eqs.~\eqref{eq:y_pdi} and \eqref{eq:y_hdi} are suppressed by a factor $\Gamma_\phi / H \sim t/\tau_\phi$. In the same figure, we also show several quantities describing the modified background cosmology (bottom right). To complement these plots, in the right panel of figure~\ref{fig:spectra}, we further show a compilation of the relevant rates for the same benchmark point.

\paragraph{Decay of the relic} Starting from the initial conditions determined in the last paragraph, we now track the decay of the relic $\phi$, as well as the subsequent neutrino cascade. Like before, we do this by following the evolution of the individual particles over small time steps $\Delta t$. Consequently, each step of the iteration can be assigned a unique set of values $T$, $T_\nu$, $t$, $a_\phi$ and $D_\nu = D_\nu(t)$, while we track the neutrinos that will be present in the next step, i.e.~at $t + \Delta t$ , in a separate list $D_\nu(t + \Delta t)$ (with $D_\nu(t + \Delta t) = \varnothing$ at the start of each step). To determine $\Delta t$ for a given temperature $T$, however, it is no longer sufficient to only consider the Hubble rate $H(T)$, since potentially relevant scattering reactions add another timescale to the problem. Therefore, we also calculate the maximal thermal scattering rate $\Gamma_\text{th}^\text{max}(T)$ and the maximal non-thermal scattering rate $\Gamma_\text{n-th}^\text{max}(T)$,\footnote{More precisely, referencing eqs.~\eqref{eq:G_th} and \eqref{eq:G_nth} below, we use $\Gamma_\text{th}^\text{max} = \max \underline{\Gamma}_{\nu\rightarrow \nu}^\text{th}$ and $\Gamma_\text{n-th}^\text{max} = \max \underline{\Gamma}_{\nu\rightarrow X}^\text{n-th}$.} and define
\begin{align} \label{eq:t_step}
    \Delta t = \frac{\varepsilon}{\max\left( \Gamma_\text{th}^\text{max} + \Gamma_\text{n-th}^\text{max}, H \right)}
\end{align}
with $\varepsilon \ll 1$. In this work, we find that $\varepsilon = 10^{-2}$ gives a good compromise between speed and accuracy. While errors on the individual quantities can be $\sim1\%-10\%$ for this choice -- specifically for small lifetimes $\tau_\phi\sim 10^{4}\,\mathrm{s}$ when many scatterings happen per Hubble time --, such differences are barely noticeable in the final plots, which span many orders of magnitude. Given a relic abundance $a_\phi$, the number $\Delta a_\phi$ of $\phi$ particles that decay during the time step $\Delta t$ is then given by
\begin{align}
    \Delta a_\phi = a_\phi \left| 1 - e^{-\Delta t/\tau_\phi} \right|\eqsp.
\label{eq:med_decay}
\end{align}
These decays lead to the two-body injection of electrons/positrons and neutrinos with energy $E_0 = m_\phi/2$ according to the branching ratios $\text{BR}_{ee}$ and $\text{BR}_{\nu\nu} \equiv 1- \text{BR}_{ee}$, respectively. The injected electron-positron pairs directly contribute to the amount of injected electromagnetic material and therefore change $S_\text{em}(t)$ in eq.~\eqref{eq:def_S_any} by (cf.~eq.~\eqref{eq:def_Sem} with $E_0 = m_\phi / 2$ and $\dot{\rho}_{e\pm}^\text{inj} \simeq -\text{BR}_{ee} \dot{\rho}_\phi = \text{BR}_{ee} m_\phi |\dot{n}_\phi|$)
\begin{align}
    S_\text{em}(t) \supset \text{BR}_{ee} \times 2 \times \frac{\Delta n_\phi}{\Delta t} = S_\text{em}^\text{dcy}
\end{align}
with $\Delta n_\phi = \Delta a_\phi n_{\phi, 0} R^3 /R_0^3 > 0$. The injected neutrinos, however, will undergo additional interactions (see below). As discussed above, to handle some of these reactions, we need to know the different neutrino spectra, meaning that all injected neutrinos must be tracked as part of $D_\nu(t+\Delta t)$. At face value, and assuming for simplicity that the relic only decays into (anti-)electron neutrinos, this would imply that for the next step we have to add
\begin{align}
    D_\nu(t + \Delta t) \supset \big[m_\phi/2: \Delta a_\nu\big] \qquad \text{with} \qquad \Delta \underline{a}_\nu \equiv \text{BR}_{\nu\nu}\times(\Delta a_\phi, 0, 0, \Delta a_{\phi}, 0, 0)^T\eqsp.\label{eq:Dnu_wrong}
\end{align}
However, since both neutrino oscillations and the EW shower due to FSR can change the composition and energy of the injected neutrinos before any scattering reactions become important, it is instructive to first handle these effects, before adding any decay products to the spectrum.

\paragraph{Final-state radiation} Immediately after getting injected, i.e.~on time-scales much smaller than those of the proceeding scattering reactions and even oscillations, the neutrinos lose some of their energy via FSR, which further leads to the injection of both EM and hadronic material. In this work, we quantify this process via the \textit{average} fractions $\zeta_\text{em}^\text{fsr}(m_\phi)$ and $\zeta_\text{hd}^\text{fsr}(m_\phi)$ of the original neutrino energy that ends up in the form of EM and hadronic material, respectively. Using these quantities, the \textit{average} energy of each injected neutrino after the shower is then given by
\begin{align} \label{eq:en_shower}
    E_0^\text{post-fsr}(m_\phi) = \frac{m_\phi}{2}\left[ 1 - \zeta_\text{em}^\text{fsr}(m_\phi) - \zeta_\text{hd}^\text{fsr}(m_\phi)\right]\eqsp.
\end{align}
The dissipated energy, in turn, gets distributed between $S_\text{em}(t)$ in eq.~\eqref{eq:def_S_any} and $S_\text{hd}(t)$ in eq.~\eqref{eq:def_Shd_any}, i.e. 
\begin{align}
    S_\text{em}(t) & \supset \text{BR}_{\nu\nu} \times \zeta_\text{em}^\text{fsr}(m_\phi) \times \frac{2\Delta n_\phi}{\Delta t} = S_\text{em}^\text{fsr} \label{eq:sh_Sem}\\
    S_\text{hd}(t) & \supset \text{BR}_{\nu\nu} \times \zeta_\text{hd}^\text{fsr}(m_\phi) \times \frac{2\Delta n_\phi}{\Delta t} = S_\text{hd}^\text{fsr}\eqsp.\label{eq:sh_Shd}
\end{align}
For correctly handling the hadronic injections, we further need to know the amount of hadrons that this process injects per time interval, i.e.~$\text{d} n_\text{hd}^\text{fsr}/\text{d} t$ in eq.~\eqref{eq:y_hdi}, as well as the corresponding distribution of kinetic energies (cf.~section~\ref{sec:hadro}). As stated above, we quantify this contribution via the simplifying assumption that all hadrons are injected with the \textit{average} kinetic energy $K_\text{hd}^\text{fsr}(m_\phi)$ from eq.~\eqref{eq:y_hdi}, while setting
\begin{align} \label{eq:dnhd_dt}
    \frac{\text{d} n_\text{hd}^\text{fsr}}{\text{d} t}(t) = \text{BR}_{\nu\nu} \times N_\text{hd}^\text{fsr}(m_\phi) \times \frac{\Delta n_\phi}{\Delta t}\eqsp,
\end{align}
where $N_\text{hd}^\text{fsr}(m_\phi)$ is the \textit{average} number of protons/neutrons that get injected by FSR following a single decay. Conceptually, the four relevant functions $\zeta_\text{em}^\text{fsr}(m_\phi)$, $\zeta_\text{hd}^\text{fsr}(m_\phi)$, $N_\text{hd}^\text{fsr}(m_\phi)$, and $K_\text{hd}^\text{fsr}(m_\phi)$ are obtained by sampling $\mathcal{O}(10^6)$ $\nu_e\bar\nu_e$ events, which are then used as an input for the simulation of the EW shower via the \texttt{VINCIA} module of \texttt{PYTHIA8.3} \cite{Sjostrand:2014zea,Bierlich:2022pfr}. All quantities are then determined as the respective averages over all simulated events.\footnote{As already indicated above, for both protons and neutrons, the quantities $N_\text{hd}^\text{fsr}$ and $K_\text{hd}^\text{fsr}$ are equal at the $\mathcal{O}(0.1\%)$ level, which is why it is justified to use identical values for both nucleons.} This procedure is analogous to the one described in section~2.1 of \cite{Hambye:2021moy}, with the only difference being that we additionally extract information on the hadronic fragments of the shower (for more information, please consult the aforementioned paper).\footnote{In practice, we also need to fix the nature of the relic to run the shower. Following the cited paper, we have decided on a vector particle. For a scalar particle, we have checked that both EM and hadronic injection vary on the level of $\mathcal{O}(1)$ which leads to mildly different bounds.} For reference, the three latter quantities are presented in figure~\ref{fig:aux} of appendix~\ref{app:figures}, while $\zeta_\text{em}^\text{fsr}(m_\phi)$ has already been shown in figure~\ref{fig:EW_shower}. Finally, let us note that the EW shower can also produce secondary neutrinos, which in principle need to be added to $D_\nu(t+\Delta t)$. However, we find that their contribution is usually much smaller than the one of the primary neutrinos, meaning that they do not lead to a significant injection of EM or hadronic energy. Consequently, we neglect them in the following.

\paragraph{Neutrino oscillations} After the EW shower from FSR, but still before any scattering reactions, the injected neutrinos further undergo oscillations. Extending the neutrino-oscillation matrix $\mathbf{M}_\text{osc}$ from eq.~\eqref{eq:osc} to six dimensions ($\mathbf{M}_\text{osc}$ acts in the same way on neutrinos and anti-neutrinos), this process changes the flavour-composition of the original injection with associated abundance $\Delta \underline{a}_\nu$ according to the relation
\begin{align} \label{eq:nu_osc}
    \Delta \underline{a}_\nu^\text{osc} = \mathbf{M}_\text{osc} \Delta \underline{a}_\nu \eqsp.
\end{align}
Consequently, to incorporate the effects of the FSR and neutrino oscillations, we replace eq.~\eqref{eq:Dnu_wrong} by
\begin{align} \label{eq:nu_spectrum}
    D_\nu(t + \Delta t) \supset \big[E_0^\text{post-fsr}(m_\phi): \Delta \underline{a}_\nu^\text{osc}\big]\eqsp.
\end{align}

\paragraph{Thermal scattering}
Given a neutrino spectrum $D_\nu(t)$, any neutrino with energy $E_i \in D_\nu(t)$ can participate in thermal scattering reactions with the background neutrinos. At this point, let us stress again that for the values of $m_\phi$ and $\tau_\phi$ considered in this work with $E_i \leq m_\phi/2$, the maximally available centre-of-mass energy $s^\text{th}_\text{max} \sim m_\phi T_\nu(\tau_\phi)$ from thermal scattering off a background neutrino with temperature $T_\nu$, is always well below the $Z$-resonance and even the $e\mu$ threshold. Consequently, given a neutrino $\nu_j$ with energy $E_i$ that scatters off any thermal background (anti-)neutrino $\nu_\text{th}$, the only available channels are $\nu_j \nu_\text{th} \rightarrow \nu \nu$ and $\nu_j \nu_\text{th} \rightarrow e^+ e^-$.\footnote{Thermal reactions therefore do not inject hadrons, meaning that $S_\text{hd}$ and $\text{d} n_\text{hd} / \text{d} t$ remain unaltered. However, if we were to consider $\tau_\phi < 10^{4}\,\mathrm{s}$, additional channels like $\nu \nu_\text{th} \rightarrow e \mu$ and $\nu \nu_\text{th} \rightarrow \pi\pi$ would eventually open up, which would lead to the injection of hadronic material and thus to a change of $S_\text{hd}$ and $\text{d} n_\text{hd} / \text{d} t$.} Using $\Gamma_0^\text{th}(E_i, T_\nu) = (7\pi/90) G_F^2 E_i T_\nu^4$ (cf.~eq.~\eqref{eq:rate_me0} without the process-dependent factor $\Sigma_\infty$), the corresponding rates are given by
\begin{align} \label{eq:gamma_thermal}
\Gamma_{\nu_j\rightarrow\nu}^\text{th}(E_i, T_\nu) \equiv \sfrac{10}{3} \Gamma_0^\text{th}(E_i, T_\nu) \qquad \text{and}\qquad
\Gamma_{\nu_j \rightarrow e}^\text{th}(E_i, T_\nu) & = F_j(E_i)\Gamma_0^\text{th}(E_i, T_\nu)\eqsp.
\end{align}
Here, $F_j(E_i)$ is the analogue of $\Sigma_\infty$ in eq.~\eqref{eq:rate_me0}, but it has been obtained by performing the integral $\int \text{d} s\, s \cdot \sigma_{ee}(s)$ for a \emph{non-vanishing} electron-mass. Consequently, $F_j \neq \text{const.}$ and $F_j(E_i \ll m_e) = 0$. Also, the index $j$ indicates that $F_j$ differs depending on the initial-state flavour (cf.~table~\ref{tab:fermirates}). In vector notation (cf.~eq.~\eqref{eq:def_vec_a}), we can therefore write
\begin{align}
    \underline{\Gamma}_{\nu\rightarrow\nu}^\text{th} \equiv \sfrac{10}{3}\Gamma_0^\text{th} (1, 1, 1, 1, 1, 1)^T \qquad\text{and}\qquad \underline{\Gamma}_{\nu\rightarrow e}^\text{th} \equiv \Gamma_0^\text{th} (F_e, F_\mu, F_\tau, F_e, F_\mu, F_\tau)^T\eqsp.\label{eq:G_th}
\end{align}
Given a neutrino with energy $E_i$, the probability for a thermal scattering reaction to happen within a time interval $\Delta t$ therefore is
\begin{align}
    \Delta \underline{P}_{\nu\rightarrow \nu/e}^\text{th}(E_i) = \underline{\Gamma}_{\nu\rightarrow \nu/e}^\text{th}(E_i) \Delta t\eqsp.\label{eq:P_th}
\end{align}
Consequently, the total number $\Delta n_{\nu\rightarrow \nu/e}^\text{th}$ of neutrinos with energy $E_i$ that scatter into other neutrinos / electron-positron pairs is given by
\begin{align}
    \Delta n_{\nu\rightarrow \nu/e}^\text{th}(E_i) = \Delta \underline{P}_{\nu \rightarrow \nu/e}^\text{th}(E_i) \cdot \underline{n}_\nu(E_i)\eqsp.
\end{align}
On the one hand, any neutrino that scatters into $e^+e^-$ transfers all of its energy $E_i$ into the electromagnetic sector, which implies that the corresponding S-factor $S_\text{em}(t)$ in eq.~\eqref{eq:def_S_any} receives a contribution of the form
\begin{align}
    S_\text{em}(t) \supset \frac{E_i}{E_0} \frac{\Delta n_{\nu\rightarrow e}^\text{th}(E_i)}{\Delta t} = \frac{E_i}{E_0} \left[ \underline{\Gamma}_{\nu \rightarrow e}^\text{th}(E_i) \cdot \underline{n}_\nu(E_i) \right] \equiv \Delta S_\text{em}^\text{th}(E_i) \qquad \forall E_i \in D_\nu(t)\eqsp.\label{eq:Sem_th}
\end{align}
Notably, such a contribution exists for each $E_i \in D_\nu(t)$, meaning that the total change of $S_\text{em}(t)$ is equal to the sum of each individual contribution, i.e.~$S_\text{em}^\text{th} = \sum_{E_i} \Delta S_\text{em}^\text{th}(E_i)$. On the other hand, any neutrino with energy $E_i$ that scatters into $\nu \nu$ effectively leads to the conversion into two, i.e.~one additional, non-thermal neutrinos with average energy $E_i/2$ and abundance $\Delta \underline{P}^\text{th}_{\nu\rightarrow\nu}(E_i) \odot \underline{a}_\nu(E_i)$,\footnote{Here, the operator $\odot$ denotes the Hadamard product, i.e.~the different entries of the two vectors are multiplied element-wise to form a new vector of the same dimension.} which --- in the next time step --- can themselves interact with the thermal neutrino background. Therefore, a full treatment would imply that the final-state particles be added to $D_\nu(t + \Delta t)$. However, we find that the contribution of these secondary neutrinos is negligible for most lifetimes considered in this work. This is because their abundance $\propto \Delta \underline{P}_{\nu\rightarrow\nu}^\text{th} \ll 1$ is sufficiently small compared to the one of neutrinos that do not scatter at all and instead undergo redshift (cf.~eq.~\eqref{eq:Dnu_rd} below). In the following, we therefore neglect such neutrinos by not adding them to $D_\nu(t+\Delta t)$.\footnote{Technically, this is only true for large lifetimes, when the number of scattering reactions per Hubble time is always $\ll 1$. However, this assumption breaks down for $\tau_\phi \lesssim 10^5\,\mathrm{s}$, and we make small errors for $\tau_\phi \rightarrow 10^4\,\mathrm{s}$. Nevertheless, even in this case, we are conservative, since additional particles would only lead to an increase of injected material. We postpone a rigorous treatment of this additional regime to future work.}

\paragraph{Non-thermal scattering}
In addition to the thermal background neutrinos, each neutrino with energy $E_i \in D_\nu(t)$ can also scatter non-thermally with any other neutrinos with energy $E_j \in D_\nu(t)$. As already stressed above, while reactions of this form are suppressed by the abundance of non-thermal neutrinos, they can still be relevant if not dominant, due to the fact that the maximal centre-of-mass energy $s^\text{n-th}_\text{max} = 4 E_i E_j$ can be much larger than $s^\text{th}_\text{max}$. In particular, the scattering can be resonantly enhanced. To calculate this effect, let us consider the scattering of two neutrinos $p,q \in \{\nu_e, \nu_\mu, \nu_\tau, \bar\nu_e, \bar\nu_\mu, \bar\nu_\tau\}$ with energies $E_i$ and $E_j$, respectively. By inserting the spectrum from eq.~\eqref{eq:def_spec} for $q$ into eq.~\eqref{eq:rate_def_gen}, we find that --- in our framework --- the scattering rate for any combinations $p,q$ is given by
\begin{align}
\Gamma_{pq}^\text{n-th}(E_i, E_j) =  \frac{n_q(E_j)}{8 E_i^2 E_j^2} \underbrace{\int_{0}^{4E_i E_j} \text{d} s\; s\cdot\sigma_{pq}(s)}_{\equiv \gamma_{pq}^\text{n-th}(E_i, E_j)}\eqsp,
\label{eq:nn_XX_nth_simp}
\end{align}
which can be interpreted as a 6x6 matrix in neutrino space. Consequently, as long as $4E_iE_j > m_Z^2$ and the scattering proceeds via an $s$-channel diagram, the corresponding rate is resonantly enhanced. As a result, Fermi theory is no longer applicable to determine the cross-section, which is why we use \texttt{MadGraph5}~\cite{Alwall:2011uj} instead for this task (cf.~figure~\ref{fig:xsec}). Using eq.~\eqref{eq:nn_XX_nth_simp}, we can calculate the rate $\underline{\Gamma}_{\nu\rightarrow X}^\text{n-th}(E_i)$ for a given neutrino with energy $E_i$ to scatter off any other available non-thermal neutrino. In vector notation, we find
\begin{align}
    \left[\underline{\Gamma}_{\nu\rightarrow X}^\text{n-th}(E_i) \right]_p = \sum_q \sum_{E_j} \Gamma_{pq}^\text{n-th}(E_i, E_j)\eqsp.\label{eq:G_nth}
\end{align}
Consequently, the scattering probability (analogues to eq.~\eqref{eq:P_th}) is given by
\begin{align}
    \Delta \underline{P}_{\nu\rightarrow X}^\text{n-th}(E_i) = \underline{\Gamma}_{\nu\rightarrow X}^\text{n-th}(E_i) \Delta t\eqsp.
\end{align}
To employ a treatment similar to the one used for the EW shower from FSR, we further need to calculate the \textit{average} energy fractions $\zeta_\text{em}^\text{n-th}(s)$ and $\zeta_\text{hd}^\text{n-th}(s)$ of injected EM and hadronic material, the \textit{average} number $N_\text{hd}^\text{n-th}(s)$ of injected protons/neutrons, as well as the \textit{average} kinetic energy $K_\text{hd}^\text{n-th}(s)$ of the injected nucleons, all of which now explicitly depend on the centre-of-mass energy $s$ for which the scattering happens. Similar to the procedure described before, we determine these quantities by using \texttt{MadGraph5} to sample $\mathcal{O}(10^6)$ events of the form $pq\rightarrow X$ with any two-particle final-state $X$ for a range of different centre-of-mass energies $s$. Afterwards, we use the resulting events as an input for \texttt{PYTHIA8.3} in order to further simulate the subsequent particle shower.\footnote{We assume that all unstable particles except neutrons decay during the shower, i.e.~we neglect any potential energy loss of these particles due to scatterings with the background plasma prior to their decay. We checked that this does not lead to a sizeable error.} The different quantities $\zeta_\text{em}^\text{n-th}(s)$ etc., are then again determined as the respective averages over all simulated events for a given value of $s$.\footnote{Since we simulate events with all possible two-body final states, this average is also an average over all possible values of $X$.} Notably, the \textit{average} kinetic energy $\tilde{K}_\text{hd}^\text{n-th}(s)$ obtained this way, is given in the centre-of-mass frame. However, we are interested in the value of this quantity in the co-moving reference frame, which we can obtain by means of the relation (cf.~appendix~\ref{app:kin})
\begin{align}
    K_\text{hd}^\text{n-th}(s) = \frac{\tilde{K}_\text{hd}^\text{n-th}(s) + m_{y}}{\sqrt{s}} ( E_i + E_j) - m_{y} \equiv y_\text{hd}^\text{n-th}(s)(E_i + E_j) - m_y\label{eq:def_K_avg}
\end{align}
with the mass $m_{y}$ of the nucleon $y \in \{p, n\}$.

At this point, it is important to note that two neutrinos with energy $E_i$ and $E_j$ can scatter with any centre-of-mass energy $0 < s < 4E_i E_j$, depending on the angle under which the collision happens. To account for this effect, we further \textit{average} all quantities over all possible values of $s$. To this end, we define the operator\footnote{Each quantity is therefore averaged twice: Once over their distribution originating from the shower for a given value of $s$ and once over all possible values of $s$ for which the particles can scatter in the first place.}
\begin{align}
    \langle Q \rangle_{pq}(E_i, E_j) = \frac{1}{\gamma_{pq}^\text{n-th}(E_i, E_j)} \int_0^{4E_i E_j} \text{d} s\; s \cdot Q_{pq}(s) \cdot \sigma_{pq}(s)\eqsp. \label{eq:def_Q_avg}
\end{align}
which determines the \textit{average} value of any quantity $Q(s)$, weighted by the probability for the scattering reaction to happens with any possible centre-of-mass energy $0 < s < 4E_i E_j$. The resulting function $\langle \zeta_\text{em}^\text{n-th}\rangle$ has already been presented in figure~\ref{fig:avg_zeta}. We refer the reader to figure~\ref{fig:aux} of appendix~\ref{app:figures} for a visualisation of the remaining quantities.
Using the definition above, we approximate the contributions to the individual $S$-factors from non-thermal scattering via the relations\footnote{For each reaction, a fraction $\langle \zeta_\text{em/hd}^\text{n-th} \rangle_{pq}(E_i, E_j)$ of the total energy $E_i + E_j$ is transformed into EM/hadronic material, while the number of reactions happening per time and comoving volume element is given by $\Gamma_{pq}^\text{n-th}(E_i, E_j) n_p(E_i)$.} (cf.~eq.~\eqref{eq:Sem_th})
\begin{align} \label{eq:nth_Sem}
    S_\text{em}(t) \supset \frac{E_i + E_j}{E_0} \sum_{p,q} \langle \zeta_\text{em}^\text{n-th} \rangle_{pq} & (E_i, E_j) \Gamma_{pq}^\text{n-th}(E_i, E_j) n_p(E_i) \equiv \Delta S_\text{em}^\text{n-th}(E_i, E_j) \\
    & \forall E_i, E_j \in D_\nu(t)\;\,\text{with}\;\, E_j \leq E_i \nonumber
\end{align}
and
\begin{align} \label{eq:nth_Shd}
    S_\text{hd}(t) \supset \frac{E_i + E_j}{E_0} \sum_{p,q} \langle \zeta_\text{hd}^\text{n-th} \rangle_{pq} &(E_i, E_j) \Gamma_{pq}^\text{n-th}(E_i, E_j) n_p(E_i) \equiv \Delta S_\text{hd}^\text{n-th}(E_i, E_j) \\
    & \forall E_i, E_j \in D_\nu(t)\;\,\text{with}\;\, E_j \leq E_i\eqsp. \nonumber
\end{align}
Here, we sum over all possible initial states $pq$, but only consider energy combinations with $E_j \leq E_i$ in order to avoid double-counting. Overall, the total change of $S_\text{em/hd}(t)$ is equal to the sum of each individual contribution, i.e.~$S_\text{em/hd}^\text{n-th} = \sum_{E_i, E_j\leq E_i} \Delta S_\text{em/hd}^\text{n-th}(E_i, E_j)$. In a similar way, the different contributions to $\text{d} n_\text{hd}^\text{n-th}/\text{d} t$ can be approximated as\footnote{We handle this quantity separately from $\text{d} n_\text{hd}^\text{fsr}/\text{d} t$, since both have different average kinetic energies.}
\begin{align} \label{eq:nth_dndt}
    \frac{\text{d} n_\text{hd}^\text{n-th}}{\text{d} t}(t) \supset \sum_{p,q} \langle N_\text{hd}^\text{n-th} \rangle_{pq} &(E_i, E_j) \Gamma_{pq}^\text{n-th}(E_i, E_j) n_p(E_i) \equiv \Delta \frac{\text{d} n_\text{hd}^\text{n-th}}{\text{d} t}(E_i, E_j) \nonumber \\
    & \forall E_i, E_j \in D_\nu(t)\;\,\text{with}\;\, E_j \leq E_i\eqsp.
\end{align}
At this point, it is important to note that each contribution $\Delta \text{d} n_\text{hd}^\text{n-th}/\text{d} t(E_i, E_j)$ fundamentally has a slightly different \textit{average} kinetic energy according to eq.~\eqref{eq:def_K_avg}, i.e.
\begin{align}
K_{\text{hd}, pq}^\text{n-th}(E_i, E_j) = \langle y_\text{hd}^\text{n-th} \rangle_{pq}(E_i, E_j) (E_i + E_j) - m_p\eqsp.
\end{align}
However, instead of handling each contribution separately, we combine them all into one contribution $\text{d} n_\text{hd}^\text{n-th}/\text{d} t = \sum_{E_i, E_j \leq E_i} \Delta \text{d} n_\text{hd}^\text{n-th}/\text{d} t(E_i, E_j)$ with the corresponding \textit{averaged} kinetic energy
\begin{align}  \label{eq:nth_avgkin}
K_\text{hd}^\text{n-th} = \frac{\sum_{E_i, E_j \leq E_i} \sum_{p,q} K_{\text{hd}, pq}^\text{n-th}(E_i, E_j) \Delta \text{d} n_\text{hd}^\text{n-th}/\text{d} t(E_i, E_j)}{\text{d} n_\text{hd}^\text{n-th}/\text{d} t}\eqsp.
\end{align}
Finally, let us note that non-thermal scattering reactions can also inject secondary neutrinos, which may undergo additional scattering reactions in the next time step. However, as before, we find that their contribution is not relevant, which is why we do not add them to $D_\nu(t + \Delta t)$. We can observe this in particular from figure~\ref{fig:spectra} (right panel): we see that even for this benchmark point which is already robustly excluded (cf.~figure~\ref{fig:flags_early}) the \emph{maximum} non-thermal scattering rate is below the Hubble rate. Thus, any neutrino injection from a non-thermal scattering process is very unlikely to undergo further interactions. We therefore estimate the uncertainty of neglecting secondary neutrinos from non-thermal interactions to be small relative to the large effect of EM and hadronic injections in these collisions. Should the mass of the relic be so high that secondary neutrinos would experience significant FSR, we are in a regime where the initial EW shower is dominant anyways so that the impact of non-thermal scattering is irrelevant.

\paragraph{Redshift}
Any particle that does not scatter within the time interval $\Delta t$, will simply redshift. The probability for this to happen is given by
\begin{align} \label{eq:redshift1}
    \Delta \underline{P}_{\nu\rightarrow \nu}^\text{rs}(E_i) = 1 - \Delta \underline{P}_{\nu \rightarrow e}^\text{th}(E_i) - \Delta \underline{P}_{\nu\rightarrow\nu}^\text{th}(E_i) - \Delta \underline{P}_{\nu \rightarrow X}^\text{n-th}(E_i)\eqsp.
\end{align}
To incorporate this process, each particle with initial energy $E_i$ is assigned a new energy
\begin{align}\label{eq:redshift2}
    E_i^\text{rs}(E_i) = E_i \times \frac{R(t)}{R(t+\Delta t)}\eqsp.
\end{align}
Finally, we add each new energy with the corresponding abundance $\underline{P}^\text{rs}_{\nu\rightarrow \nu}(E_i)\odot \underline{a}_\nu(E_i)$ to the spectrum $D_\nu(t+\Delta t)$,
\begin{align}
    D_\nu(t + \Delta t) \supset \big[ E_i^\text{rs}(E_i): \Delta \underline{P}_{\nu\rightarrow \nu}^\text{rs}(E_i)\odot \underline{a}_\nu(E_i) \big] \qquad \forall E_i \in D_\nu(t)\eqsp.\label{eq:Dnu_rd}
\end{align}
Consequently, all of these particles will be processed again in the next time step, albeit with a slightly smaller energy. As a reminder, let us note that we do not take into account the energy redistribution of neutrino-to-neutrino scatterings, and consequently the redshift is the only mechanism to populate different energies in the spectrum.

\paragraph{The next iteration step}
After handling all processes, we update all relevant quantities in order to prepare the next iteration step. For this, we set $t \rightarrow t + \Delta t$ with
\begin{itemize}\label{it:update}
    \item $T(t + \Delta t) \simeq T(t) + \frac{\text{d} T}{\text{d} t}(t)\Delta t$
    \item $T_\nu(t+ \Delta t) \simeq T_\nu(t) + \frac{\text{d} T_\nu}{\text{d} t}(t)\Delta t$
    \item $a_\phi(t + \Delta t) \simeq \max\{a_\phi(t) - \Delta a_\phi(t), 0\}$
\end{itemize}
In the last step, we made sure that $a_\phi < 0$ never happens numerically. Moreover, we replace $D_\nu(t) \rightarrow D_\nu(t + \Delta t)$ and update the energy density of neutrinos by means of the following relation
\begin{align}
\rho_\nu(t + \Delta t) \simeq \underbrace{\rho_\nu(t) \left[ \frac{R(t)}{R(t + \Delta t)} \right]^4}_{\text{Redshift}} \underbrace{+ m_\phi \Delta a_\phi(t)}_{\phi\;\text{decay}} \underbrace{- E_0 S_\text{em}(t) \Delta t}_{\text{EM inj.}} \underbrace{- E_0 S_\text{hd}(t) \Delta t}_{\text{hadronic inj.}}\eqsp.\label{eq:rhonu_upd}
\end{align}
Afterwards we repeat the procedure with the new set of quantities $t$, $T$, $T_\nu$, $a_\phi$, $\rho_\nu$, and $D_\nu$. Finally, we stop the iteration once $T < T_f \equiv 10^{-2}T(\tau_\phi)$, i.e.~long after the decay of the relic particle has concluded.

\paragraph{Post-processing}
At the end of the iterative procedure, we still have to calculate some derived quantities. Assuming that we stop at a temperature $T_f$, we first calculate the effective number of additional neutrino species via the equation
\begin{align}
    \Delta \Neff = \frac{\rho_\nu(T_f)}{\rho_\text{ref}(T_f)}
\end{align}
with $\rho_\text{ref}(T_f) = 2 \cdot \sfrac78 \cdot \sfrac{\pi^2}{30} \cdot  (\sfrac{4}{11})^{\sfrac43} \cdot  T_f^4$.
Afterwards, we use this result to determine the best-fit value for the baryon-to-photon ratio $\eta$ according to eq.~\eqref{eq:def_eta}.

Finally, let us note that we technically also calculate the parameter $f_\phi$ only at the end of the cascade, since there might not be a one-to-one relation between $f_\phi$ and $m_\phi Y_\phi$ due to significant entropy injection. In this case 
\begin{align}
    f_\phi = m_\phi Y_\phi s_\text{sm}(T_\text{cmb}) / \rho_\text{dm}(T_\text{cmb})\eqsp,
\end{align}
with the SM entropy density $s_\text{sm}$ at the CMB temperature $T_\text{cmb}$ and the corresponding energy density of DM $\rho_\text{dm}(T_\text{cmb}) = 8.095895\times 10^{-47} \times \Omega_\text{dm} h^2$. However, for all parameter points considered in this work, we find that $s_\text{sm}(T_\text{cmb}) / \rho_\text{dm}(T_\text{cmb}) = \text{const.} \approx 2.306\times10^9$, meaning that $f_\phi \propto m_\phi Y_\phi$ and consequently $f_\phi$ is already known from the start for a given value of $m_\phi Y_\phi$. Based on this observation, in the following, we will therefore often use $f_\phi$ and $m_\phi Y_\phi$ interchangeably to parameterise the pre-decay abundance of the decaying relic, noting that $Y_\phi$ generally is a more fundamental input parameter.

\section{Results and Discussion}\label{sec:results}

In this section, we present the limits derived from our numerical framework discussed above. Here, we mainly consider the case $\phi \to \bar{\nu}_e \nu_e$ with $\text{BR}_{ee} = 0$; however, we will later also comment on other flavors and cases with $\text{BR}_{ee} \neq 0$. Consequently, the relevant parameter space is three-dimensional, featuring $f_\phi$, or equivalently $m_\phi Y_\phi$, $m_\phi$, and $\tau_\phi$. To provide maximal coverage of the parameter space, our discussion below is split into three sections covering the three possible combinations of two parameters each.

\subsection{The $\tau_\phi - f_\phi \;(m_\phi Y_\phi)$ parameter plane
}\label{sec:plane1}

Let us first discuss the resulting limits in the $\tau_\phi-f_\phi$ parameter plane for a selection of fixed masses $m_\phi$. Since we already discussed the two particular choices $m_\phi = 100\,\mathrm{GeV}$ and $m_\phi = 500\,\mathrm{GeV}$ at the end of section~\ref{sec:basics} (cf.~figure~\ref{fig:flags_early}), we will refrain from a redundant discussion of the individual contributions to the total limits, and instead focus on the general trend of how the limits change with $m_\phi$.

\begin{figure}[t]
    \centering
        \includegraphics[width=0.75\textwidth]{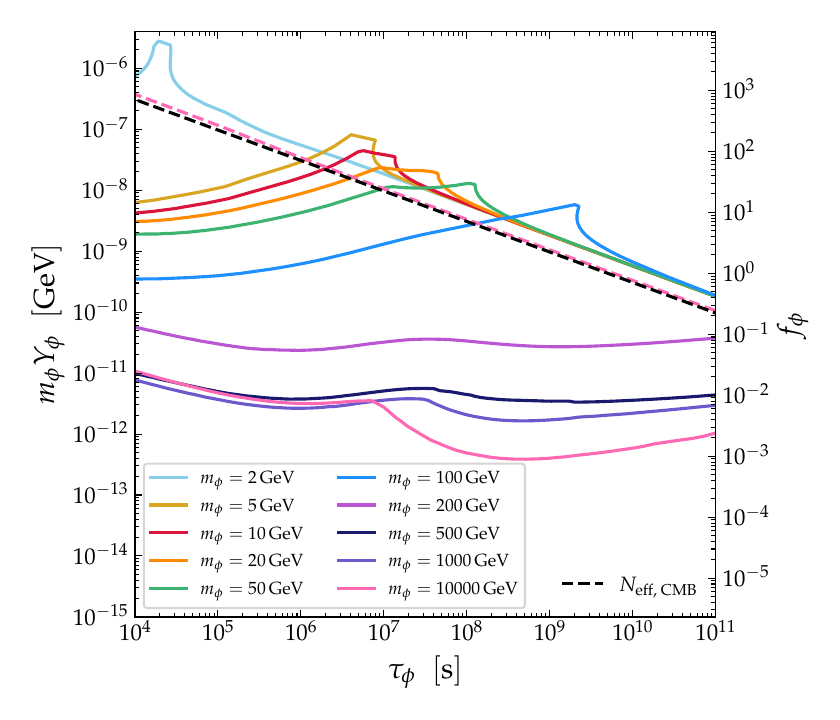}
    \caption{Overall BBN constraints originating from the decay $\phi\rightarrow \nu_e \bar\nu_e$ for different masses $m_\phi$ (solid, different colours). In addition, we also show the corresponding $\Neff$ constraint for $m_\phi = 10\,\mathrm{TeV}$ (pink dashed), noting that this constraint only deviates from the simple power law from eq.~\eqref{eq:neff_num} described in appendix~\ref{app:neff} (black dashed) for large masses and small lifetimes.}
    \label{fig:mphi_all}
\end{figure}

In figure~\ref{fig:mphi_all}, we show the resulting constraints for different relic masses between $2\,$GeV and $10\,$TeV (solid, different colours). For smaller masses, i.e.~$m_\phi \lesssim 2\,\mathrm{GeV}$, we find that the limits become subleading to the ones from $\Neff$ and are therefore irrelevant. This is due to the fact that \textit{(i)} protons and neutrons become kinematically unavailable, and \textit{(ii)} photodisintegration becomes less efficient. In fact, the threshold for the production of nucleons -- which induce the leading constraints for large parts of parameter space -- is roughly $m_\phi \sim 2\,$GeV. Moreover, when it comes to the efficiency of photodisintegration, we find that the EW shower from FSR is suppressed below the EW scale, while thermal collisions at the GeV-scale only produce very few electron-positron pairs, even for high temperatures (cf.~figure~\ref{fig:EW_shower}). Additionally, even though the fraction of injected EM material (cf.~figure~\ref{fig:avg_zeta}) due to non-thermal collisions is still relatively high for $\sqrt{s}=\mathcal{O}(\text{GeV})$, the corresponding cross-section (cf.~figure~\ref{fig:xsec}) is already deep within the effective Fermi regime and therefore quickly decouples for small neutrino energies and thus relic masses. Consequently, for such small masses, we find an absence of efficient disintegration reactions, meaning that the light-element abundances are mainly changed by the modified cosmology, i.e.~due to the change in the baryon-to-photon ratio. We therefore refrain from considering masses below $m_\phi = 2\,\mathrm{GeV}$.

At this point, let us quickly note that all bounds that are driven by scattering reactions, decouple for sufficiently large lifetimes due to dilution effects, which implies that the modified background cosmology inevitable takes over the total limit at some point. In particular, this becomes apparent by the fact that the constraints for all masses $m_\phi \leq 200\,\mathrm{GeV}$ asymptotically approach a line parallel to the one from the CMB $\Neff$ limit (dashed), albeit weaker by a factor $\sim 2$. Regarding the $\Neff$ constraint, we show two different lines: one for the ``naive'' expectation (dashed black), assuming that all neutrinos end up in the form of additional radiation, as well as the consistently-calculated one for $m_\phi = 10\,\mathrm{TeV}$ (dashed pink), in which case a relevant portion of the neutrino energy ends up in the form of EM or hadronic radiation, thus leading to a slightly smaller energy density of neutrinos at recombination and therefore to weaker limits. The $\Neff$ constraints for all other masses lie between these two cases. In general, we find that for larger values of $m_\phi$, the resulting constraints outperform the ones from $\Neff$ for a larger range of lifetimes. Also, in the transition region, where the bounds transition from being dominated by the background cosmology to being dominated by disintegration reactions (e.g.~at $\tau_\phi \sim 10^9\,\mathrm{s}$ for $m_\phi = 100\,\mathrm{GeV}$), we find a weakening of the limits due to a generic cancellation of both effects, as already discussed in section~\ref{sec:basics}.

Starting at small masses, as soon as kinematically available, the hadronic injections from non-thermal scattering come to rapidly dominate the bounds for small lifetimes, thus pushing them beyond the ones from $\Neff$. This effect becomes apparent from the massive jump between $m_\phi = 2\,\mathrm{GeV}$ (light blue) and $m_\phi=5\,\mathrm{GeV}$ (gold). However, beyond this boost, a further increase in mass up to $m_\phi \sim 50\,\mathrm{GeV}$ (green) only leads to a mild further strengthening of the limits, as the energy $E < m_\phi/2$ of the injected neutrinos remains below the $Z$-resonance. However, when going past the resonance, i.e.~from $m_\phi = 50\,\mathrm{GeV}$ (green) to $m_\phi = 100\,\mathrm{GeV}$ (blue), the limits strengthen by another factor $\sim 10$, showcasing the fact that the hadronic injections from non-thermal scattering receive a massive resonance boost. However, beyond this threshold, the rate for non-thermal interactions is no longer expected to increase drastically, since the corresponding cross-section (cf.~figure~\ref{fig:xsec}) saturates. This effect can also be observed when comparing the left and right panels of figure~\ref{fig:flags_early} in section~\ref{sec:basics}. Let us further stress that the bounds from hadrodisintegration do not become constant for $\tau_\phi < 10^4\,\mathrm{s}$, but are expected to vanish for $\tau_\phi \lesssim 10^2\,\mathrm{s}$, when all injected hadrons are efficiently stopped before initiating a hadronic cascade.\footnote{Note that for $\tau_\phi < 10^{2}\,\mathrm{s}$, other effects like the pion-induced interconversion of protons and neutrons become important.}

Notably, pushing the mass beyond the EW scale, further enables efficient EM/hadronic injection due to FSR (also cf.~figure~\ref{fig:EW_shower}), which causes another jump between $m_\phi = 100\,\mathrm{GeV}$ (blue) and $m_\phi=200\,\mathrm{GeV}$ (purple) (also cf.~figure~\ref{fig:200GeV} for the individual contributions for $m_\phi = 200\,\mathrm{GeV}$, similar to figure~\ref{fig:flags_early}). In fact, for sufficiently large masses, $m_\phi > 200\,\mathrm{GeV}$ (purple), the limits originating from FSR outperform the ones from $\Neff$ for the full range of lifetimes considered in this work. For even larger masses, the constraints still slightly strengthen with mass, but also start to slowly saturate towards the TeV-scale. For such masses, the limits become slightly stronger for intermediate lifetimes, $\tau_\phi\sim 10^8\, \mathrm{s}-10^{10}\,\mathrm{s}$, causing a small bump in an otherwise fairly flat limit. This can be understood from the fact that -- in this regime -- the EM contribution outperforms the (only mildly lifetime-dependent) hadronic one. 

For the remainder of this section, let us discuss the asymptotic behaviour of the different limits. To this end, let us first fix the mass and the abundance of $\phi$, thus quantifying the asymptotic behaviour for large lifetimes. In this case, when increasing the lifetime -- at some point -- the hadronic injections will consist solely of protons, since the neutrons with lifetime $\tau_n \approx 880\,\mathrm{s} \ll \tau_\phi$ effectively decay immediately. In addition, for sufficiently large lifetimes, the energy-loss mechanism of protons becomes inefficient due to the low energy of the background particles, meaning that all protons enter the hadrodisintegration processes with their (potentially high) initial injection energy. As long as the hadrodisintegration rate is still above the Hubble rate, the hadronic cascade will therefore proceed until all protons (as well as other fragments) are scattered down to energies beyond which they can no longer disintegrate $^4$He. As a result, the different parameters $\xi_X^p$, i.e.~the numbers of additional nuclei produced due to the injection of a \emph{single} proton, become independent of the injection time. Moreover, the number of injected protons only depends on the initial abundance $\phi$, i.e.~-- for a fixed abundance -- there is no dilution effect and the same amount of nucleons is injected for each individual decay, independent of the time. Therefore, the limits arising from hadronic injections are expected to approximately reach a plateau at large lifetimes. Considering EM injections instead, photodisintegration is well parameterised by the universal spectrum (cf.~eq.~\eqref{eq:universal}), which is always temperature dependent. Therefore, the asymptotic behaviour is non-trivial, and numerically we find that the limits from EM injections weaken for sufficiently large lifetimes.

Additionally, we can also quantify the asymptotic behaviour for large masses -- for which FSR again dominates -- by fixing the lifetime and the abundance of $\phi$. Increasing $m_\phi$ for fixed $\tau_\phi$ and $f_\phi \propto m_\phi Y_\phi$, leads to a linear reduction of the ``no-decay'' number density $\tilde{n}_\phi \propto f_\phi/m_\phi$ of $\phi$ (cf.~eq.~\eqref{nphit}). For the terms on the right-hand side of eq.~\eqref{eq:y_pdi} describing photodisintegration, this loss in number density is compensated for by an increase of $f_{\gamma, \text{univ}}$ with $E_0 \sim m_\phi$ (cf.~eq.~\eqref{eq:universal}), such that $f_{\gamma, \text{univ}} \propto S_\text{em} E_0 \sim \tilde{n}_\phi \zeta_\text{em}^\text{fsr}(m_\phi) E_0 \sim f_\phi m_\phi^{-1} \zeta_\text{em}^\text{fsr}(m_\phi) m_\phi \sim f_\phi \zeta_\text{em}^\text{fsr}(m_\phi)$. Hence, for large masses, the limits still increase but only slowly according to the (mild) increase of $\zeta_\text{em}^\text{fsr}$ with mass (cf.~figure~\ref{fig:EW_shower}). Specifically, this behaviour is relevant for $\tau_\phi \gtrsim 10^{7}\,\mathrm{s}$, for which the EM component of FSR dominates the limits for large masses. However, for the terms on the right-hand side of eq.~\eqref{eq:y_hdi} describing hadrodisintegration, such a compensation does not exist for large $m_\phi$. In fact, both $N_\text{hd}^\text{fsr}$ and thus $\text{d} n_\text{hd}^\text{fsr} / \text{d} t$ (cf.~figure~\ref{fig:aux}) as well as $\xi_X^y$ (cf.~figure~\ref{fig:xis}) only mildly (but not linearly) depend on $\sqrt{s} \sim E \sim m_\phi$, i.e.~not enough to counteract the loss of abundance. Once the loss in sensitivity due to the smaller abundance overtakes the increase in sensitivity due to the slightly larger values of $N_\text{hd}^\text{fsr}$ and $\xi_X^y$, the limits therefore undergo a turn-around, i.e.~they start to weaken with mass after an initial phase of strengthening. Indeed, this behaviour is relevant for $\tau_\phi \lesssim 10^{7}\,\mathrm{s}$, in which case the limits are dominated by the hadronic component of FSR. Specifically, the results obtained above are well represented in the figure (cf.~the change between $m_\phi=1\,\mathrm{TeV}$ and $m_\phi = 10\,\mathrm{TeV}$) and also in agreement with the results of \cite{Kawasaki:2018}. 

Based on the discussion above, we naturally find that the EM contribution from FSR slowly grows towards an asymptotic value (cf.~figure~\ref{fig:EW_shower}). In particular, this contribution is a good proxy for the full limits, as shown in figure~\ref{fig:zeta_lim} of appendix~\ref{app:figures}.

\subsection{The $m_\phi-\tau_\phi$ parameter plane}

\begin{figure}[t]
    \centering
    \includegraphics[width=0.75\textwidth]{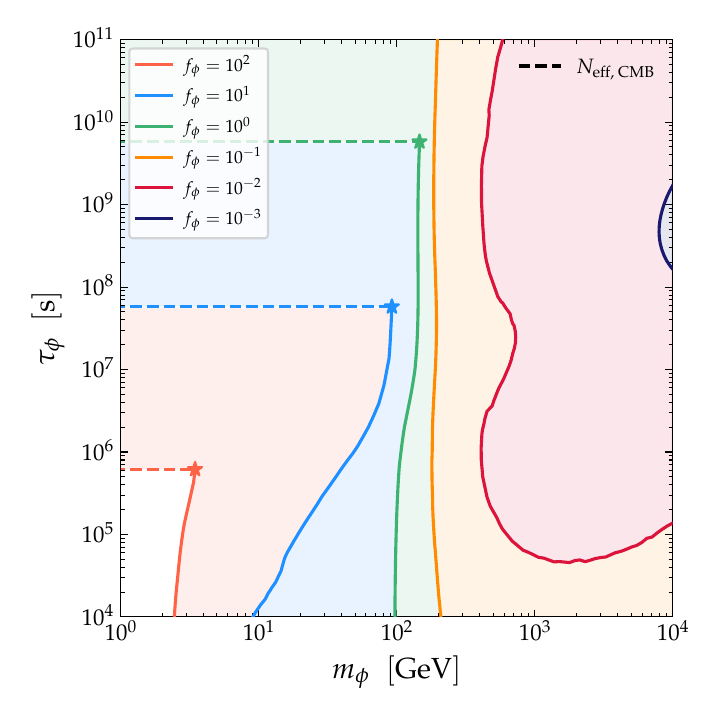}
    \caption{
    Overall BBN constraints originating from the decay $\phi\rightarrow \nu_e \bar\nu_e$ for different initial abundances $f_\phi$ (solid, different colours). In addition, we also show the corresponding $\Neff$ constraints (dashed, different colours), as well as the point beyond which the latter ones become dominant (star). For clarity, we further shade the excluded part of parameter space.
    }
    \label{fig:fdm_all}
\end{figure}

We now turn to the discussion of the $m_\phi-\tau_\phi$ parameter plane for a selection of fixed abundances. In figure~\ref{fig:fdm_all}, we show the corresponding results for six different choices of $f_\phi$ (different colours). As already discussed above, the $\Neff$ bound is effectively independent of the particle mass and therefore manifests itself as a horizontal line, becoming stronger for a larger abundance. We show this bound as a dashed line, whereas the ones stemming from BBN are shown as solid lines. Moreover, the intersection of both constraints is marked by a star and the excluded region is shaded for clarity. We see that abundances above $f_\phi=10^2$ are largely excluded for $m_\phi \gtrsim 1$~GeV, except for a small region, for which EM and hadronic injections are suppressed. Overall, reducing the initial abundances of $\phi$ relaxes the resulting limits. We find that for every abundance, there exists a value of $\tau_\phi$ beyond which the $\Neff$ constraints start to dominate, which is due to the different redshift behaviour of matter ($\phi$) and radiation ($\nu$). The parameter space above this line is fully excluded. For smaller lifetimes, the constraints are dominated by NBBN. Moreover, we find that small abundances are excluded only for large masses and intermediate lifetimes, as expected from the results in figure~\ref{fig:mphi_all}.

\subsection{The $m_\phi-f_\phi$ parameter plane}

\begin{figure}[t]
   \centering
   \includegraphics[width=0.75\textwidth]{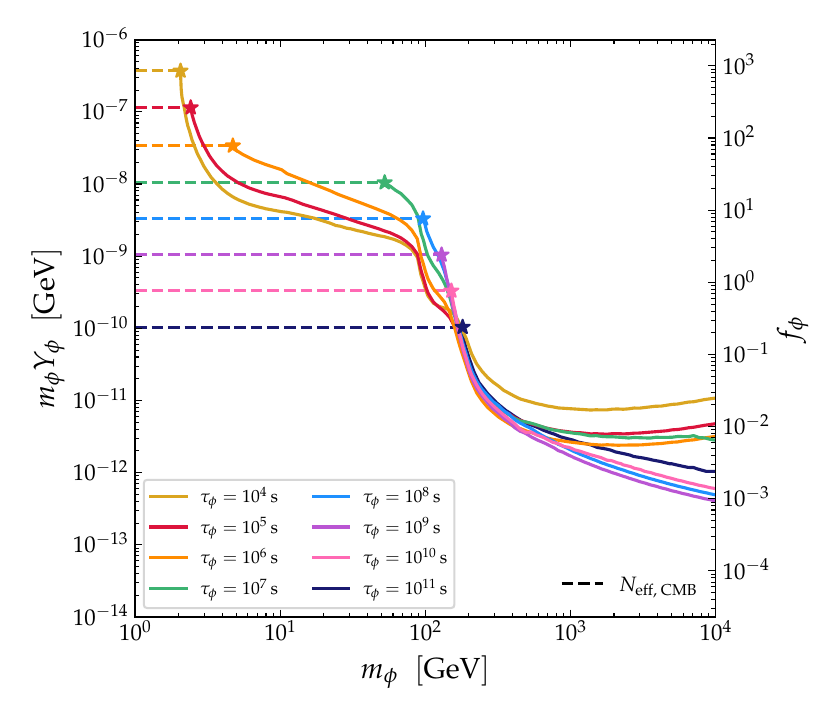}
   \caption{
   Overall BBN constraints originating from the decay $\phi\rightarrow \nu_e \bar\nu_e$ for different lifetimes $\tau_\phi$ (solid, different colours). In addition, we also show the corresponding $\Neff$ constraints (dashed, different colours), as well as the point beyond which the latter ones become dominant (star).}
   \label{fig:tau_all}
\end{figure}

Let us now discuss the $m_\phi-f_\phi$ parameter plane for a selection of fixed lifetimes. In figure~\ref{fig:tau_all}, we show the resulting constraints for eight different choices of $\tau_\phi$ (different colours). Here, the total BBN bounds obtained in this work are shown as solid lines, while the ones originating from $\Neff$ are shown with dashed lines. Again, the transition points between the two are marked by a star. Note that we do not show $m_\phi < 1\,$GeV, since all bounds are weaker than the one from $\Neff$ below this threshold, in accordance with our previous discussion. We find that particles with larger lifetimes require larger masses in order to outperform the $\Neff$ limits. Moreover, there exist two points beyond which the limits strengthen substantially. The first one is at $m_\phi \sim 2\,\mathrm{GeV}$ when hadronic final states becomes kinematically available. The second one is for masses close to the EW scale, at which point \textit{(i)} the $Z$-resonance can be hit during non-thermal scattering and \textit{(ii)} injection due to FSR becomes efficient. We observe that for $m_\phi \gtrsim 500\,\mathrm{GeV}$ and $\tau_\phi \gtrsim 10^8\,\mathrm{s}$, the limits are generically well-described by effectively only considering the EM component of the EW shower from FSR (cf.~figure~\ref{fig:zeta_lim} and appendix~\ref{app:figures} for a more detailed discussion). However, for smaller lifetimes, the hadronic component of the EW shower from FSR is dominant for heavy relics, which explains the different asymptotic behaviour.

\subsection{Results for other neutrino flavours}

\begin{figure}[t]
    \centering
    \includegraphics[width=0.75\textwidth]{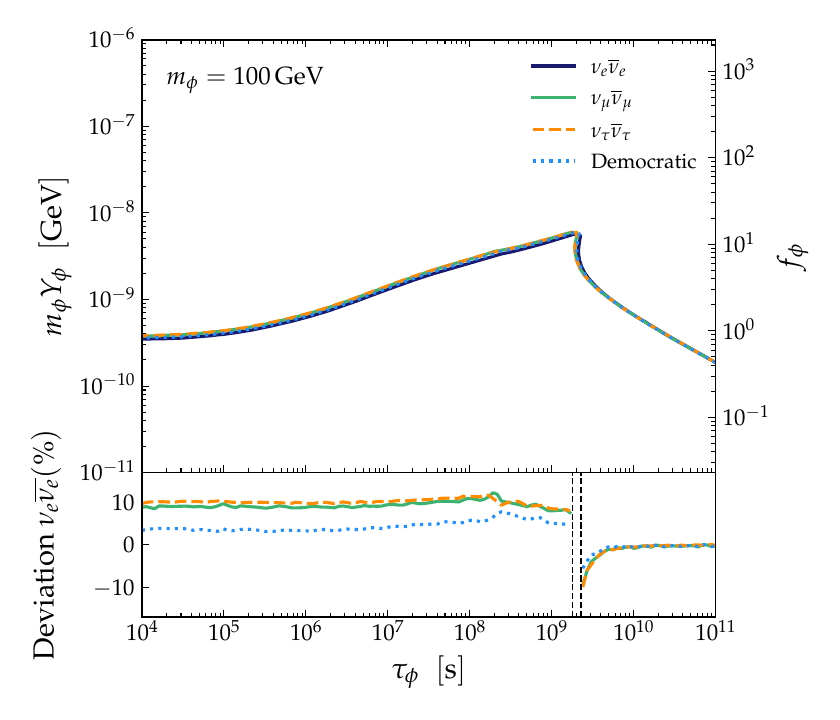}
    \caption{Absolute and relative comparison of the total BBN limits for $m_\phi = 100 \mathrm{GeV}$, assuming different initial neutrino flavours in the decay. For the relative comparison, we leave out a narrow region where there is no one-to-one correspondence between $\tau_\phi$ and $m_\phi Y_\phi$.}
    \label{fig:fdm_all6}
\end{figure}

Let us now briefly discuss how the limits change when other neutrino flavours are initially injected in the $\phi$ decay. 
In figure~\ref{fig:fdm_all6}, we show the resulting constraints in the  $\tau_\phi - f_\phi$ plane for $m_\phi = 100\,\mathrm{GeV}$, and \textit{(i)} the three cases of single-flavour injections $\phi \to \bar{\nu}_e \nu_e$ (dark blue), $\phi \to \bar{\nu}_\mu \nu_\mu$ (green), $\phi \to \bar{\nu}_\tau \nu_\tau$ (orange), as well as \textit{(ii)}  the ``democratic case'' in which all neutrino-flavours are injected with equal branching ratios. We find that deviations among the different cases are very small, meaning that the bounds for $\phi \to \bar{\nu}_e \nu_e$ (which we mostly show throughout this paper) are also valid for the other cases to a very good approximation.

\subsection{Results for $\text{BR}_{ee} \neq 0$}\label{sec:br}

\begin{figure}[t]
   \centering
   \includegraphics[width=0.75\textwidth]{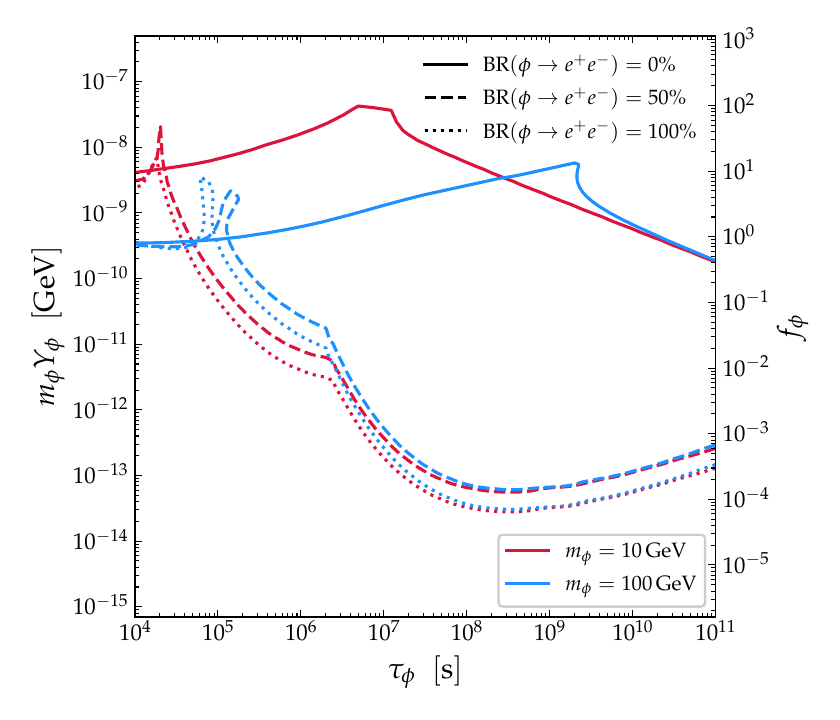}
   \caption{Comparison of the resulting BBN limits for $m_\phi \in \{10, 100\}\,\mathrm{GeV}$ (red, blue) and different values of $\text{BR}_{ee}$ (different linestyles). Specifically, we show the three cases $\text{BR}_{ee} = 0$ (solid), $\text{BR}_{ee} = 0.5$ (dashed), and $\text{BR}_{ee} = 1$ (dotted). }
   \label{fig:mixed}
\end{figure}

Up to this point, we have tried to keep this work as model-agnostic as possible. However, one may wonder how the resulting limits for $\phi \rightarrow \nu\nu$ compare to those for $\phi \phi \rightarrow e^+ e^-$, specifically since $\text{SU}(2)$ invariance in the SM naively enables both channels with equal branching ratios. In figure~\ref{fig:mixed}, we therefore present the resulting limits for $m_\phi \in \{10, 100\}\,\mathrm{GeV}$ (red, blue) and different choices of $\text{BR}_{ee} \neq 0$. Specifically, we show the case of pure neutrino injection with $\text{BR}_{ee} = 0$ (solid), pure EM injection with $\text{BR}_{ee} = 1$ (dotted), and mixed injections with $\text{BR}_{ee} = 0.5$ (dashed). As expected, pure EM injection yield stronger limits for most lifetimes, since the direct injection of EM material (dotted) leads to much more efficient photodisintegration.

However, for small lifetimes, electromagnetic injection ceases to be effective and (even suppressed) hadronic injections become relevant. In this regime pure neutrino injections (solid) lead to comparably strong constraints, mainly due to the presence of hadronic injections arising from non-thermal scatterings. In addition, neutrinos also profit from the \emph{delayed} injection of EM/hadronic material during the neutrino cascade, meaning that a relic with $\tau_\phi = 10^4\,\mathrm{s}$ still injects a significant amount of material at $t \gg \tau_\phi$ due to the fact that scattering is rare in large regions of the parameter space. Therefore, neutrinos can push beyond the usual cutoff at $\tau_\phi \sim 10^{4}\,\mathrm{s}$ ($\tau_\phi \sim 10^{2}\,\mathrm{s}$) below which the constraints from photodisintegration (hadrodisintegration) vanish. For the case of electrons it turns out that contributions from hadronic FSR are still relevant even for these rather small masses due to the emission of virtual photons -- a contribution which is not present for the neutrinos.\footnote{We thank the anonymous referee for pointing out the relevance of this contribution.} Note that there now also exists a transition region between $\tau_\phi = 10^4\,\mathrm{s}$ and $\tau_\phi = 10^5\,\mathrm{s}$, for which a cancellation between the deuterium production due to the hadronic injection and the deuterium destruction due to the EM injection weakens the limits from electron injections significantly.

When considering the $\text{SU}(2)$ symmetric case (dashed), the limits are typically weaker than the ones for pure injections by a factor 2. Again there is a cancellation among the EM and hadronic contributions. Note however that the cancellation is displaced significantly w.r.t.\ the case of $\text{BR}_{ee}=1$ for $m_\phi=100\,$GeV 
as the cancellation is very sensitive to the relative strength of hadronic and EM contributions.
Note that for much larger masses, $m_\phi \gg 100\,\mathrm{GeV}$, and for large lifetimes, the limits for pure neutrino and pure EM injections are expected to exhibit a very similar behaviour, since -- in this case -- also the limits for $\text{BR}_{ee} = 0$ are dominated by EM injections, specifically from FSR. Consequently, the ratio of both limits is then equal to the ratio of the injected EM energy. Since for $\text{BR}_{ee}=1$ basically $100\%$ of the energy is released in the form of EM material, this ratio is given by the red line in figure~\ref{fig:EW_shower}. Consequently, the ratio approaches a value close to unity asymptotically for large masses, implying that the bounds for pure neutrino injections are only marginally weaker in this regime. For smaller lifetimes, the same statement is not true, since hadronic injections complicate the matter. In conclusion, we therefore find that neutrinos still lead to important constraints, even if the relic can also directly decay into electron-positron pairs. However, there also exist well-motivated models for which $\text{BR}_{ee}$ is naturally either small or zero, some of which are discussed in the following section.

\subsection{A note on model-building}
While the previous section discusses a valid point, injected neutrinos do not necessarily imply an equal amount or even any primary charged leptons. Let us briefly discuss some potential scenarios, for which a relic $\phi$ can dominantly decay into two neutrinos.
A well known example of such a scenario is the Majoron framework.
In its standard minimal version, the key interaction is of the form  $y_N \phi \overline{N^c} N$
with a sterile neutrino $N$ and $\phi= s e^{i\theta}$, where $\theta$ is the pseudoscalar Majoron field, a pseudo-Goldstone boson resulting from the breaking of a global Abelian $B-L$ symmetry by the vacuum expectation value of the field $s$ (see \cite{Chikashige:1980ui,Schechter:1981cv} and e.g.~\cite{Garcia-Cely:2017oco,Frigerio:2011in}).
The decay into light neutrinos proceeds through seesaw mixing between the heavy sterile neutrino and ordinary neutrinos at a rate $\Gamma_\phi\sim m_\theta y_N^2 m_\nu^2/(16 \pi m_N^2)$, with $m_\nu$ being the neutrino mass scale. In this scenario, a suppressed decay into $e^+e^-$ is also induced at one loop, with an extra relative factor $\sim (m_e^2 m_N^2/m_W^4)/(4\pi^2)^2$ in the decay width. Specifically, this decay channel is relatively enhanced by the mass of the seesaw-state, but involves extra chirality flip and loop suppression factors, so that it is suppressed for $m_N \lesssim 10^9 \; \mathrm{GeV}$. The range of lifetimes considered in this work, $\tau_\phi = (10^4 - 10^{12})$~s, typically  requires a seesaw scale $m_N\sim (10^4 - 10^8~\hbox{GeV}) \cdot y_N$. Obviously, in this case the Majoron is not the DM particle, as often considered, but instead corresponds to the unstable relic $\phi$ assumed in this work. Overall, we see that for such a model, the BBN bounds on the neutrino channel are stronger than the BBN ones on the charged lepton channel in a significant part of the parameter space. 

Another option is to consider the coupling of a gauge boson to two sterile neutrinos, also followed by seesaw neutrino mixings into ordinary neutrinos \cite{Coy:2020wxp}. In this case, both the decay into light neutrinos and into $e^+e^-$ are suppressed by four powers of the heavy sterile neutrino, with an extra loop suppression for the charged lepton channel, so that here too the primary decay channel is into light neutrinos.

Yet another straightforward example of a scenario leading to a decay into neutrinos, without -- in this case -- any production of primary charged leptons, is to consider the source particle as the neutral component of a scalar triplet of hypercharge 2, $\Delta=(\delta^{++},\delta^+,\delta^0)$. In this case, the two charged scalar triplet components, which are heavier due to EW symmetry breaking, disappear quickly from the thermal bath by decaying into the neutral component. The remaining neutral component decays into $\nu\nu$ from a $\Delta L L$ Yukawa interaction. 
The relic density of such triplets, prior to their decay, is the one of a hypercharge 2 scalar triplet DM candidate, see \cite{Cirelli:2005uq,Hambye:2009pw}, that is to say of a typical WIMP. From the EW annihilations, the relic density is below (above) the observed DM relic density for masses below (above) 2.1 TeV \cite{Bottaro:2022esy} (and can also be below it around the Higgs boson resonance). Scalar quartic couplings can further increase the annihilation, i.e.~suppress the relic density \cite{Hambye:2009pw}. 
In this framework, there is no decay into charged leptons due to lepton-number conservation.

Phenomenologically very similar to the decays of a relic into two neutrinos as studied in this work are decays into one neutrino and one dark-sector state. In the latter case, it is natural that no associated charged leptons are produced. An example of such a setup would be a supersymmetric sneutrino, which decays into one neutrino and one gravitino, a scenario previously studied in~\cite{Kanzaki:2007pd}. 
Another option is to consider the decay of a scalar or a gauge boson into two not so heavy sterile neutrinos $\nu_s$. In this case, the leading decay, if kinematically allowed, would be $\phi \to \bar{\nu}_s \nu_s$, but mixing with SM neutrinos would also induce e.g.~$\phi \to \bar{\nu}_s \nu$, which is much less suppressed than the decay into charged leptons. In this case, there would be a strong limit from $\Neff$ considerations, but for sufficiently large mixings, the limits derived in this work may also be relevant. If the sterile neutrino mass is close to the mass of $\phi$, the decay $\phi \to \bar{\nu}_s \nu$ could also be the leading decay channel.

\section{Conclusion}\label{sec:conc}

In this work, we have performed a detailed study of the decay of a relic particle $\phi$ into neutrinos, $\phi \rightarrow \nu \bar{\nu}$, concentrating on lifetimes larger than about $10^4$~seconds. At this time, primordial nucleosynthesis has basically completed, but late-time injections of EM or hadronic material may still disintegrate the light elements previously synthesised. The case of neutrino injections is of particular interest, as it is expected to be less constrained compared to injection of other SM particles. We have shown that, although these constraints are generically weaker, they are still much stronger than naively expected, covering large parts of the parameter space defined by the mass $m_\phi$, the lifetime $\tau_\phi$, and the abundance $f_\phi$ of the relic particle $\phi$.

Once produced, the neutrinos interact very little with the abundant photons, baryons, and electrons that are present at the time. However, they can still interact significantly with the numerous background neutrinos that are present in the primordial plasma (\emph{thermal scattering}). Moreover, two injected neutrinos can also interact among each other (\emph{non-thermal scattering}). While the injection of electron-positron pairs due to thermal scattering can lead to relevant limits, we find that non-thermal scattering further results in hadronic injections, which we show to be very important, as they dominate the resulting limits by several orders of magnitude in significant regions of parameter space. Specifically, this effect dominates the constraints for masses up to $\sim 150$~GeV and (depending on the mass) lifetimes up to $10^7-10^9$~s (cf.~figures~\ref{fig:flags_early} and \ref{fig:mphi_all}). In fact, hadronic injections from non-thermal scattering are dominant in this region for three main reasons: 
\begin{enumerate}
    \item[a)] non-thermal scatterings, unlike thermal ones, profit from the fact that they involve two energetic neutrinos in the initial state, leading to a larger centre-of-mass energy, thus allowing the production of heavier particles due to a larger cross-section from resonant scattering close to the $Z$-pole;
    \item[b)] the nucleons produced via non-thermal scattering feature a hadrodisintegration cross-section for destroying helium-4 (and producing deuterium) that is two orders of magnitude larger than the one for photodisintegration of deuterium;
    \item[c)] helium-4 is more than three orders of magnitude more abundant than deuterium.
\end{enumerate} 
Instead, for masses up to $\sim 100$~GeV and larger lifetimes, the dominant constraint on the light-element abundances stems from the modification of the baryon-to-photon ratio $\eta$. However, in this regime, the CMB $\Neff$ bounds dominate over the ones from BBN. In between these two regimes, there is a cancellation of both effects, which leads to weaker constraints. Interestingly, for small lifetimes, the constraints coming from neutrino injections can be more stringent than the ones coming from electron-positron injection, as the effects of a pure EM cascade is suppressed at early times.

For masses beyond the EW scale, the effects of the EW shower from FSR quickly become dominant. In this region, we further find that -- for lifetimes that are not too large -- the induced hadrons lead to an improvement of the constraints by several orders of magnitude compared to the injection of only EM material. 
In fact, hadrons dominate the limits for lifetimes up to $10^7$~s. For larger lifetimes, photodisintegration from EM material due to FSR dominates. Both EM and hadronic contributions from FSR lead to an upper bound on the source-particle abundance, which is relatively independent of the lifetime and mass, namely $f_\phi\leq 10^{-2}-10^{-3}$.

The derivation of our results was made possible by a new, Monte-Carlo inspired probabilistic approach, which can be used to track the relevant part of the evolving phase-space distribution of the injected neutrinos. This is essential in order to faithfully evaluate the effects of non-thermal scattering, which previously had been largely ignored in the literature due to its computationally challenging nature. In this context, an important, simplifying insight is that any neutrino that scatters once, no longer has to be tracked as part of the spectrum. This is particularly true in case the corresponding interaction rates are smaller than the Hubble rate, so that neutrinos typically scatter only once (if at all). 

In this work, we have concentrated on a ``model-agnostic'' approach regarding the late-time injection of energetic neutrinos and
included generic effects which had previously been neglected in the literature. We have briefly commented on a number of different particle-physics scenarios, which result in a predominant injection of neutrinos (possibly in addition to truly decoupled states) and therefore are expected to lead to constraints that are very similar to the ones presented in this work. Overall, we found that, even for this seemingly innocent case, the resulting constraints from photo- and hadrodisintegration can be very stringent, much stronger than the indirect ones from a new radiation component parameterised by $\Neff$.

\acknowledgments
We would like to thank the authors of~\cite{Chang:2024mvg} for communications early in the project, and N. Grimbaum-Yamamoto for discussions.
This work is funded by the Deutsche Forschungsgemeinschaft (DFG) through Germany's Excellence Strategy --- EXC 2121 ``Quantum Universe'' --- 390833306. JF is supported by an ERC StG grant (“AstroDarkLS”, Grant No. 101117510).
The work of TH and MH is supported by the Belgian IISN convention 4.4503.15 as well as by the Brussels laboratory of the Universe - BLU-ULB.

\newpage
\appendix

\section{A simplified calculation of the CMB $\Neff$ limit}\label{app:neff}

In this appendix, we present a simplified calculation of the CMB $\Neff$ constraint within our scenario.
In the main text and specifically for eq.~\eqref{eq:neff}, we assumed that $\phi$ had already decayed fully into neutrinos. However, to track the cosmological evolution and to estimate the resulting limit, it is useful to also take into account the energy density in $\phi$ prior to its decay. Using both the energy density $\rho_\nu^\text{n-th}(t)$ of the ultra-relativistic non-thermal neutrinos, as well as the energy density $\rho_\phi(t)$ of the non-relativistic relic, we can define a combined DM energy density $\rho_\text{DM}(t)$ by means of the simplified relation
 \begin{align}   
   \rho_{\nu}^\text{n-th}(t) + \rho_\phi(t) \equiv &\begin{cases}
   f_\phi \rho_\text{DM}(t) & \quad\text{for}\quad t\leq \tau_\phi\\
   f_\phi \rho_\text{DM}(t)\left(\frac{R(\tau_\phi)}{R(t)}\right) & \quad\text{for}\quad  t>\tau_\phi
   \end{cases}\eqsp.
\end{align}
Taking into account the transition from radiation to matter domination at the time $t_\text{eq}$ of matter-radiation equality, we find at the time $t_\text{rec}$ of recombination
\begin{align} 
 \rho_{\nu}^\text{n-th}(t_\text{rec})&\simeq f_\phi\rho_\text{DM}(t_\text{rec})\left(\frac{R(\tau_\phi)}{R(t_\text{eq})}\right)\left(\frac{R(t_\text{eq})}{R(t_\text{rec})}\right)\simeq f_\phi\rho_\text{DM}(t_\text{rec})\frac{\tau_\phi^{1/2} t_\text{eq}^{1/6}}{t_\text{rec}^{2/3}}\eqsp.
\end{align}
Using this relation, we can derive an approximate limit on the combination $f_\phi\tau_\phi^{1/2}$, assuming that none of the neutrinos dissipate any of their energy in the form of EM/hadronic material.\footnote{In our numerical analysis, we take into account that a fraction of the energy ends up in the form of EM/hadronic material.} Taking for simplicity $N_\text{eff,sm} = 3$, since the difference to the actual value is much smaller than what is allowed by the limit on $\Delta \Neff$, and by approximating $T_\nu=(4/11)^{1/3}T$, we find
\begin{align}
& \Delta \Neff \simeq \frac{\rho_{\nu}^\text{n-th}(t_\text{rec})}{2\frac{7}{8}\frac{\pi^2}{30}T_{\nu,\text{rec}}^4} <0.29\\
\Rightarrow \quad & f_\phi \tau_\phi^{1/2} \lesssim 1.4\cdot 10^{6}\,\text{s}^{1/2} \frac{\rho_{\nu}^\text{th}}{3\rho_\text{DM}}\Big|_{t=t_\text{rec}} = 1.4\cdot 10^{6}\,\text{s}^{1/2} \frac{\Omega_{\nu}^\text{th}h^2}{3\Omega_\text{DM}h^2}\Bigg|_{t=t_\text{today}}z_\text{rec}\simeq 7\cdot 10^4\,\text{s}^{1/2} \nonumber \\
\Rightarrow \quad & f_\phi \left(\frac{\tau_\phi}{10^4\,\text{s}}\right)^{1/2}\simeq 700\;.
\end{align}
Here, $t_\text{eq}\approx 50000\,$yrs, $t_\text{rec}\approx 380000\,$yrs, and $z_\text{rec}\approx 1100$. Moreover, we used the fact that $2\frac{7}{8}\frac{\pi^2}{30}T_{\nu,\text{rec}}^4\simeq \rho_{\nu}^\text{th}(t_\text{rec})/3$ which can then be related to $\Omega_{\nu}^\text{th}h^2$ \cite{Planck:2018vyg}. In fact, this approximation is very close to what is obtained by a full numerical treatment, i.e.
\begin{align}
    f_\phi \left(\frac{\tau_\phi}{10^4\,\text{s}}\right)^{1/2}\simeq 680\eqsp,\label{eq:neff_num}
\end{align}
which we use e.g.~for the black line in figure~\ref{fig:mphi_all}.

Finally, let us note that as part of our numerical analysis, we calculate a more consistent value of $\Neff$, taking into account also the fact that the non-thermal neutrinos lose some of their energy in the form of EM/hadronic material. Eq.~\eqref{eq:neff_num} is therefore used only for comparison and reference.

\section{On the relevance of neutrinodisintegration reactions}\label{app:deplete}

In this appendix, we show that the abundance of injected neutrinos is not significantly impacted by deep inelastic scattering (DIS) between the neutrinos and the background nucleons.\footnote{For neutrino energies beyond a few GeV, DIS scattering dominates over elastic scattering.} To this end, we compare the interaction rate of this process $\sim \langle \sigma_n v\rangle n_n$ with the one for scattering off thermal background neutrinos $\sim \langle \sigma_\nu v\rangle n_\nu$, i.e.
\begin{align}
    \frac{\langle \sigma_n v\rangle n_n}{\langle \sigma_\nu v\rangle n_\nu}\sim \frac{10^{-38}\,\text{cm}^2\frac{E}{\text{GeV}}\eta T^3}{G_F^2 E T^4}
    \sim \frac{10^{-11}\eta}{10^{-10}\frac{T}{\text{GeV}}}\sim  10^{2}\eta\frac{\text{MeV}}{T}\sim  10^{-8}\frac{\text{MeV}}{T}\eqsp.
\end{align}
Here, we utilised the DIS cross-section from \cite{Formaggio:2012cpf,SajjadAthar:2022pjt} as well as $n_n\sim \eta T^3$. In conclusion, we find that -- for all relevant temperatures $T \lesssim 10^{-2}\,\mathrm{MeV}$ --, it is indeed safe to ignore the scattering reactions between non-thermal neutrinos and the background nuclei. However, we note that similar to the quasi-elastic scattering studied in \cite{Chang:2024mvg}, sizeable neutrino injections during thermal BBN could change this conclusion. Nevertheless, this problem is irrelevant for this work specifically, since we  focus on $\tau_\phi \geq 10^4\,$s.

\section{A simplified analytical expression for the neutrino spectrum}
\label{app:ana}

In this appendix, we briefly discuss a simplified analytical expression for the neutrino spectrum that we found during the early stages of the project. After developing a numerical implementation which is both efficient and more precise, we refrained from using this expression in more detail. However, we still find that this expression can give some instructive insights, specifically in the context of showcasing the agreement between the numerical and the analytical result in figure~\ref{fig:spectra}.

In the following, we sketch the different steps that are required to derive this simplified analytical approximation, thereby highlighting all important approximations. Assuming that all scattering rates are negligible w.r.t.~the Hubble expansion, and by using $H \simeq 1/2t$, The evolution of the neutrino spectrum is determined by the following simplified Boltzmann equation
\begin{align}
    \frac{\partial f_\nu}{\partial t}-\frac{E}{2t}\frac{\partial f_\nu}{\partial E} \simeq \frac{4\pi^2}{E^2}\frac{n_\phi}{\tau_\phi}\delta(E-E_0)\eqsp.
\end{align}
Here, we used the fact that the initial particle $\phi$ is non-relativistic and features a lifetime $\tau_\phi$ as well as an abundance $n_\phi$, the latter of which is defined in eq.~\eqref{nphit}. Moreover, $E_0 = m_\phi/2$. This equation can be solved analytically, which yields
\begin{align}
    f_\nu(t,E) \simeq \frac{8\pi^2}{E_0^5}\frac{n_\phi\left(t E^2/E_0^2\right)}{\tau_\phi}t E^2\Theta(E_0-E)\Theta\left(E-E_0\sqrt{t_0/t}\right)\eqsp.
\end{align}
In this expression, the upper limit in energy, as indicated by the first Heaviside function, is determined by the maximally possible neutrino energy $E_0=m_\phi/2$. The lower limit, as indicated by the second Heaviside function, is less trivial and accounts for the ``starting time'' $t_0$ of the neutrino injection. Technically, such a starting time is in fact artificial, since decays always happen in some capacity, even well before $t=\tau_\phi$. Here, and specifically for figure~\ref{fig:spectra}, we set $t_0=\tau_\phi/\Lambda$ as in section~\ref{sec:MC} for consistency. However, let us note that even in a more realistic scenario, the interactions of the neutrinos with the background will eventually deplete the spectrum, thus leading to an effective cutoff, a feature that can e.g.~be observed in the full numerical result in figure~\ref{fig:spectra}. Finally, let us note that the spectrum further depends on the number density $n_\phi$ evaluated at $t (E/E_0)^2$. This factor naturally shuts off the injections due to the exponential suppression of the relic abundance beyond its lifetime. Additionally, the redshift factor in the argument ensures that only those neutrinos injected at a later time suffer from this suppression, while neutrinos injected at earlier times have already lost some of their energy $E$ due to redshift, i.e.~$E\ll E_0$, and therefore are not suppressed.

\section{A more detailed look at the hadronic cascade}
\label{app:hadro}

We provide a concise summary of the physical processes driving the hadronic cascade in this appendix. In combination with the Boltzmann equation in eq.~\eqref{eq:y_nth}, the numerical framework presented in this section forms the basis for the implementation of the hadrodisintegration reactions in \texttt{v2.0.0-dev} of \texttt{ACROPOLIS}. In particular, we focus on the calculation of the parameters $\xi_X^y[n_j](T, K)$ for $y \in \{ p, n \}$ and $X \in \{p, n, \text{D}, \text{T}, {}^3\text{He}, {}^4\text{He} \}$.

In a nutshell, energetic hadrons injected into the thermal plasma can scatter off the background particles, thus initiating a hadronic cascade. In this context, the most important processes are
\begin{enumerate}
    \item EM scattering with background particles, i.e.~primarily electrons $e^-$ and photons $\gamma$, which leads to continuous energy loss,
    \item decay of the unstable hadrons, i.e.~primarily neutrons in our case,
    \item hadronic scattering with background nuclei, i.e.~primarily protons $p_\text{BG}$ and helium-4 nuclei $\alpha_\text{BG}$.
\end{enumerate}
Notably, when describing these processes, we can safely neglect the expansion of the Universe, since the Hubble rate is always much smaller than the rates of any of the processes above.\footnote{We estimate this by comparing the Hubble rate $H\sim 1\,\text{s}^{-1} (T/\text{MeV})^2$ with a typical nuclear interaction rate $\Gamma\sim n_p \sigma_\text{nuc}\sim 10^8\,\text{s}^{-1}(T/\text{MeV})^3 \gg H$. Consequently, for any temperature before recombination, the hadronic cascade is well approximated as instantaneous w.r.t.~to cosmic expansion.}

Before delving into a detailed description of the various processes, let us first point out some important approximations that we employ as part of our calculation. In \hyperlink{cite.Kawasaki:2005}{KKM05}, one of the main goals was to derive strictly conservative constraints originating solely from the destruction of $^4$He and the subsequent production of D, T and $^3$He. Since we follow this publication closely, many of the following simplifications are also based on this principle. However, it is worth noting that -- in the larger context of this work -- the injection of energetic nucleons is only one of the important effects, and the final light-element abundances are also impacted by a modified background cosmology and EM injections. As discussed already in the main text, hadronic and EM injections can have opposite effects and thus lead to cancellation effects, which can make some exclusion bounds overly optimistic and no longer conservative. However, in practise, we find that such cancellations are only relevant for very small parts of parameter space, meaning that the following approximations can still largely be considered conservative:

\begin{itemize}[label={}]
    \item \textbf{Target particles.} As shown in \hyperlink{cite.Kawasaki:2005}{KKM05}, the nuclear cross-sections are of similar orders-of-magnitude for all light elements, meaning that the corresponding scattering probabilities differ only due to the different abundances of the targets. Therefore, we consider only the two most abundant nuclei as targets, i.e.~$p_\text{BG}$ and $\alpha_\text{BG}$. The impact of any other element is generically suppressed by their small abundance.
    \item \textbf{Projectile particles.} The most important hadrons that are injected by non-thermal scattering of the neutrinos are $p$ and $n$, with additional (high-energy) nucleons being produced during the various steps of the hadronic cascade. Heavier final-state nuclei resulting from the cascade often have a smaller kinetic energy based on the prescription used for the energy transfer (also cf.~appendix~C of \hyperlink{cite.Kawasaki:2005}{KKM05}). Their impact on the subsequent evolution of the cascade is therefore subleading compared to the one of neutrons and protons. As a result, we assume that only protons and neutrons act as projectile particles. However, in accordance with \hyperlink{cite.Kawasaki:2005}{KKM05}, we do use a simplified treatment to estimate if the produced nuclei survive or are disintegrated by either a CMB photon or a background proton (see discussion below). If the survival condition is not met, we assume that the particle is broken up into its separate nucleons with equal momenta, e.g.~$p_\text{BG} + \text{T} \rightarrow p + 2n$, with the latter ones being reinjected into the cascade if their kinetic energy is above $20\,$MeV.
    \item \textbf{Energy range.} Experimental measurements for the cross-sections of hadronic processes where the kinetic energy of the projectile is higher than $\sim$ 20 GeV are not available, except for some data on $pp$ and $np$ scatterings.\footnote{In general, we use the nuclear cross-sections presented in \hyperlink{cite.Kawasaki:2005}{KKM05}.} However, based on the available data, \hyperlink{cite.Kawasaki:2005}{KKM05} estimates the cross-sections for high-energy hadronic scattering to become approximately constant beyond $20\,$GeV, an approximation that we also employ in this work. Nevertheless, we note that the energy of nucleons injected from FSR and/or non-thermal scattering is usually only a few \% of the initial energy and thus -- for the parameters considered in this work -- rarely much above the extrapolation threshold.
    \item \textbf{Isospin symmetry.} Even below $\sim 20\;$GeV, some relevant experimental data is not available. Therefore, we invoke isospin symmetry to approximate the missing cross-sections whenever necessary and possible.\footnote{EM corrections are typically subleading in the energy range of interest.} More precisely, we consider all processes shown in table~\ref{tab:hadronic},\footnote{Note that we denote charged pions, neutral pions, and multiple pions in the final state all by $\pi$ for simplicity.} but use separate cross-sections for protons and neutrons only for the reactions in the upper part of the table, while using the same (isospin-symmetric) cross-section for protons and neutrons in the lower part. Although a comprehensive analysis of the full nuclear interaction network would yield more precise predictions for the effects of hadrodisintegration, we still expect that all dominant effects are captured well within this approximation. In particular, since we anyway assume a 20\% uncertainty on the nuclear reaction rates.
\end{itemize}

\subsection{Parametrisation of the hadron spectrum}
\label{sec:para_spec}
In this section, we introduce the scheme that we use to track the spectrum/amount of the individual nuclei that are present during the cascade. Here, the numerical framework is inspired by the discussion in appendix~D of \hyperlink{cite.Kawasaki:2005}{KKM05}. First, we introduce a vector $\underline{A}$ of the form
\begin{align}
    \underline{A} = \begin{pmatrix} 
    \underline{N}_p \\
    \underline{N}_n \\
    \Delta N_p \\
    \Delta N_n \\
    \Delta N_\text{D} \\
    \Delta N_\text{T} \\
    \Delta N_{{}^3\text{He}} \\
    \Delta N_{{}^4\text{He}}
    \end{pmatrix}\eqsp.
\end{align}
In this expression, $\underline{N}_p$ and $\underline{N}_n$ are $N$-dimensional vectors describing the distribution of protons and neutrons over $N$ energy bins that are equidistant in log-space, e.g.~$N_{p, i}$ is the number of protons in the $i$th energy bin. Here, the minimal kinetic energy $K_\text{min} \approx 20\,\mathrm{MeV}$ equals the threshold energy for the disintegration of helium-4, and the maximal kinetic energy $K_\text{max}$ corresponds to the highest possible injection energy of the hadrons. Moreover, the six different values $\Delta N_X$ count the relative number of $X$ particles that have been produced/destroyed due to the hadronic injection. Consequently, $\underline{A}$ is a $2N + 6$ dimensional vector. Note that $\Delta N_p$ and $\Delta N_n$ must be considered in addition to the spectra $\underline{N}_p$ and $\underline{N}_n$, as they also track nucleons with kinetic energy $K < K_\text{min}$. \emph{In the following, we label the different entries starting at an index 0 as this directly maps to the numerical implementation, and we use the same convention for the indices of any of the matrices discussed below.}

The evolution of $\underline{A}$ throughout the cascade is then handled by two evolution matrices, i.e.~one matrix $\mathbf{M}_\text{eloss}$ describing the energy loss of the available hadrons up until their next scattering reaction, and one matrix $\mathbf{M}_\text{scat}$ describing \emph{one} of those proceeding hadronic scattering reactions. After \emph{one} step of the cascade, we thus have $\underline{A}_0 \rightarrow \mathbf{M}_\text{scat} \mathbf{M}_\text{eloss} \underline{A}_0$ for some initial vector $\underline{A}_0$. As an example, let us assume that a single proton with kinetic energy $K$ is injected into the plasma. The initial vector $\underline{A}_0$ would then only have two non-zero entries, namely $\Delta N_p = 1$ and $N_{p, j} = 1$, where $j$ denotes the energy bin to which $K$ belongs to. During the cascade, i.e.~by applying $\mathbf{M}_\text{scat} \mathbf{M}_\text{eloss}$ repeatedly, many other entries of $\underline{A}$, including $\Delta N_\text{D}$ etc., will assume non-zero values. Ultimately, the calculation converges, leading to a final vector $\underline{A}_f = \lim_{n\rightarrow \infty}(\mathbf{M}_\text{scat} \mathbf{M}_\text{eloss})^n \underline{A}_0$, which can be used to extract $\xi_X^y$. We discuss the calculation of the two transition matrices in the respective sections below.

\subsection{Electromagnetic energy loss}\label{sssec:eloss}

\begin{table}[t]
    \centering
    \begin{tabular}{|c|c|c|c|}
        \hline
        Process & $i=n$ & $i=p$ & Reaction Type \\
        \hline
        (i, $p_{BG}$; 1) & $n + p_{BG} \rightarrow n + p$ & $p + p_{BG} \rightarrow p + p$ & elastic \\
        (i, $p_{BG}$; 2) & $n + p_{BG} \rightarrow n + p + \pi$ & $p + p_{BG} \rightarrow p + p + \pi$ & inelastic \\
        (i, $p_{BG}$; 3) & $n + p_{BG} \rightarrow n + n + \pi$ & $p + p_{BG} \rightarrow p + n + \pi$ & inelastic \\
        (i, $p_{BG}$; 4) & $n + p_{BG} \rightarrow p + p + \pi$ & $p + p_{BG} \rightarrow n + p + \pi$ & inelastic \\
        (i, $p_{BG}$; 5) & $n + p_{BG} \rightarrow p + p + \pi$ & $p + p_{BG} \rightarrow n + n + \pi$ & inelastic \\
        \hline
    \end{tabular}

\vspace{10pt}
    \centering
    \begin{tabular}{|c|c|c|c|}
        \hline
        Process & $i=n$ & $i=p$ & Reaction Type \\
        \hline
        (i, $\alpha$; 1) & $n + \alpha_{BG} \rightarrow n + \alpha$ & $p + \alpha_{BG} \rightarrow p + \alpha$ & elastic \\
        (i, $\alpha$; 2) & $n + \alpha_{BG} \rightarrow \text{D} + \text{T}$ & $p + \alpha_{BG} \rightarrow \text{D} + {}^3\text{He}$ & inelastic \\
        (i, $\alpha$; 3) & $n + \alpha_{BG} \rightarrow 2n + {}^3\text{He}$ & $p + \alpha_{BG} \rightarrow p + n + {}^3\text{He}$ & inelastic \\
        (i, $\alpha$; 4) & $n + \alpha_{BG} \rightarrow p + n + \text{T}$ & $p + \alpha_{BG} \rightarrow 2p + \text{T}$ & inelastic \\
        (i, $\alpha$; 5) & $n + \alpha_{BG} \rightarrow n + 2\text{D}$ & $p + \alpha_{BG} \rightarrow p + 2\text{D}$ & inelastic \\
        (i, $\alpha$; 6) & $n + \alpha_{BG} \rightarrow p + 2n + \text{D}$ & $p + \alpha_{BG} \rightarrow 2p + n + \text{D}$ & inelastic \\
        (i, $\alpha$; 7) & $n + \alpha_{BG} \rightarrow 2p + 3n$ & $p + \alpha_{BG} \rightarrow 3p + 2n$ & inelastic \\
        (i, $\alpha$; 8) & $n + \alpha_{BG} \rightarrow n + \alpha + \pi$ & $p + \alpha_{BG} \rightarrow p + \alpha + \pi$ & inelastic \\
        \hline
    \end{tabular}
    \caption{\textbf{Top:} Hadronic processes involving a background proton $p_\text{BG}$. \textbf{Bottom:} Hadronic processes involving a background helium-4 nuclei $\alpha_\text{BG}$. Both tables are taken from KKM05; however, following KKMT18, we added additional inelastic processes in the upper part.}
    \label{tab:hadronic}
\end{table}

For the calculation of the hadronic cascade, it is of particular importance to perform a careful analysis of the energy loss of the injected hadrons due to EM interactions (point 1 above): If the hadrons lose much of their energy before scattering hadronically, they might no longer have enough energy to initiate any disintegration reactions. However, if their energy loss is small, they can instead scatter energetically with the background nuclei, thus potentially destroying them in the process. In this subsection, we therefore first discuss the relevant processes leading to such EM energy loss. An injected hadron $H_i$ can interact with the background electrons and photons, mainly via:
\begin{itemize}
    \item Coulomb/magnetic moment scattering $H_i + e^\pm \rightarrow H_i + e^\pm$;
    \item Compton scattering $H_i + \gamma \rightarrow H_i + \gamma$;
    \item Bethe-Heitler (BH) scattering $H_i +\gamma \rightarrow H_i + e^+ + e^- $;
    \item Photo-pion process $H_i + \gamma \rightarrow H_i' + \pi$.
\end{itemize}
Here, $H_i'$ is a potentially different hadron.
For charged hadrons, e.g.~protons, EM energy loss is dominated by Coulomb scattering, whereas
for neutral hadrons, e.g.~neutrons, magnetic-moment interactions are most relevant.
In general, we can express the energy-loss rate as (cf.~appendix~B of \hyperlink{cite.Kawasaki:2005}{KKM05} for a collection of the individual expressions)
\begin{equation}
     \frac{\text{d}E_{H_i}}{\text{d}t} = \left( \frac{\text{d}E_{H_i}}{\text{d}t} \right)_\text{Coulomb} + \left( \frac{\text{d}E_{H_i}}{\text{d}t} \right)_\text{Compton} + \left( \frac{\text{d}E_{H_i}}{\text{d}t} \right)_\text{BH} + \left( \frac{\text{d}E_{H_i}}{\text{d}t} \right)_\text{photo-pion}\eqsp.
     \label{eq:def_dEdt}
\end{equation}
Following \hyperlink{cite.Kawasaki:2005}{KKM05}, we then define the quantity
\begin{equation}
\label{eq:loss_before_scatt}
    R_{H_i+A_j\rightarrow A_k} \left( E_{H_i}^{(\text{in})}, E'_{H_i}; T \right) \equiv \int_{E_{H_i}^{(\text{in})}}^{E'_{H_i}} \Gamma_{H_i+A_j\rightarrow A_k}\left( \frac{\text{d}E_{H_i}}{\text{d}t}\right)^{-1} \text{d}E_{H_i}\eqsp,
\end{equation}
which can be interpreted as the number of hadronic interactions from table~\ref{tab:hadronic} of the type $H_i+A_j\rightarrow A_k$ that happen during the time it takes for $H_i$ to electromagnetically downscatter from the initial energy $E_{H_i}^{\text{(in)}}$ to the final energy $E'_{H_i}$. Here, $A_{j/k}$ are generic initial and final states of the hadronic interactions in table~\ref{tab:hadronic}, and $\Gamma_{H_i+A_j\rightarrow A_k}$ are the corresponding interaction rates. Since we only consider $p_\text{BG}$ and $\alpha_\text{BG}$ to be relevant targets, we have $A_j\in \{p_\text{BG},\alpha_\text{BG}\}$. Moreover, for the hadronic cascade $H_i\in \{p, n\}$, whereas $H_i\in \{p, n, \text{D}, \text{T}, {}^3\text{He}, {}^4\text{He}\}$ for the calculation of the survival probability (see below). Finally, a list of all possible (multi-particle) states $A_k$ is encoded in table~\ref{tab:hadronic}, which labels interactions as $(i,j;k)$. Note that the notation used in eq.~\eqref{eq:loss_before_scatt} is somewhat simplified, since only $T$ is specified as an external parameter. However, in reality, the hadronic cascade also depends on the primordial helium-4 abundance $\mathcal{Y}_p$ and the baryon-to-photon ratio $\eta$ via the interaction rates $\Gamma_{H_i + A_j \rightarrow A_k} \propto n_{A_j}$.\footnote{In terms of final quantities $\xi_X^y[n_j](T, K)$, this manifests in terms of the dependence on $[n_j]$, i.e.~$Y_j = n_j/n_b$ with the baryon number density $n_b \sim \eta n_\gamma$.} We keep this dependence implicit in our notation to avoid visual clutter.

Using eq.~\eqref{eq:loss_before_scatt}, we estimate the total energy loss by summing over all hadronic processes, i.e.
\begin{equation}\label{eq:individual_scattering}
    R^{H_i} \left( E_{H_i}^{(\text{in})}, E'_{H_i}; T \right) \equiv \sum_{j,k} R_{H_i+A_j\rightarrow A_k} \left( E_{H_i}^{(\text{in})}, E'_{H_i}; T \right)\eqsp.
\end{equation}
Here, the quantity $R^{H_i} ( E_{H_i}^{(\text{in})}, E'_{H_i}; T )$ parameterises the total number of hadronic scattering reactions that occur before the initial energy $E_{H_i}^{\text{(in)}}$ of the hadron $H_i$ is reduced to the final energy $E_{H_i}'$ due to EM interactions with the thermal plasma. This parameter allows determining whether an energetic hadron is stopped before engaging in hadronic interactions. On the one hand, if $R^{H_i} ( E_{H_i}^{(\text{in})}, E_{H_i}^\text{th}; T ) \lesssim 1$ with the smallest threshold energy $E_{H_i}^\text{th}$ of any disintegration reaction involving $H_i$, the hadron $H_i$ is stopped before being able to induce any hadrodisintegration reaction. On the other hand, if $R^{H_i} ( E_{H_i}^{(\text{in})}, E_{H_i}^{\text{th}}; T ) > 1$,\footnote{In this case, the interpretation as the number of interactions starts to break down, since hadronic scattering will also contribute to the energy loss.} the hadrons are not fully stopped and can therefore alter the primordial light-element abundances. Using $R^{H_i}$, we can therefore determine the energy $\Tilde{E}^{(R=1)}_{H_i}$ of the energetic hadron $H_i$ before it engages in disintegration reactions, i.e.~by solving
\begin{equation}
\label{eq:en_scattering}
    R^{H_i}\left( E_{H_i}^{(\text{in})}, \Tilde{E}^{(R=1)}_{H_i}; T \right) = 1\eqsp.
\end{equation}
For stable hadrons, the interaction rate coincides with the scattering rate, $\Gamma_{H_i+A_j\rightarrow A_k}=n_{A_j}\sigma_{H_i+A_j\rightarrow A_k}\beta_{H_i}$. However, for unstable hadrons like neutrons, there exists an additional contribution from the decay rate, e.g.~$\Gamma_{n\rightarrow p}=(\gamma_n \tau_n)^{-1}$ with $\gamma_n=E_n/m_n$ for neutrons.\footnote{In \hyperlink{cite.Kawasaki:2005}{KKM05}, the neutron decay is incorporated into the energy loss instead; however, we checked that this leads to only small numerical deviations in the final result.}

Finally, the equations presented above can be used to construct the matrix $\mathbf{M}_\text{eloss}$ by implementing the following procedure: For any given entry $j < N$ ($N \leq j < 2N$) of $\underline{A}$, we set $H_i = p$ ($H_i = n)$ and $E_{H_i}^\text{(in)} = K_j$, where $K_j$ is the central value of the kinetic energy bin corresponding to the entry $j$.\footnote{Note that for $j' < N$ both $j=j'$ and $j=j' + N$ share the same value $K_{j'} = K_{j'+N}$, with $A_{j'}$ ($A_{j'+N})$ counting the number of protons (neutrons) with this energy.} Afterwards, we solve eq.~\eqref{eq:en_scattering} to obtain $\tilde{E}_{H_i}^{(R=1)} \equiv K_k$, corresponding to a different energy bin $k(j)$, which explicitly depends on $j$, i.e.~the energy and identity of the incoming particle, but also other parameters like the temperature $T$ (cf.~eq.~\eqref{eq:en_scattering}). Based on this result, the corresponding entries of the transfer matrix are then given by
\begin{align}
    (\mathbf{M}_\text{eloss})_{ij} = \delta_{i, k(j)} \qquad \text{for} \quad j < 2N \quad \text{if} \quad K_k > K_\text{min}\eqsp.
\end{align}
In addition, since the EM energy loss does not lead to the production or destruction of any nuclei, we have
\begin{align}
    (\mathbf{M}_\text{eloss})_{ij} = \delta_{i,j} \qquad \text{for} \quad 2N \leq j < 2N + 6\eqsp.
\end{align}
Note that the so-defined matrix is strictly triangular, since the energy-loss mechanism never increases the energy of the initial particle.

\begin{figure}[!t]
 \centering
    \begin{subfigure}{0.47\linewidth}
      \centering
      \includegraphics[width=0.72\linewidth, trim= 3cm 0 2cm 2cm]{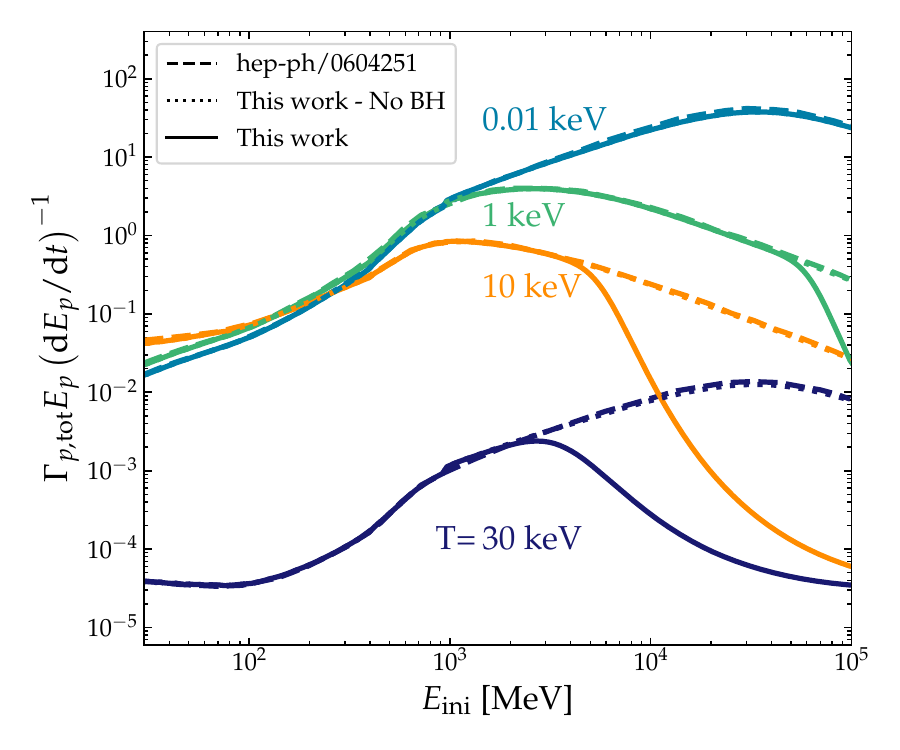}
    \end{subfigure}
    \begin{subfigure}{0.47\linewidth}
      \centering
      \includegraphics[width=0.72\linewidth, trim= 2cm 0 3cm 2cm]{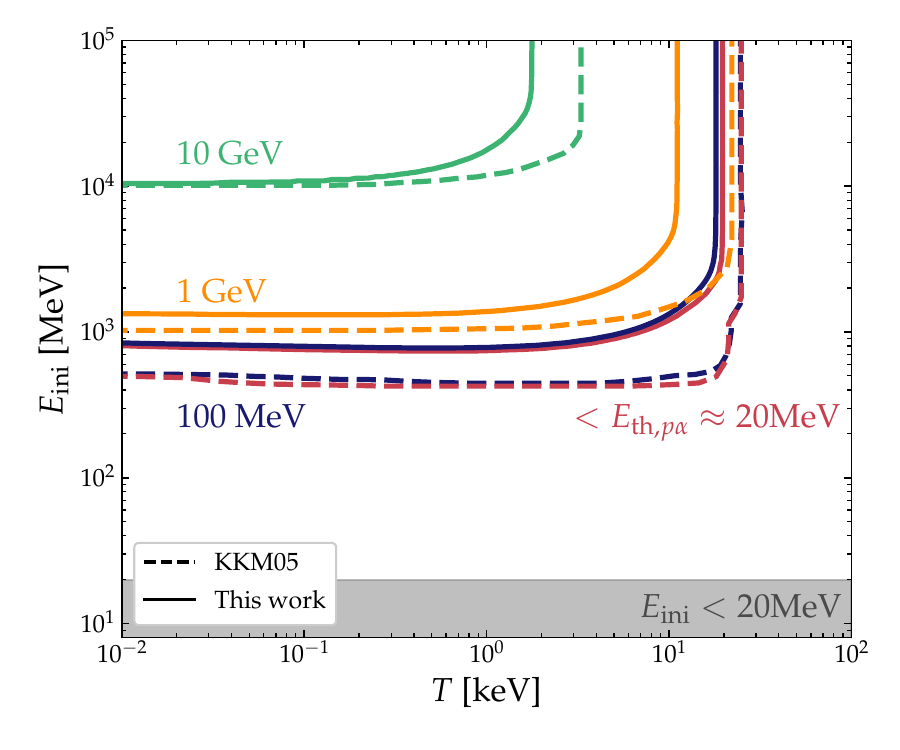}
    \end{subfigure}
\caption{\textbf{Left:} The energy-loss term in eq.~\eqref{eq:def_dEdt} as obtained from our implementation based on KKM05 (dash-dotted). For comparison, we also show the results from figure~1 of \cite{Jedamzik:2006} (dashed), which we are able to reproduce with good precision. For higher energies, the agreement gets spoiled, since the Bethe-Heitler process becomes dominant, which has been ignored in \cite{Jedamzik:2006} (see the solid line). 
\textbf{Right:} Reproduction of figure~15 in~KKM05. The discrepancy is more pronounced for smaller energies and higher temperatures.} 
\label{fig:p_eloss}
\end{figure}

\paragraph{A note on \cite{Kawasaki:2005}:}
As it turns out, we do not seem to be able to reproduce the stopping power of charged particles shown in figures~15-20 of \hyperlink{cite.Kawasaki:2005}{KKM05}. This is despite the fact that the corresponding expressions used to calculate the stopping power in  \hyperlink{cite.Kawasaki:2005}{KKM05} are in agreement with other literature results, e.g.~in \cite{Jedamzik:2006}. In figure~\ref{fig:p_eloss} (right), we illustrate this difference by showing the analogue of figure~15 in \hyperlink{cite.Kawasaki:2005}{KKM05}, i.e.~the contours of constant $\tilde{E}_p^{(R=1)}$ from eq.~\eqref{eq:en_scattering} for protons as a function of the temperature $T$ and the injection energy $E_\text{ini} = E_p^\text{(in)}$. Most notably, we find that the minimum starting energy a particle can have in order to interact before dropping below the disintegration threshold for helium-4 is significantly higher in our calculation.

To analyse and understand this discrepancy, we have cross-checked our own implementation of $\text{d} E_p / \text{d} t$ from eq.~\eqref{eq:def_dEdt} with the findings of \cite{Jedamzik:2006}, in particular figure~1 of that work. To this end, in figure~\ref{fig:p_eloss} (left), we show a comparison between our results and those from \cite{Jedamzik:2006} for the quantity $\Gamma_{p, \text{tot}} E_p (\text{d} E_p / \text{d} t)^{-1}$ where $\Gamma_{p, \text{tot}}$ is the total scattering rate of protons. We find that our implementation, which after all is based on \hyperlink{cite.Kawasaki:2005}{KKM05}, accurately matches the results of \cite{Jedamzik:2006}, except for the high-energy regime, for which the authors of \cite{Jedamzik:2006} do not take into account the contribution from Bethe-Heitler scattering. However, artificially removing this terms from our calculation, again leads to a good match between the two results.\footnote{Note the visible (but numerically irrelevant) bump in the curves around $1\,$GeV, which comes from transition between the ultra- and non-relativistic regime in the expressions of \hyperlink{cite.Kawasaki:2005}{KKM05}.} Overall, we therefore conclude that our implementation is sound, while the discrepancy most likely comes from a difference in the Coulomb terms, as this would explain why our results coincide with those of \hyperlink{cite.Kawasaki:2005}{KKM05} in the case of neutron. However, since our results for $\text{d} E_p / \text{d} t$ agree with those in \cite{Jedamzik:2006} -- while \hyperlink{cite.Kawasaki:2005}{KKM05} does not show a similar plot -- we trust our own implementation and therefore use it for this work.

\subsection{Hadronic scattering reactions}

As already discussed earlier, if $R^{H_i} ( E_{H_i}^{(\text{in})}, E_{H_i}^\text{th}; T ) > 1$, hadronic scattering reactions become relevant. In this case, the $\alpha_\text{BG}$ targets can get destroyed (cf.~table~\ref{tab:hadronic}), thus producing additional D, T and/or $^3$He nuclei, as well as additional protons and neutrons, which are potentially able to destroy further elements in the next cascade step. To fully consider this effect, it is imperative to follow the different daughter particles throughout the hadronic cascade. In this section, we provide a brief description of the numerical treatment that we employ for this task. For more details, we again refer the reader to the reference papers \hyperlink{cite.Kawasaki:2005}{KKM05} and \hyperlink{cite.Kawasaki:2018}{KKMT18}.

Ultimately, the full treatment of the hadronic cascade proceeds in two distinct steps. At first, protons and neutrons loose part of the energy due to EM interactions until $R^{H_i}=1$.\footnote{Notably, some particles are slowed down below threshold without ever starting a cascade.} Afterwards, the different projectile particles with energy $E_i$ undergo scattering reactions with the possible target particles, whereby the probabilities $P_{(i,j;k)}(E_i;T)$ for each of the processes $(i,j,k)$ in table~\ref{tab:hadronic} to happen is proportional to the corresponding interaction rate, i.e.
\begin{equation}\label{eq:prob}
    P_{(i,j;k)}\left( E_i;T\right) \equiv \frac{\Gamma_{H_i+A_j\rightarrow A_k}\left( E_i;T\right)}{\sum_{A_j = p_\text{BG}, \alpha_\text{BG}} \sum_{A_k} \Gamma_{H_i+A_j\rightarrow A_k}\left( E_i;T\right) + \Gamma_{H_i\rightarrow \dots}(E_i; T)}\eqsp.
\end{equation}
Here, $k$ labels the set of final-state particles, while $i$ and $j$ label the initial-state projectile and target particle, respectively. Note that we also include the decay rate $\Gamma_{H_i\rightarrow \dots}$ in the denominator, which, however, is only non-zero for unstable particles, i.e.~neutrons. The nuclear cross-sections driving these rates can be found in \hyperlink{cite.Kawasaki:2005}{KKM05}.

For describing the redistribution of energy due to these processes, which is imperative for calculating $\mathbf{M}_\text{scat}$, we track the spectra of all final-state particle $A_{k,l}$, where the index $l$ differentiates the different nuclei in the multi-particle state $A_k$. To this end, for a given initial state, we calculate the energies of the resulting final-states particles according to the prescription used in appendix~C of \hyperlink{cite.Kawasaki:2005}{KKM05}. Notably, if $A_{k,l}$ is neither a proton nor a neutron, it is important to check if the produced nucleus survives. To this end, we perform a simplified test by evaluating the equations
\begin{equation}
\label{eq:survival}
    E_{A_{k,l}}^{(R=1)}\leq E_{A_{k,l}}^{\text{th},p}\qquad \text{and} \qquad
    \sqrt{3T E_{A_{k,l}}}\leq Q_{A_{k,l}}\eqsp.
\end{equation}
Here, $E_{A_{k,l}}$ is the energy of the particle $A_{k,l}$, $Q_{A_{k, l}}$ is the threshold energy for photodisintegration of $A_{k, l}$ due to a CMB photon and $E_{A_{k,l}}^{\text{th},p}$ is the threshold energy for hadrodisintegration of $A_{k, l}$ due to a background proton.
If at least one of these conditions is violated, we assume that the nucleus is fully destroyed and we re-inject its nucleons into the cascade assuming equipartition of their momentum. Consequently, the survival probability of $A_{k, l}$ is given by
\begin{equation}
    P^\text{surv}_{A_{j,k}}(E_{A_{j,k}};T)=\begin{cases}
        0\quad,\qquad \text{if eq.}~\eqref{eq:survival}~ \text{is violated}\\
        1\quad,\qquad \text{otherwise}
    \end{cases}\eqsp.
    \label{eq:def_Psurv}
\end{equation}
As a result, even if the final-state particles in table~\ref{tab:hadronic} do not contain neutrons or protons, the reinjection of nucleons due to the destruction of heavier nuclei can still give rise to additional projectile particles.

Bases on this setup, we calculate the different entries $(\mathbf{M}_\text{scat})_{ij}$ of the transition matrix in the following way
\begin{enumerate}
    \item Split the transition matrix into different components for the different nuclear reactions, weighted by their respective scattering probability, i.e.
    \begin{align}
        \mathbf{M}_\text{scat} = \sum_{i, j, k} P_{(i,j;k)} \mathbf{M}_\text{scat}^{(i,j;k)}\eqsp.
    \end{align}
    \item For a given process $(i,j;k)$, consider each energy bin $\alpha$ with central kinetic energy $K_\alpha$ for the projectile particle and calculate the spectra $f_p(K; K_\alpha)$ and $f_n(K; K_\alpha)$ of the outgoing protons and neutrons according to appendix~C of \hyperlink{cite.Kawasaki:2018}{KKMT18}. Afterwards, set the different entries $(\mathbf{M}_\text{scat}^{(i,j;k)})_{\beta\alpha}$ with $K_\text{min} < K_\beta < K_\alpha$ such that the entries $\underline{N}_p$ and $\underline{N}_n$ of $\underline{A}' = \mathbf{M}_\text{scat}^{(i,j;k)} \underline{A}$  -- with $\underline{A}$ encoding the state directly after the injection of a single proton or neutron with energy $K_\alpha$ -- are distributed according to the spectra $f_p(K_\beta; K_\alpha)$ and $f_n(K_\beta; K_\alpha)$. Additionally, set the entries $(\mathbf{M}_\text{scat}^{(i,j;k)})_{\beta\alpha}$ with $2N \leq \beta < 2N + 2$ such that $\Delta N_p$ and $\Delta N_n$ of $\underline{A}'$ represent the change in the number of protons and neutrons.
    \item For each process $(i,j;k)$, also calculate the energies of the outgoing D, T, ${}^3$He, and ${}^4$He. For each nucleus that survives (cf.~eq.~\eqref{eq:survival}), set the entries of $(\mathbf{M}_\text{scat}^{(i,j;k)})_{\beta\alpha}$ with $2N + 2 \leq \beta < 2N + 6$ such that $\Delta N_\text{D}$ etc.~reflect the change in nuclei (including the destruction of a potential $\alpha_\text{BG}$ in the initial state). Instead, if the nucleus does not survive, ensure that the individual nucleon fragments are represented in the different entries of $\underline{A}'$.
\end{enumerate}
Notably, this is only a sketch for the calculation of $\mathbf{M}_\text{scat}$. For more details on the exact implementation, we also refer the reader to the publicly-available source code of \texttt{v2.0.0-dev} of \texttt{ACROPOLIS}.

\subsection{Extracting $\xi_X^y$}

Following the previous discussion, we are interested in extracting the parameters $\xi_X^y$ for the Boltzmann equation~\eqref{eq:y_hdi}.

As already mentioned before, we can use $\mathbf{M}_\text{eloss}$ and $\mathbf{M}_\text{scat}$ to calculate the final vector $\underline{A}_f$ at the end of the cascade, i.e.
\begin{align}
    \underline{A}_f =\lim_{n\to\infty} \left(\mathbf{M}_\text{scat}\mathbf{M}_\text{eloss}\right)^{n}\underline{A}_0\equiv \mathbf{T}\underline{A}_0\eqsp.
\end{align}
Here, $n$ denotes the number of cascade steps and $\underline{A}_0$ is the initial vector describing the injection (cf.~section~\ref{sec:para_spec}). Since both $\mathbf{M}_\text{eloss}$ and $\mathbf{M}_\text{scat}$, are strictly triangular, the same is true also for the final transfer matrix $\mathbf{T}$. For $n\rightarrow \infty$, the matrix $\mathbf{T}$ is mostly zero, except for the entries $T_{ij}$ with $i \in [2N,\dots, (2N+6)-1]$ and $j \in [1,\dots,2N-1]$ (\emph{remember that we start indexing at 0}). In practice, we set $n$ to a sufficiently large number, such that all other entries might not be zero but negligible. In fact, since the matrix multiplication is very fast, we can easily ensure convergence.\footnote{For best performance, note that
\begin{equation}   
\left(\mathbf{M}_\text{scat}\mathbf{M}_\text{eloss}\right)^{2^{n}}=\left[\left(\mathbf{M}_\text{scat}\mathbf{M}_\text{eloss}\right)^{2}\right]^{n}\eqsp,
\end{equation}
meaning that $n$ multiplications are sufficient to simulate  $2^{n}$ cascade steps.} 

Given the transfer matrix $\mathbf{T}$, we can then extract the different parameters $\xi_X^y$ from eq.~\eqref{eq:y_hdi}. More precisely, we can identify (omitting the $T$ dependence)
\begin{align}
    \xi^{p}_{p}(K_i)= T_{2N+0, 2i} \quad &,\quad \xi^{n}_{p}(K_i)= T_{2N+0, 2i+N} \\
    \xi^{p}_{n}(K_i)= T_{2N+1, 2i} \quad &,\quad \xi^{n}_{n}(K_i)= T_{2N+1, 2i+N} \\
    \xi^{p}_\text{D}(K_i)= T_{2N+2, 2i} \quad &,\quad \xi^{n}_\text{D}(K_i)= T_{2N+2, 2i+N} \\
    \xi^{p}_\text{T}(K_i)= T_{2N+3, 2i} \quad &,\quad \xi^{n}_\text{T}(K_i)= T_{2N+3, 2i+N} \\
    \xi^{p}_{{}^3\text{He}}(K_i)= T_{2N+4, 2i} \quad &,\quad \xi^{n}_{{}^3\text{He}}(K_i)= T_{2N+4, 2i+N} \\
    \xi^{p}_{{}^4\text{He}}(K_i)= T_{2N+5, 2i} \quad &,\quad \xi^{n}_{{}^4\text{He}}(K_i)= T_{2N+5, 2i+N}\eqsp.
\end{align}
Here, $i \in [0, \dots, N-1]$ runs over the different energy bins with central energy $K_i$. Using this identification, we can implement the hadronic cascade into \texttt{ACROPOLIS} by means of eq.~\eqref{eq:y_nth}. For consistency, we further explicitly check that baryon number is conserved. More precisely, we demand
\begin{equation}
    \sum_X B_X\xi^{y}_X = 1\eqsp,
\end{equation}
with the baryon number $B_X$ of the nucleus $X$ and $y\in \{n, p\}$, since the relation must be fulfilled for both nucleons separately. This equation is based on the interpretation of $\xi^y_X$ as the number of produced/destroyed nuclei $X$ per single injection of the initial nucleon $y$.

\begin{figure}[!t]
 \centering
    \begin{subfigure}{0.47\linewidth}
      \centering
      \includegraphics[width=0.72\linewidth, trim= 3cm 0 2cm 2cm]{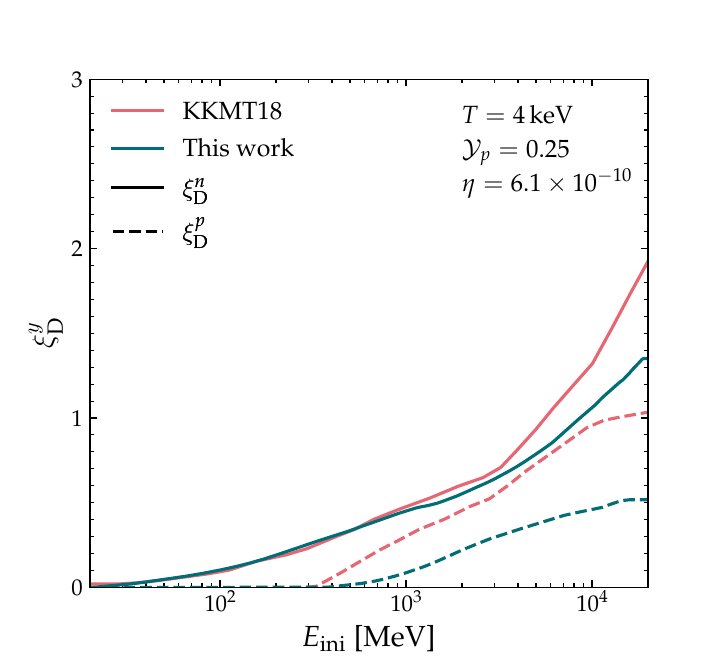}
    \end{subfigure}
    \begin{subfigure}{0.47\linewidth}
      \centering
      \includegraphics[width=0.72\linewidth, trim= 2cm 0 3cm 2cm]{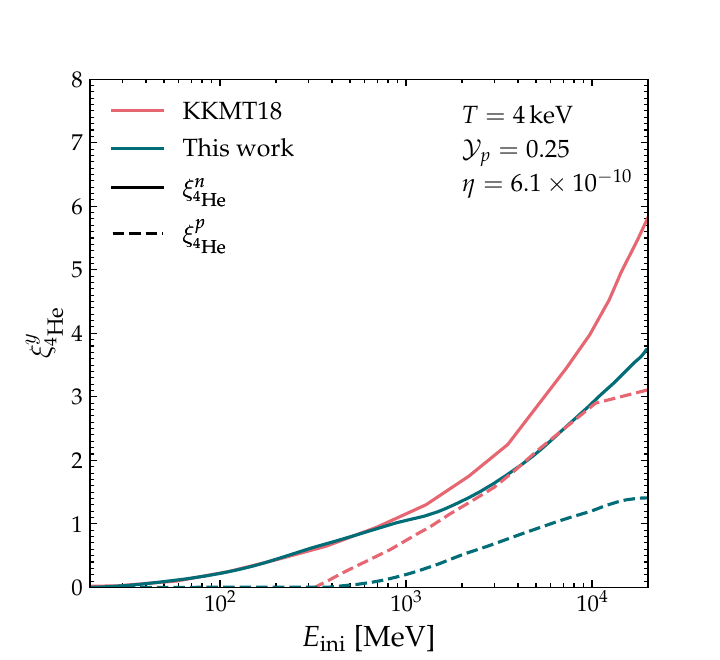}
    \end{subfigure}
    \caption{\textbf{Left:} Deuterium yield (blue-green) per injected baryon for fixed temperature compared to the literature results (red). \textbf{Right:} The same on the left but for $^4$He. Note that in this panel, we have flipped the sign of $\xi^y_{{}^4\text{He}}$ in order for it to be positive.}
    \label{fig:xis}
\end{figure}

To conclude this part of the appendix, let us briefly discuss the results we obtain for one benchmark value of \hyperlink{cite.Kawasaki:2018}{KKMT18}, i.e.~for $T=4\,$keV, $\eta=6.1\cdot 10^{-10}$, and $\mathcal{Y}_p=0.25$. Here, we focus on the production of D and the destruction of $^4$He, since we expect these contributions to have the largest impact on our final results. In the left part of figure~\ref{fig:xis}, we show the deuterium yield (blue-green) per injected hadron for both neutrons (solid) and protons (dashed), and compare it to the corresponding results in \hyperlink{cite.Kawasaki:2018}{KKMT18} (red). We find excellent agreement for the case of injected neutrons in the low-energy regime. For neutrons of higher energies as well as for injected protons in general, we instead observe small deviations, due to the discrepancy we find in the handling of the energy loss (cf.~figure~\ref{fig:p_eloss}). This discrepancy becomes especially clear for the case of injected protons, for which the aforementioned difference in the Coulomb term yields a much stronger stopping power in the context of our own calculation. Therefore, injected protons are significantly less efficient in causing hadrodisintegration reactions and higher energies are required to cause a non-zero impact on the nuclear abundances. For neutrons of small energy, the discrepancy in the energy-loss terms only plays a sub-leading role, as they mainly interact via magnetic-moment scattering. Consequently, the (otherwise leading) Coulomb term is irrelevant. However, at higher energies, conversions from protons into neutrons (also cf.~the upper part of table~\ref{tab:hadronic}) become important (as also analysed in \hyperlink{cite.Kawasaki:2018}{KKMT18}), thus leading again to deviations in the results. Specifically, the protons produced by these conversion processes will experience the same increased stopping power that already caused the deviations in the case of injected protons. Therefore, the efficiency of neutron injection also decreases for higher energies, for which the inelastic (conversion) processes become more pronounced as shown in figures~11 and 12 of \hyperlink{cite.Kawasaki:2005}{KKM05}. This argument is fairly general and therefore directly applies also to the destruction of $^4$He, as shown in the right part of figure~\ref{fig:xis}. For better comparability, we have flipped the sign of our $\xi^y_{{}^4\text{He}}$ in this plot, since it is negative for our convention while \hyperlink{cite.Kawasaki:2018}{KKMT18} defines it as positive.

Finally, let us note that we employ one last approximation as part of the numerical implementation. In general, $\xi_X^y[n_j]$ depends non-linearly on the number density of the available background nuclei $n_j$. However, we have confirmed that to a reasonable approximation, some combinations $\xi_X^y[n_j]/n_A$ are basically constant. In fact, this statement is true specifically for $A = p$ in case $X \in \{p, n\}$ and $A={}^4\text{He}$ if $X\in \{\text{D}, \text{T}, {}^3\text{He}, {}^4\text{He}\}$. Therefore, we can approximate
\begin{align}
    \xi_X^y[n_j](T, K) \simeq \frac{\xi_X^y[n_j^0](T, K)}{n_{A(X)}^0} \times n_{A(X)}(T)\eqsp,
\end{align}
with $A(X) = p$ if $X \in \{p, n\}$ and $A(X) = {}^4\text{He}$ otherwise, which transforms eq.~\eqref{eq:y_hdi} from a non-linear to a linear differential equation. Here, $n_{j/A(X)}^0$ are the corresponding initial abundances at the start of NBBN. Using this approximation, within \texttt{ACROPOLIS} eq.~\eqref{eq:y_hdi} can therefore be solved in the same manner as eq.~\eqref{eq:y_pdi} (see the \texttt{ACROPOLIS} manual for more information).

\section{Boosting between the cosmic rest frame and the centre-of-mass frame}\label{app:kin}

In this appendix, we derive the formula used in eq.~\eqref{eq:def_K_avg}. To this end, we introduce two 4-momenta for the two initial-state neutrinos of the interaction, which we parametrise by the corresponding scattering angle $\theta$ between both neutrinos in the cosmic rest frame, i.e.
\begin{align}
    p_1=\begin{pmatrix}
    E_1\\0\\0\\E_1
    \end{pmatrix}\quad \text{and}\quad 
    p_2=\begin{pmatrix}
    E_2\\0\\E_2\sin\theta\\E_2\cos\theta
    \end{pmatrix}\eqsp.
\end{align}
Here, $E_1$ and $E_2$ are the energies of the initial-state neutrinos. At this point, it is useful to introduce the Mandelstam variable
\begin{align}
    s=(p_1+p_2)^2=2E_1E_2\left(1-\cos\theta\right)\quad\Leftrightarrow\quad \cos\theta=1-\frac{s}{2E_1E_2}\;,
\end{align}
for $m_\nu = 0$, which is independent of the reference frame.\footnote{In our convention, scattering with $\theta=0$ is impossible and therefore not considered in the following.}
Given the 4-momenta from above, the total 4-momentum of the system in the cosmic rest frame is given by
\begin{align}
    p_1+p_2=\begin{pmatrix}
    E_1+E_2\\0\\E_2\sin\theta\\E_1+E_2\cos\theta
    \end{pmatrix}\eqsp.
\end{align}
Since we handle non-thermal scattering in the centre-of-mass (CM) frame, we have to boost $p_1+p_2$ into this exact frame in order to use the results of \texttt{PYTHIA8.3}.
To this end, we first determine the corresponding Lorentz transformation $\Lambda_z$ in $z$-direction. Specifically, the boost velocity $\beta_z$ in this direction is given by
\begin{align}
    \beta_z  =\frac{E_1+E_2\cos\theta}{E_1+E_2}\eqsp,
\end{align}
which yields
\begin{align}
    \Lambda_z(p_1+p_2) =\begin{pmatrix}
    \sqrt{\left(1-\cos\theta\right)E_2\left(2E_1+E_2\left(1+\cos\theta\right)\right)}\\
    0\\
    E_2\sin\theta \\
    0
    \end{pmatrix}\;.
\end{align}
Similarly, for the Lorentz transformation $\Lambda_y$ in $y$-direction the boost is given by
\begin{align}
    \beta_y =\sqrt{\frac{(1+\cos\theta)E_2}{2E_1+E_2(1+\cos\theta)}}\eqsp,
\end{align}
as it leads to the desired result
\begin{align}
   \Lambda_y \Lambda_z(p_1+p_2) &=\begin{pmatrix}
   \sqrt{2E_1E_2\left(1-\cos\theta\right)}\\
    0\\
    0 \\
    0
    \end{pmatrix}\eqsp.
\end{align}
We now consider the 4-momentum of a generic baryon in the centre-of-mass frame resulting from non-thermal scattering, i.e.
\begin{align}
    p_{b,\text{cm}}=\begin{pmatrix}
    E_b\\p_b\sin\theta_b\cos\phi_b\\p_b\sin\theta_b\sin\phi_b\\p_b\cos\theta_b
    \end{pmatrix}\eqsp.
\end{align}
Here, $E_b$ and $p_b$ are the energy and momentum of the baryon, while $\theta_b$ and $\phi_b$ are two angles describing its traversal direction. By applying the inverse transformation $\Lambda_z^{-1} \Lambda_y^{-1}$, we obtain the corresponding 4-momentum in the cosmic rest frame,
\begin{align}
    p_{b,\text{lab}}=\Lambda_z^{-1}\Lambda_y^{-1}p_{b,\text{cm}}\eqsp,
\end{align}
which still depends on the angles defined in the CM frame. Assuming an isotropic distribution of injected baryons, we perform the average
\begin{align}
    \notag \langle p_{b,\text{lab}}\rangle&=\frac{1}{4\pi}\int_{-1}^{1}\int_0^{2\pi}\text{d}\cos\theta_b \text{d}\phi_b\; \Lambda_z^{-1}\Lambda_y^{-1}p_{b,\text{cm}}\\ \notag
    &=\frac{\Lambda_z^{-1}\Lambda_y^{-1}}{4\pi}\int_{-1}^{1}\int_0^{2\pi}\text{d}\cos\theta_b \text{d}\phi_b\; p_{b,\text{cm}}\\
    &=\Lambda_z^{-1}\Lambda_y^{-1} \begin{pmatrix}E_b\\0\\0\\0\end{pmatrix}\;,
\end{align}
where we have used the linearity of the Lorentz transformation. Ultimately, we are only interested in the temporal component of this vector,
\begin{align}
    E_{b,\text{lab}}=\frac{E_1+E_2}{\sqrt{s}}E_b\;,
\end{align}
which is the exact expression used in eq.~\eqref{eq:def_K_avg}.

\section{A collection of supplementary figures}
\label{app:figures}

\begin{figure}[!t]
    \centering
    \includegraphics[width=0.75\textwidth]{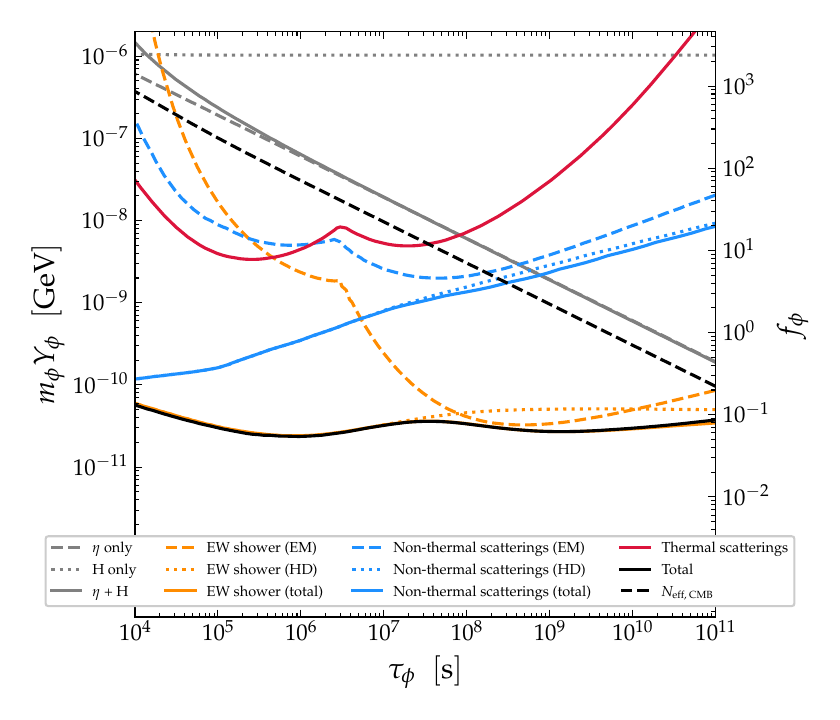}
    \caption{Similar to figure~\ref{fig:flags_early}, but for $m_\phi=200\,$GeV.}
    \label{fig:200GeV}
\end{figure}

\begin{figure}[!t]
    \centering
    \includegraphics[width=0.75\textwidth]{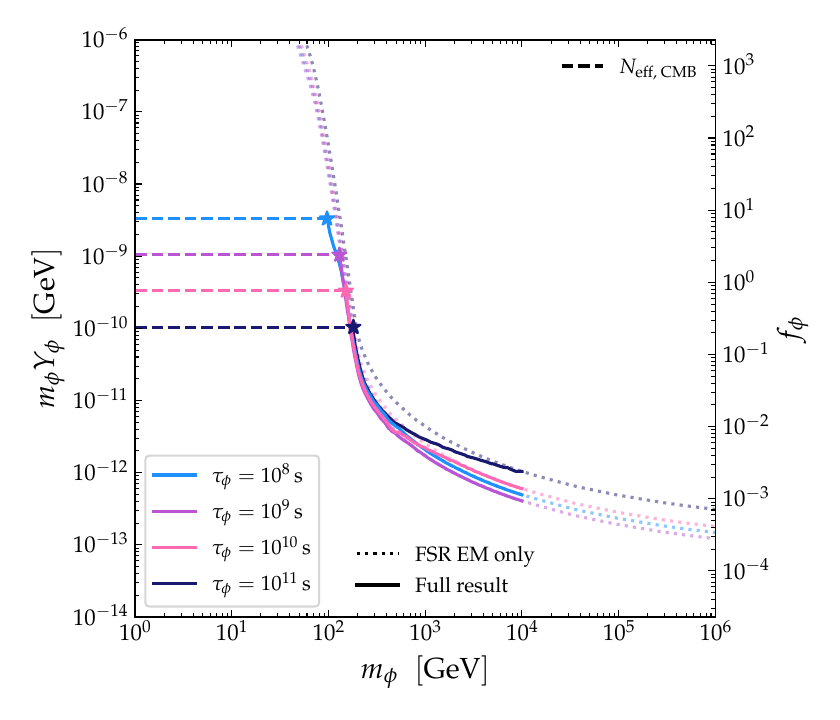}
    \caption{Comparison between the full numerical result for different (large) lifetimes with the one for direct $e^+ e^-$ injections from FSR, i.e.~suppressed by a factor $\zeta_\text{em}^\text{fsr}(m_\phi)$.}
    \label{fig:zeta_lim}
\end{figure}

\begin{figure}[!t]
 \centering
    \begin{subfigure}{0.49\textwidth}
      \centering
      \includegraphics[width=\textwidth]{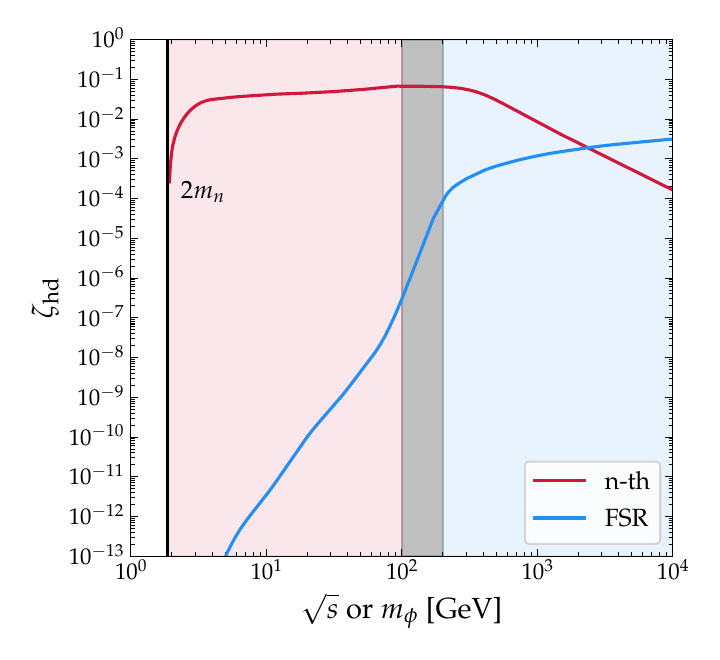}
    \end{subfigure}
    \begin{subfigure}{0.49\textwidth}
      \centering
      \includegraphics[width=\textwidth]{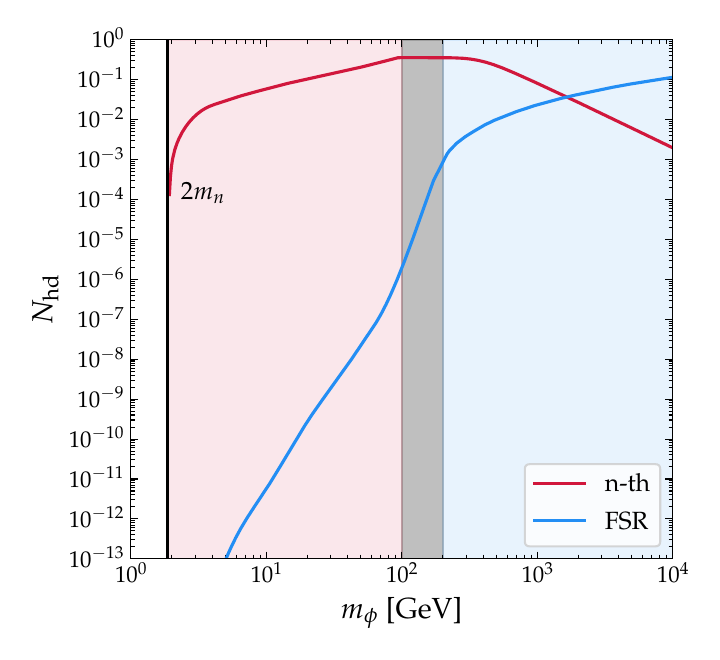}
    \end{subfigure}
    \begin{subfigure}{0.49\textwidth}
      \centering
      \includegraphics[width=\textwidth]{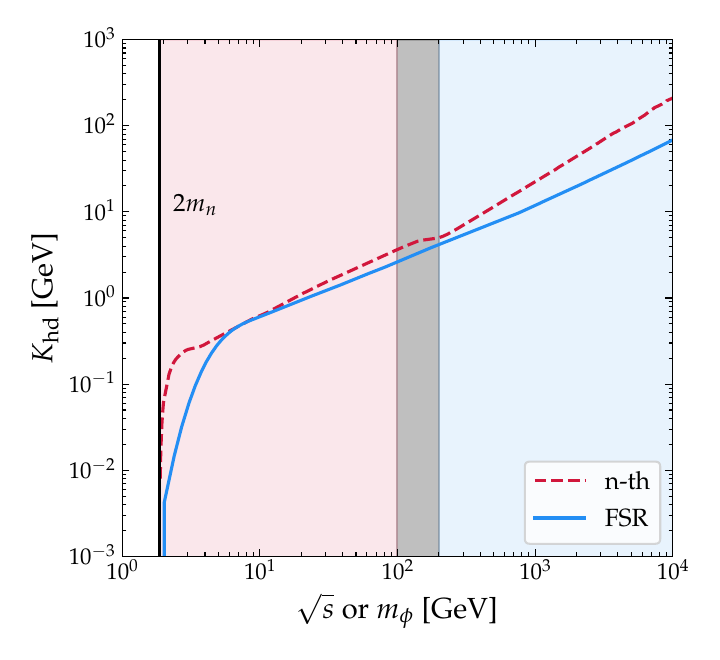}
    \end{subfigure}
\caption{Additional quantities describing the hadronic injection (cf.~section~\ref{sec:MC}), i.e.~$\zeta_\text{hd}$ (upper left), $N_\text{hd}$ (upper right), and $K_\text{hd}$ (lower centre), for non-thermal scattering (red, $x=\sqrt{s}$) and FSR (blue, $x=m_\phi$). In the case of non-thermal scattering, the $y$-axis for a given quantity $Q_\text{hd}$ technically shows the averaged quantity $\langle Q_\text{hd}^\text{n-th} \rangle$. We shade the regions of energy/mass where a given effect is dominant with the corresponding colour, with the grey band indicating the transition region where both effects can be equally important. Note that for $K_\text{hd}$ from non-thermal scattering, we show the value in the centre-of-mass frame, since the corresponding value in the cosmic rest frame explicitly depends on the energies of the initial-state particles (cf.~eq.~\eqref{eq:def_K_avg}).}
\label{fig:aux}
\end{figure}

In this appendix, we collect some figures that supplement our selection of plots in the main text. While we briefly comment on each of them here, we refrain from a complete discussion due to the auxiliary nature of these figures.

In figure~\ref{fig:200GeV}, we show the analogue of figure~\ref{fig:flags_early} but for intermediate case $m_\phi = 200\,\mathrm{GeV}$. For this mass, the resulting constraints are just at the transition between being dominated by non-thermal scattering for smaller masses and FSR for higher masses. 
Qualitatively, we find it to be very similar to the cases that have already been discussed in greater detail in section~\ref{sec:first_look}.

In figure~\ref{fig:zeta_lim}, we compare in the $m_\phi-f_\phi$ plane \textit{(i)} the numerical results obtained in this work (full lines), and \textit{(ii)} the resulting EM limits by assuming only FSR (dotted lines). We find that for particles heavier than the EW scale that are sufficiently long-lived, case \textit{(ii)} provides an excellent approximation, as all other effects are subleading.\footnote{This statement is mildly lifetime-dependent, because the EM part of the EW shower from FSR has a maximum contribution at around $\tau_\phi \sim 10^8$-$10^9\,$s and for values close to this peak, the dominance of the limits happens at even smaller masses. This becomes apparent by comparing the lines of $\tau_\phi=10^9\,$s and $\tau_\phi=10^{11}\,$s.} Moreover, we extend the line from case \textit{(ii)} to masses beyond the ones considered in this work, in order to provide a projection for even heavier long-lived particles.

Finally, in figure~\ref{fig:aux}, we show some of the auxiliary quantities that have been used in section~\ref{sec:MC}. More precisely, we visualise  $\zeta_\text{hd}$ (upper left), $N_\text{hd}$ (upper right), and $K_\text{hd}$ (lower centre) originating from both non-thermal scattering (red, $x = \sqrt{s}$) and FSR (blue, $x=m_\phi$). In general, we find that $\zeta_\text{hd}^\text{fsr}$ and  $N_\text{hd}^\text{fsr}$ behave qualitatively similar to $\zeta_\text{em}^\text{fsr}$ from figure~\ref{fig:EW_shower}. In addition, the average kinetic energies $K_\text{hd}^\text{fsr}$ and $\langle K_\text{hd}^\text{n-th} \rangle$ grow roughly linearly beyond the nucleon threshold. Instead, for the quantities $\langle \zeta_\text{hd}^\text{n-th} \rangle$ and $\langle N_\text{hd}^\text{n-th} \rangle$ from non-thermal scattering, we observe larger contributions around the EW scale with a strong decline beyond it. 

\clearpage

\bibliographystyle{JHEP}
\bibliography{biblio.bib}
\end{document}